\documentclass[12pt]{article}
\pdfoutput=1
\usepackage{comment}
\usepackage{jheppub}
\usepackage{xcolor,enumitem}
\usepackage{mathrsfs,multirow}
\usepackage{pgfplots}
\pgfplotsset{compat=1.17}
\usetikzlibrary{decorations.pathreplacing,calc}

\newcommand\shihab[1]{\textcolor{red}{(Shihab: #1)}}
\newcommand\arash[1]{\textcolor{red}{(Arash: #1)}}

\author[a]{Arash Arabi Ardehali,}
\author[b]{Mathieu Boisvert,}
\author[b]{Shehab Hossam Fadda$\,$}

\affiliation[a]{Physics Department, Sharif University of Technology,\\
Azadi Avenue, Tehran, Iran}

\affiliation[b]{C.N. Yang Institute for Theoretical Physics, Stony Brook University,\\ Stony Brook, NY 11794, USA}

\emailAdd{a.a.ardehali@gmail.com}\emailAdd{mathieu.boisvert@stonybrook.edu}\emailAdd{shihab.fadda@stonybrook.edu}

\title{Cardy limit of the 3d superconformal index}

\date{}

\begin{document}

\abstract{We study the superconformal index \(Z(q)\) of 3d \(\mathcal{N}=2\) gauge theories in Cardy-like limits \(\beta = \log \tfrac{1}{q} \to 0^+\) using its closed-form expression as a basic hypergeometric integral, extending techniques recently developed for elliptic hypergeometric integrals in the 4d \(\mathcal{N}=1\) context. For theories with vectorlike matter content we find: 
\begin{itemize}
  \item \textbf{First sheet (\(q \to 1\))}: 
  \[
    Z(q) \sim \beta^{-\#},
  \]
  where the exponent \(\#\) is determined by a \emph{multiscale decomposition} of the BPS moduli space (appearing in the Coulomb branch localization formula) into near zones, Gaussian zones, and an outer patch.
  \item \textbf{Second sheet (\(q \to e^{2\pi i}\))}:  
  \[
    Z(q) \sim e^{\#/ \beta}.
  \]
  Here, the long-standing puzzle of apparent gauge-enhancing saddles is resolved (in the absence of Chern--Simons couplings) via a novel \emph{Lorentzian factorization} formula that establishes complete screening.
\end{itemize}

A key insight is the use of \emph{Poisson resummation}, which streamlines the asymptotic analysis, sharpens the link to Kaluza--Klein effective field theory, and provides a dual description of parts of the BPS moduli space in terms of punctured surfaces.

The Lorentzian factorization formula also emerges from Poisson resummation, though applied after a contour crossing in moduli space. This suggests that by relating the Lorentzian factorization formula---originating on the Coulomb branch---to earlier vortex-antivortex factorization results on the Higgs branch, one can obtain a correspondence between Coulomb branch monopoles and Higgs branch vortices, mediated by the 2d duality implemented through Poisson resummation.}

\maketitle

\section{Introduction and summary}\label{sec:intro}

Over the past decade, substantial progress has been made in understanding Cardy-like limits \(\beta := \log \tfrac{1}{q} \to 0^+\) \cite{Cardy:1986ie} of superconformal indices in 4d supersymmetric QFTs.\footnote{See \cite{Ardehali:2021irq} and references therein.} Beyond their intrinsic mathematical interest \cite{Rains:2006dfy,van2007hyperbolic,ArabiArdehali:2018ngl}, the results have found applications ranging from supersymmetric gauge dynamics (e.g.~\cite{Aharony:2013dha,Buican:2015hsa,ArabiArdehali:2019zac,Amariti:2024bdd,Nakanishi:2024tnj}) to holography (e.g.~\cite{ArabiArdehali:2014otj,ArabiArdehali:2014cey,Choi:2018hmj,Honda:2019cio,ArabiArdehali:2019tdm,Cabo-Bizet:2019osg,Kim:2019yrz,ArabiArdehali:2019orz}) and the SCFT/VOA correspondence (e.g.~\cite{Neitzke:2017cxz,Dedushenko:2019yiw,ArabiArdehali:2023bpq,ArabiArdehali:2024ysy,ArabiArdehali:2024vli}). In this work, we take the first steps toward a comparably precise analysis of the superconformal index \(Z(q)\) of 3d \(\mathcal{N}=2\) gauge theories. This has been a long-standing challenge in supersymmetric QFT.

The 3d superconformal index is defined as \cite{Bhattacharya:2008zy,Kim:2009wb,Imamura:2011su,Kapustin:2011jm,Dimofte:2011py,Aharony:2013kma}
\begin{equation}
    Z(q):=\mathrm{Tr}^{}_{S^2}(-1)^{2j} q^{R/2+j}.\label{eq:index_def_intro}
\end{equation}
The trace is over the $S^2$ Hilbert space, $j$ is the angular momentum quantum number, and $R$ is the $U(1)_R$ charge. For simplicity, our main focus will be on rank-one cases where the gauge group is $G=U(1)$ or $\mathrm{SU}(2)$, but most of our methods can be applied to theories with more general compact gauge groups as well.

\subsection*{Motivation and plan of the paper}

Cardy-like limits of superconformal indices in diverse dimensions are of interest for at least two research directions. The first is supersymmetric gauge dynamics on $\mathbb{R}^{d-1}\times S^1$. Since the index can be thought of as a BPS partition function on $S^{d-1}\times S_{\beta}^1$, and since it depends only on the ratio of the radii $\frac{\beta/2\pi}{ r_{S^{d-1}}}$, the Cardy limit $\beta\to0$ is equivalent to the decompactification limit of the $(d-1)$-sphere. One thus heuristically expects connections between the Cardy limit of the index and the dynamics on $\mathbb{R}^{d-1}\times S^1$. The natural lens through which to see the link is Kaluza-Klein effective field theory (KK EFT), arising from harmonic expansion of the fields around $S^1.$ This connection has been fleshed out and used to great effect in the context of the 4d index and its relation with gauge dynamics on $\mathbb{R}^3\times S^1$; see e.g.~\cite{Aharony:2013dha,Aharony:2013kma,ArabiArdehali:2015ybk,ArabiArdehali:2019zac,ArabiArdehali:2024ysy,ArabiArdehali:2026kvt}. In the context of the 3d index and its relation with gauge dynamics on $\mathbb{R}^2\times S^1$ \cite{Aharony:2017adm}, a comprehensive picture has remained elusive. An outstanding challenge in this context has been the determination of \emph{dominant field configurations} in the Cardy limit. This is a particularly interesting problem for periodic boundary conditions around the circle, corresponding to the ``1st sheet'' of the index, i.e. $Z(q)$ itself.

The second research direction is holographic microstate counting. Asymptotic degeneracy of (high quantum number) BPS states can be extracted from a Legendre transform of the Cardy (or small $\beta$) limit of the index, and compared with the Bekenstein-Hawking entropy of bulk BPS black holes via Boltzmann's formula. See \cite{Choi:2019zpz,Bobev:2019zmz,Benini:2019dyp,Nian:2019pxj,Choi:2019dfu,GonzalezLezcano:2022hcf,Bobev:2022wem,BenettiGenolini:2023rkq,Bobev:2024mqw} for examples in AdS$_4$/CFT$_3$. Since the bulk black holes correspond to non-periodic boundary conditions around the circle, these holographic studies often feature the 3d superconformal index on its ``2nd sheet'', i.e.~$Z(q\,e^{2\pi i})$, which is the BPS partition function with $U(1)_R$-twisted boundary conditions around $S_\beta^1$. An outstanding challenge in this context has been the presence of certain \emph{apparent gauge-enhancing saddles} (with coincident eigenvalues of the holonomy matrix) in the Cardy limit of the index, that are incompatible with expected asymptotics.\footnote{See footnote~7 in \cite{Choi:2019zpz}. See also \cite{Jain:2013py} for related discussions in a non-supersymmetric context.}

We address both of the outstanding challenges mentioned above. The first will be addressed through multiscale analysis: the moduli space of BPS field configurations is {divided} into patches where analytic control is possible locally, and then the patches are {conquered one at a time}. The biggest (and simplest) patch, which corresponds to the BPS configurations where there is no light charged matter field in the 2d KK EFT on $S^2$, will be called \emph{the gauge patch}. Poisson resummation on the gauge patch yields integrals over \emph{punctured surfaces}, which can be analyzed using the multidimensional stationary phase method. The rest of the patches, corresponding to the \emph{punctures}, will be treated less systematically here, but still reliably, at least in the examples considered. The dominant field configurations in the small-circle limit correspond to the patches that dominate the index as $\beta\to0.$

The second challenge will be addressed through a combination of techniques involving contour deformation, Poisson resummation, and Lorentzian factorization, proving that the problematic saddles of the 2nd-sheet index are exponentially suppressed (or ``non-perturbatively screened'').

Powerful tools hence emerge for asymptotic analysis of the 3d superconformal index in Cardy-like limits, both on the 1st and 2nd sheet. We suspect that our Lorentzian factorization formula, in an interplay with the vortex-antivortex factorization results of \cite{Hwang:2012jh,Hwang:2015wna,Hwang:2017kmk,Crew:2020psc,Bullimore:2020jdq}, can find further applications in non-perturbative studies of 3d $\mathcal{N}=2$ gauge theories, perhaps shedding further light on the monopole-vortex relations in~\cite{Aharony:1997bx,Intriligator:2013lca,Intriligator:2014fda}. We also anticipate our multiscale approach to be useful in analyzing supersymmetric gauge dynamics on $\mathbb{R}^2\times S^1$, and in studying the dynamics of various 2d $\mathcal{N}=(2,2)$ gauge theories arising from circle-compactified 3d QFTs, analogously to the 4d$\,\to\,$3d story in \cite{ArabiArdehali:2019zac,ArabiArdehali:2024ysy,ArabiArdehali:2024vli,ArabiArdehali:2026kvt}.

In the following, we first give a detailed summary of our tools and results. Our terminology is explained in Subsection~\ref{subsec:terminology}. A detailed comparison between our results and those of the previous literature, as well as a highlight of some important open problems is given in Subsection~\ref{subsec:previous_work}. Section~\ref{sec:generalities} delves into the multiscale analysis of the 3d index in Cardy-like limits and makes contact with the 2d Kaluza-Klein EFT. In Section~\ref{sec:example} various rank-one examples are treated using the machinery developed in Section~\ref{sec:generalities}. The examples are chosen such that they all have a simple dual without gauge fields. The dual index can thus be analyzed with elementary tools. This provides a crucial benchmark for our gauge-theory analysis and serves to calibrate our novel techniques. In the non-abelian examples, we will see that the gauge-enhancing saddles on the 2nd sheet are incompatible with the expected asymptotics. This motivates our development of the Lorentzian factorization technique in Section~\ref{sec:factorization&screening}. The factorization formula will then be used to demonstrate the full screening of gauge-enhancing saddles of typical 2nd-sheet indices in non-chiral theories. Section~\ref{sec:open_problems} contains a conceptual overview, a technical overview, and seven concrete problems formulated to chart some of the most interesting territories for future exploration. Appendix~\ref{app:AA} contains derivations of the main asymptotic estimates used in our work. It also reviews the multidimensional stationary phase method, which we apply to the surface integrals arising from Poisson resummation.

\subsection*{Starting point: the sum-integral expression for the index}

The $S^2\times S^1$ superconformal index of a rank-one 3d $\mathcal{N}=2$ gauge theory takes the form
\begin{equation}
    Z(q\,e^{2\pi i\nu})=\frac{1}{|W|}\sum_{m\in\mathbb{Z}}\int_{-\frac{1}{2}}^{\frac{1}{2}}\mathrm{d}x\ f_\nu(u;\beta)\,\tilde{f}_\nu(\bar{u};\beta)\,,\label{eq:Z_intro}
\end{equation}
on the $(\nu+1)^{\text{th}}$ sheet. Here $x$ parametrizes the gauge holonomies around $S^1$, and $m$ is the magnetic flux on $S^2.$ The order of the Weyl group of the gauge group ($=\!1$ for~$U(1)$ and $=\!2$ for $\mathrm{SU}(2)$) is denoted $|W|.$ The functions $f_\nu,\,\tilde{f}_\nu$ are special functions whose explicit form we suppress in this section (see Eqs.~\eqref{eq:CS_fac}--\eqref{eq:Ch_fac}). The variable $u$, which parametrizes the moduli-space of BPS field configurations on $S^2\times S^1$, is defined as
\begin{equation}
    u:=x-i\frac{m\beta}{4\pi},
\end{equation}
where $\beta$ is the ratio of the circumference of $S^1$ to the radius of $S^2$. Note that $u$ is valued in a complex strip, to be called \emph{the $u$-strip}, because $\mathrm{Re}\,u=x\in(-\frac{1}{2},\frac{1}{2}]$. Defining
\begin{equation}
    y:=e^{2\pi iu}=e^{2\pi ix+\frac{m\beta}{2}},
\end{equation}
we may alternatively speak of \emph{the $y$-plane}.

\subsection*{Main challenge: multiscale structure of the integrand in the Cardy limit}

Since we want to take the $\beta\to0$ limit inside the integrand of \eqref{eq:Z_intro}, we need asymptotic estimates for $f_\nu,\,\tilde f_\nu$ that are uniformly valid throughout the domain $x\in(-\frac{1}{2},\frac{1}{2}].$ (There is multiscale behavior in $m$ too, but that is straightforward to deal with, especially by taking advantage of the holomorphy in $u$. See the discussion below and in Section~\ref{sec:generalities}.)

We give such a uniform estimate in Section~\ref{sec:generalities} (Eq.~\eqref{eq:index_Cardy_master}, based on \eqref{eq:qPoch_master_estimate_body}). However, it is too complicated to be useful for standard methods of asymptotic analysis, such as the saddle-point technique. The reason is that the integrand (assuming for now that $m\sim\mathcal{O}(\beta^0)$) has singular behavior near specific $x$, where charged matter fields become light in the 2d Kaluza-Klein EFT. Denote these specific $x$ by $x_{p_j}$, with the subscript $j$ to emphasize that there can be several of them. The uniform estimate is complicated because it simultaneously encodes the rich physics near the singularities and the simple physics far from them. Because of this multiscale structure of the uniform estimate, we can not apply standard tools (such as the saddle-point method or the stationary phase technique) to integrate and sum over it.

Such situations are common in multiscale analysis---or \emph{boundary layer theory}. To make progress, one usually decomposes the domain into smaller patches (or layers) that are under better analytic control locally. One thus gives up uniform control over the whole domain, and instead aims at uniform control over smaller patches. This is in line with the philosophy of effective field theory. The local estimates often correspond to keeping the fields that are light in the patch and integrating out the fields that are heavy. Section~\ref{subsec:Cardy_estimates} is largely dedicated to fleshing out this interplay between multiscale analysis and effective field theory.

We refer to the subset of $(-\frac{1}{2},\frac{1}{2}]$ where $x$ is at least $\epsilon$ away (with $0<\epsilon\ll1$) from all $x_{p_j}$ as \emph{the gauge patch}. Nowhere on this patch there is any light charged matter field in the 2d KK EFT: the 2d field content is pure super-glue, possibly with uncharged light matter multiplets. The $\epsilon$ neighborhoods of $x_{p_j}$ will be called \emph{matter patches}, since they support light, charged matter fields in the 2d EFT. 

The multiscale structure in $m$ can now be incorporated by taking advantage of the holomorphy in $u$: the matter patches are complexified into small disks.\footnote{To see why the $\epsilon$ intervals complexify into disks (and not thin strips for instance), one has to understand the large-$m$ regime. This will be discussed in Section~\ref{sec:generalities}.} These disks will be called \emph{punctures} and denoted $p_j$. The gauge patch itself complexifies into a \emph{punctured $u$-strip} denoted $S_p$ (or alternatively a \emph{punctured $y$-plane} denoted $\mathbb{C}_p$).

On the punctured strip, the functions $f_\nu,\,\tilde{f}_\nu$ simplify in the $\beta\to0$ limit as\footnote{For simplicity, we assume in this section that our 3d theory has no Cartan-neutral chiral multiplets $\Phi$ with $R$-charge $r_\Phi\in\frac{2}{\nu }\mathbb{Z}$, otherwise the RHS should by multiplied by $Z_{S^2}^{\rho=0,\frac{\nu r_\Phi}{2}\in\mathbb{Z}}$ in \eqref{eq:Z_rho=0}, and also $g$ as defined in \eqref{eq:I&g_def} acquires $\beta$ dependence through $\Omega^L$ in \eqref{eq:W_Omega_simpl_out}. See Section~\ref{subsec:appetizer} for an example.} (suppressing $\nu$ and $q\,e^{2\pi i\nu}$ to avoid clutter):
\begin{equation}
    Z^{}_{S_p}=\frac{1}{|W|}\sum' \int'\mathrm{d}x\ e^{i\frac{I(u,\bar u)}{\beta}}\,g(u,\bar u)\ \big(1+\mathcal{O}(\beta)\big)\,,\label{eq:Z_Sp_intro}
\end{equation}
with the primes indicating the excisions of the $\epsilon$ punctures around $x_{p_j}$. The functions $I,\,g$ are real-valued, and can be expressed in terms of the effective twisted superpotential $W$ and the effective dilaton $\Omega$ of the 2d KK EFT on $S^2$ as
\begin{equation}
    i\frac{I(u,\bar u)}{\beta}=W(u;\beta)-\overline{W(u;\beta)},\qquad \mathrm{d}x\ g(u,\bar u)=\frac{\mathrm{d}\frac{2\pi x}{\beta}}{2\pi}\ e^{-i\pi\big(\Omega(u;\,\beta)-\overline{\Omega(u;\,\beta)}\big)}.\label{eq:I&g_def}
\end{equation}
That the supersymmetric localization of a 2d $\mathcal{N}=(2,2)$ gauge theory with these $W,\,\Omega$ on $S^2$ yields \eqref{eq:Z_Sp_intro} \cite{Benini:2012ui,Aharony:2017adm}, establishes the connection between the Cardy limit and the KK EFT on the punctured strip (or on the complexified gauge patch).

There remain the punctures (or the complexified matter patches). These are more difficult to analyze. The source of difficulty is that \emph{the punctures themselves involve multiple scales}. For example, as we shall see in Section~\ref{sec:generalities}, a disk of radius $\sim\mathcal{O}(\beta)$ around $x_{p_j}$ (to be called a ``near zone'') contributes a different $\beta$ scaling to the index than a layer of size $\mathcal{O}(\beta^{1/2})$ surrounding it (to be called a ``Gaussian zone''). This calls for a further decomposition of the punctures into subzones that are under better analytic control. Physically, this corresponds to isolating the scales at which different mechanisms (or terms in the effective action of the 2d EFT) are dominant. Mathematically, this is an instance of identifying the important scales in singular perturbation problems through the principle of dominant balance \cite{miller2006applied}. In this work, we take the following preliminary step toward a full multiscale resolution of the punctures.

\subsection*{Novel idea 1: charting the scale hierarchy inside the punctures}

We break up the punctures into two zones:
\begin{equation*}
    \begin{split}
        &\text{\textbf{\small\emph{near zone}} \qquad\qquad\qquad$|u-x_{p_j}|<\frac{\beta\Lambda}{2\pi}$,}\\
        &\text{{\small intermediate zone}\qquad\ \ $\frac{\beta\Lambda}{2\pi}\le|u-x_{p_j}|<\epsilon$},
    \end{split}
\end{equation*}
with an arbitrary but fixed $\Lambda\gg1.$ We conjecture that a puncture $p_j$ contributes negligibly (in the sense that it never dominates Cardy-like asymptotics of the index) except from its near zone and the Gaussian subset of its intermediate zone
\begin{equation*}
    \hspace{-1.3cm}\text{\textbf{\small\emph{Gaussian zone}} {\small($\subset\, $intermediate zone)}\qquad\hspace{.2cm} $\epsilon_2\sqrt{\beta}<|u-x_{p_j}|<\Lambda_2\sqrt{\beta}$,}
\end{equation*}
where $\epsilon_2\ll1$ and $\Lambda_2\gg1$. 

Physically, the near zone is where the light charged 2d chirals in the KK EFT have $\mathcal{O}(1)$ masses from the familiar Higgs mechanism on the Coulomb branch. The effective description is therefore through 2d $\mathcal{N}=(2,2)$ Coulomb branch dynamics with the inclusion of the light charged chirals. The Gaussian zone is where the 2d photon multiplet mass term descending from the 3d tree-level CS coupling becomes $\mathcal{O}(1)$. The effective description is hence through the dynamics of a 2d massive photon multiplet coupled to charged chirals that now have $\mathcal{O}(1/\sqrt\beta)$ masses and can be integrated out.
If there is no tree-level CS coupling, there is no new physics associated with the Gaussian zone, and the near zone would be the only relevant region of the puncture.

Our conjecture relies on the observation that the intermediate zone hosts no other scale with new physics besides the Gaussian zone. The examples studied in Section~\ref{sec:example} also corroborate the conjecture.

As we will see in our case study of the $U(1)_{1/2}+\Phi_{+1}$ theory (i.e. the electric side of elementary mirror symmetry), in non-vectorlike theories new scales might be associated to the part $|u-x_{p_j}|\sim\frac{\beta}{\log\frac{1}{\beta}}$ of the near zone where the running FI term of the 2d EFT becomes $\mathcal{O}(1)$, and the part $m\sim\frac{\log\frac{1}{\beta}}{\beta}$ (or $|\mathrm{Im}u|\sim\log\frac{1}{\beta}$) of the complexified outer patch (or the punctured $u$-strip) where the EFT twisted superpotential becomes $\mathcal{O}(1)$. This motivates additional decomposition of the near zones and the punctured $u$-strip, but we leave that to future work.




\subsection*{Novel idea 2: Poisson resummation outside the punctures}

With the integrand under better control by simplified estimates appropriate to each scale, it remains to integrate and sum over it. 

We do this using Poisson resummation on the punctured strip and the Gaussian zones. Near zones, on the other hand, contain (gamma function) poles and are therefore not amenable to Poisson resummation. However, their contribution is often simple enough that can be analyzed directly on a case-by-case basis, at least on the first sheet of our various examples.

Postponing the discussion of the near zones and Gaussian zones to Sections~\ref{sec:generalities}, \ref{sec:example}, we keep our focus on the punctured strip in this introductory outline. As explained in Section~\ref{sec:generalities}, Poisson resummation allows on the punctured strip the replacement:
\begin{equation}
    \sum_m' \int'\mathrm{d}x\ \longleftrightarrow\ \frac{2\pi i}{\beta}\sum_{w\in\mathbb{Z}}\int_{S_p}\mathrm{d}u\wedge\mathrm{d}\bar{u}\ e^{-\frac{4\pi^2}{\beta}w(u-\bar u)}.\label{eq:punctured_PR_intro}
\end{equation}
We interpret $w$ as labeling the winding sectors of the dual-photon twisted chiral multiplet in the 2d KK EFT. Equation~\eqref{eq:Z_Sp_intro} thus turns into:
\begin{equation}
    Z_{S_p}=\frac{1}{|W|}\frac{2\pi i}{\beta}\sum_{w\in\mathbb{Z}}\int_{S_p}\mathrm{d}u\wedge\mathrm{d}\bar{u}\ e^{i\,\frac{I^{(w)}(u,\bar u)}{\beta}}\,g(u,\bar u)\ \big(1+\mathcal{O}(\beta)\big)\,,\label{eq:Z_Sp_PRd_intro}
\end{equation}
where (defining $W^{(w)}(u;\beta):=W(u;\beta)-4\pi^2wu/\beta$)
\begin{equation}
    I^{(w)}(u,\bar u):=-i\beta\big(W^{(w)}(u;\beta)-\overline{W^{(w)}(u;\beta)}\big)=I(u,\bar u)+4\pi^2 i\,w(u-\bar u)\,.
\end{equation}

The surface integrals in \eqref{eq:Z_Sp_PRd_intro} are now amenable to the multidimensional stationary phase method. It turns out that only a finite number of the winding sectors host stationary-phase loci. Therefore the infinite sum over $w$ (unlike the original sum over $m$) is not a hindrance to our asymptotic analysis!

The stationary-phase loci are the critical points of $I^{(w)}(u,\bar u)$. These correspond to the (perturbative) Bethe vacua of the 2d KK EFT. See Section~\ref{sec:generalities}.

\subsection*{Asymptotics on the 1st and 2nd sheet}

Label the connected components of the set of stationary-phase loci of \eqref{eq:Z_Sp_PRd_intro} by $u_k^\ast,w_k^\ast$. The subscript $k$ is there to emphasize that there may be several of them.

Going through multiple examples in Section~\ref{sec:example}, we find that \textbf{\small on the 1st sheet} (i.e. considering $Z(q)$ itself) typically:
\begin{itemize}
    \item[1a)] either $I(u;\bar u)=0$, so that only the $w=0$ sector has stationary phase, albeit lacking an exponential integrand altogether and displaying only power-law growth;
    \item[1b)] or $I(u;\bar u)$ is non-trivial but $I^{(w_k^\ast)}(u_k^\ast,\overline{u_k^\ast})=0,$ so that the stationary-phase growth ends up being again power-law.
\end{itemize}
In the five examples studied in Section~\ref{sec:example}, scenario 1a arises on the 1st sheet of non-chiral theories (SQED with $N_f=1$ or $3$, and $\mathrm{SU}(2)$ SQCD with $N_f=3$), and scenario 1b on the 1st sheet of chiral theories (the electric sides of the elementary mirror symmetry and the duality appetizer).

The punctured strip hence typically\footnote{We say \emph{typically} because, as discussed at the end of Section~\ref{subsec:mirror_sym}, sufficiently chiral gauge theories can display exponential growth even on their first sheet. The 4d analog of this is the fact that in sufficiently chiral gauge theories the first sheet asymptotics can be enhanced from the typical $e^{\#/\beta}$ to $e^{\#/\beta^2}$ (see e.g.~\cite{Cassani:2021fyv}).} yields power-law growth in $\frac{1}{\beta}$ on the 1st sheet. The punctures also yield power-law growth in all our examples, sometimes dominating the index and sometimes not; see Table~\ref{tab:patches_intro}.

\begin{table}[t]
    \centering
    \begin{tabular}{|l|c|c|c|c|}
    \hline
    theory&sheet& 
    \begin{tabular}{@{}c@{}}puncture \\ (near zone)\end{tabular} & 
    \begin{tabular}{@{}c@{}}puncture \\ (Gaussian zone)\end{tabular} & 
    \begin{tabular}{@{}c@{}}punctured \\ $y$-plane \end{tabular} \\ \hline \hline

    \multirow{2}{*}{SQED} & \textcolor{blue}{1\textsuperscript{st}} & -- & -- & \textcolor{blue}{$\times$} \\ \cline{2-5}
    & 2\textsuperscript{nd} & -- & -- & $\times$ \\ \hline

    \multirow{2}{*}{$U(1)_{+1/2} + \Phi_{+1}$} & \textcolor{blue}{1\textsuperscript{st}} & -- & -- & \textcolor{blue}{$\times$} \\ \cline{2-5}
    & 2\textsuperscript{nd}  & -- & -- & $\times$ \\ \hline

    \multirow{2}{*}{SQED$_{N_f=3}$} & \textcolor{blue}{1\textsuperscript{st}} & \textcolor{blue}{$\times$} & -- & -- \\ \cline{2-5}
    & 2\textsuperscript{nd} & -- & -- & $\times$ \\ \hline

    \multirow{2}{*}{$\mathrm{SU}(2)~\text{SQCD}_{N_f=3}$} & \textcolor{blue}{1\textsuperscript{st}} & \textcolor{blue}{$\times$} & -- & -- \\ \cline{2-5}
    & 2\textsuperscript{nd} & -- & -- & $\times$ \\ \hline

    \multirow{2}{*}{$\mathrm{SU}(2)_{-2}+ \Phi_{\square\!\square}$} & \textcolor{blue}{1\textsuperscript{st}} &  -- &\textcolor{blue}{$\times$} & -- \\ \cline{2-5}
    & 2\textsuperscript{nd} & -- & -- & $\times$ \\ \hline
    \end{tabular}
    \caption{The dominant patch in various rank-one examples.}
    \label{tab:patches_intro}
\end{table}

The \emph{typical growth on the 1st sheet is thus power-law} in $\frac{1}{\beta}$. The exponent has to be determined on a case-by-case basis in the competition between the punctured strip and the punctures.

\textbf{\small On the 2nd sheet} (i.e. considering $Z(q\,e^{2\pi i})$) we find that typically $I(u,\bar u)$ is non-trivial, and that the punctured strip hosts isolated stationary-phase points with $I^{(w_k^\ast)}(u_k^\ast,\overline{u_k^\ast})\neq 0.$ In our examples, as seen in Table~\ref{tab:patches_intro}, we find that the punctures are always negligible on the second sheet. The \emph{typical growth on the 2nd sheet is thus exponential} in $\frac{i}{\beta}$.

\vspace{.2cm}
In non-abelian cases, however, extra care is needed when
\begin{itemize}
    \item[2a)] either $u_k^\ast=0$ or $\pm\frac{1}{2}$, so that the stationary-phase point is gauge enhancing;
    \item[2b)] or $\mathrm{Im}u_k^\ast=\pm\infty,$ so that the stationary-phase point corresponds to infinite flux.
\end{itemize}
In both cases $g(u_k^\ast,\overline{u_k^\ast})=0,$ implying that the growth in \eqref{eq:Z_Sp_PRd_intro} is suppressed at least by additional powers of $\beta$. In our examples in Section~\ref{sec:example}, based on the easy asymptotics of the dual side, we expect that the suppression is actually beyond all orders in $\beta$.

Establishing such beyond-all-order suppressions directly in gauge theory is not an easy task with common techniques and has been an outstanding challenge in earlier literature. See for example the discussion in footnote~7 of \cite{Choi:2019zpz}.

We shall demonstrate such ``non-perturbative'' (or beyond-all-order) screenings in Section~\ref{sec:factorization&screening} for non-chiral theories---extension to chiral theories remains a major open problem. The key tool is Lorentzian factorization.

\subsection*{Novel idea 3: Poisson resummation on crossed (non-semiclassical) contours and non-perturbative saddle screenings from Lorentzian factorization}

We argue in Section~\ref{sec:factorization&screening} that, \emph{for non-chiral theories}, 
in \eqref{eq:Z_intro} we can replace
\begin{equation}
    \int_{-\frac{1}{2}}^{\frac{1}{2}}\mathrm dx\ \longrightarrow \int_{\ \overset{\text{\LARGE$|$}}{}\hspace{-.14cm}{\text{\large$\sqcup$}}\!\!\!\overset{\text{\LARGE$|$}}{}}\ \mathrm dx\,,
\end{equation}
by extending the integration contour vertically, from the two ends at $x=\pm\frac{1}{2}$, to $\mathrm{Im}x=+\infty$. The extended contour runs counter-clockwise. This extension is possible because $i$) the two half-infinite contours cancel each other due to their opposite direction and the $x\to x+1$ periodicity; $ii$) the integrand is exponentially suppressed as $\mathrm{Im}x\to+\infty$, for any $m$.

The residue theorem allows, \emph{on the second sheet}, a further replacement\footnote{This is the step that fails on the first sheet, as explained in Section~\ref{subsubsec:cut_obstruction} (unless suitable flavor fugacities are turned on). An implicit assumption we will be making throughout this paper, which in particular allows taking this step on the second sheet, is that all the (gauge-) charged 3d chiral multiplets have $R$-charge $0<r_\Phi<1$. The range $1<r_\Phi<2$ can be covered by straightforward modifications of our arguments. Charged chiral multiplets with $R$-charge either $1$ or outside the interval $(0,2)$ can presumably be treated via analytic continuation, but we do not attempt that here.\label{fn:crossing}}
\begin{equation}
     \int_{\ \overset{\text{\LARGE$|$}}{}\hspace{-.14cm}{\text{\large$\sqcup$}}\!\!\!\overset{\text{\LARGE$|$}}{}}\ \mathrm dx\ \longrightarrow\  \int_{\ \text{\LARGE$|$}\hspace{-.23cm}\text{\large$\downarrow$}\ \text{\LARGE$|$}\hspace{-.22cm}\text{\large$\uparrow$}}\ \ \mathrm dx\,.\label{eq:cont_crossing}
\end{equation}
We take the downward contour (labeled $\ell$) to coincide with the $\mathrm{Re}\,x=0$ axis, and the upward contour (labeled $r$) to coincide with the $\mathrm{Re}\,x=\frac{1}{2}$ line. We refer to these as ``{crossed}'' contours. The $\ell$ (resp. $r$) contour intersects the semiclassical contour $x\in(-\frac{1}{2},\frac{1}{2}]$ only at $x=0$ (resp. $x=\frac{1}{2}$). 

Next we implement the sum over $m$ via Poisson resummation:
\begin{equation}
    \sum_{m\in\mathbb{Z}}\int_{\ \text{\LARGE$|$}\hspace{-.23cm}\text{\large$\downarrow$}\ \text{\LARGE$|$}\hspace{-.22cm}\text{\large$\uparrow$}} \mathrm dx\ f(u;\beta)\,\tilde{f}(\bar{u};\beta)=\sum_{w\in\mathbb{Z}}\int_{-\infty}^\infty\mathrm{d}m\int_{\ \text{\LARGE$|$}\hspace{-.23cm}\text{\large$\downarrow$}\ \text{\LARGE$|$}\hspace{-.23cm}\text{\large$\uparrow$}} \mathrm dx\ f(u;\beta)e^{\frac{-4\pi^2}{\beta}wu}\ \tilde{f}(\bar{u};\beta)e^{\frac{4\pi^2}{\beta}w\bar u}\,.
\end{equation}

To arrive at factorizable surface integrals, we use crossed (Wick-rotated) variables: $x=ix^c$ on the $\ell$ contour, and $x=\frac{1}{2}+ix^c$ on the $r$ contour. We can then write
\begin{equation}
\begin{split}
    \int_{\ \text{\LARGE$|$}\hspace{-.23cm}\text{\large$\downarrow$}\ \text{\LARGE$|$}\hspace{-.22cm}\text{\large$\uparrow$}}\ \ \mathrm dx\ f(u;\beta)e^{\frac{-4\pi^2}{\beta}w u}\ \tilde{f}(\bar{u};\beta)e^{\frac{4\pi^2}{\beta}w\bar u}&=-i\int_{-\infty}^\infty\mathrm dx^c\ f(iu^-;\beta)e^{\frac{-4\pi^2}{\beta}w iu^-}\ \tilde{f}(iu^+;\beta)e^{\frac{4\pi^2}{\beta}wi u^+}\\
    &\hspace{-3cm}+i\int_{-\infty}^\infty\mathrm dx^c\ f(iu^-+\frac{1}{2};\beta)e^{\frac{-4\pi^2}{\beta}w (iu^-+\frac{1}{2})}\ \tilde{f}(iu^++\frac{1}{2};\beta)e^{\frac{4\pi^2}{\beta}w(i u^++\frac{1}{2})},
    \end{split}
\end{equation}
where $u^\pm:=x^c\pm\frac{\beta m}{4\pi}$. Adapting the measure to the Lorentzian coordinates $u^\pm$:
\begin{equation}
    \mathrm{d}m\,\mathrm{d}x^c=\frac{2\pi}{\beta}\mathrm{d}u^+\mathrm{d}u^-,
\end{equation}
the integrals now completely factorize
\begin{equation}
    Z(q\,e^{2\pi i})=-\frac{2\pi i}{\beta}\frac{1}{|W|}\sum_{w\in\mathbb{Z}}\Big(B^{(w)}_\ell(q)\,\widetilde{B^{(w)}_\ell}(q)-(\ell\leftrightarrow r)\Big),\label{eq:fully_factorized_intro}
\end{equation}
with the holomorphic blocks $B^{(w)},\widetilde{B^{(w)}}$ defined as
\begin{equation}
    B^{(w)}_\ell(q):=\int_{-\infty}^\infty\mathrm{d}u^- \,f(iu^-;\beta)\,e^{-\frac{4\pi^2}{\beta}iwu^-},\quad \widetilde{B^{(w)}_\ell}(q):=\int_{-\infty}^\infty\mathrm{d}u^+ \,\tilde f(iu^+;\beta)\,e^{\frac{4\pi^2}{\beta}iwu^+},
\end{equation}
\begin{equation}
    B^{(w)}_{\overset{}{r}}(q):=\!\int_{-\infty}^\infty\!\mathrm{d}u^- f(iu^-+\frac{1}{2}\,;\beta)\,e^{-\frac{4\pi^2}{\beta}iwu^-},\quad \widetilde{B^{(w)}_{\overset{}{r}}}(q):=\!\int_{-\infty}^\infty\!\mathrm{d}u^+ \tilde f(iu^++\frac{1}{2}\,;\beta)\,e^{\frac{4\pi^2}{\beta}iwu^+}.
\end{equation}

The Lorentzian factorization formula \eqref{eq:fully_factorized_intro} turns out to quickly establish the non-perturbative screening of gauge-enhancing saddles outside punctures. As explained in Section~\ref{sec:factorization&screening}, the gauge-enhancing saddles (at $x^\ast=0$ or $\frac{1}{2}$) on the punctured plane arise only in winding sectors where the integrand of the corresponding block ($B^{(w^\ast)}_\ell$ if $x^\ast=0$ and $B^{(w^\ast)}_r$ if $x^\ast=\frac{1}{2}$) is odd under reflection across the saddle. The block then vanishes due to its odd integrand, establishing that the gauge-enhancing saddles on the punctured plane do not contribute to $Z(q\,e^{2\pi i})$ at all.

We also give an argument in Section~\ref{sec:factorization&screening} that indicates the beyond-all-orders (in $\beta$) suppression of the infinite-flux saddles using \eqref{eq:fully_factorized_intro}. Interestingly, the argument relies on the $\ell$ and $r$ terms in \eqref{eq:fully_factorized_intro} canceling each other in the appropriate winding sectors.


\subsection*{Analogies and contrasts with the Cardy limit of the 4d index}

Readers familiar with the Cardy limit of the 4d superconformal index may find it instructive to compare our results for the 3d index with those of the 4d context.

Holographic considerations suggest \cite{Cabo-Bizet:2018ehj,Bobev:2019zmz} that superconformal indices display their fastest growth (associated with bulk black holes) on the 2nd sheet. We hence begin by comparing the 2nd-sheet results. Typically, one has
\begin{equation}
\begin{split}
    Z_{4d}(q\,e^{2\pi i})\sim\int\mathrm{d}x\ e^{i\frac{Q_2(x)}{\beta^2}} &\sim e^{i\frac{\#}{\beta^2}},\ \ \ \ \ \ \ \ \ \ \ \ \ \ \ \ \ \ \ \\
    Z_{3d}(q\,e^{2\pi i})\sim\sum_{w}\iint\mathrm{d}u\wedge\mathrm{d}\bar u\ e^{i\frac{I^{(w)}(u,\bar u)}{\beta}}&\sim e^{i\frac{\#}{\beta}}.
    \end{split}\label{eq:2nd_sheet_asy_intro}
\end{equation}
Since we expect local holomorphy in $q,$ these relations amount to exponential growth for suitably complexified $\beta$. In the 4d context $Q_2(x)$ is a piecewise quadratic function whose critical points correspond to the perturbative vacua on $\mathbb{R}^3\times S^1$ \cite{ArabiArdehali:2026kvt} (see also \cite{DiPietro:2016ond}). In the 3d context $I^{(w)}(u,\bar u)\propto W^{(w)}(u;\beta)-\overline{W^{(w)}(u;\beta)}$, so the critical points of $I^{(w)}$ correspond to the perturbative (Bethe) vacua on $\mathbb{R}^2\times S^1$ \cite{Aharony:2017adm}.

In the 4d setting, the exponent $\#$ in \eqref{eq:2nd_sheet_asy_intro} receives a universal contribution $\sim \frac{\pi^3}{6}(\mathrm{Tr}R-\mathrm{Tr}R^3)$ from a small neighborhood the origin $x=0$ on the moduli-space of holonomies \cite{Kim:2019yrz,Cabo-Bizet:2019osg}. This sets the leading asymptotics for suitably complexified $\beta$ (such that $e^{i\frac{\#}{\beta^2}}$ amounts to growth).\footnote{See also \cite{Cassani:2021fyv,Ohmori:2021dzb} for a viewpoint where 't~Hooft anomalies in the KK EFT take center stage.} In the 3d context, the universal contribution to the second-sheet index from small neighborhoods of gauge-enhancing stationary-phase points seems to be zero! (We demonstrate this for non-chiral theories, and our duality appetizer example suggests this is true more generally.) Therefore, the exponential growth of the index must arise from other saddles. There does not appear to be any universal saddle that sets the leading asymptotics for suitably complexified $\beta$ as in the 4d context, although for certain holographic models a universal asymptotics is expected to emerge at large $N$ \cite{Bobev:2019zmz,Choi:2019dfu}, with $F^{}_{S^3}$ setting the exponential growth.

On the 1st sheet we typically have
\begin{equation}
\begin{split}
    Z_{4d}(q)&\sim e^{\frac{\#}{\beta}},\\
    Z_{3d}(q)&\sim \frac{1}{\beta^{\#}}\,.
    \end{split}
\end{equation}
The exponents $\#$ on the 1st sheet do not appear to be governed by universal formulae, and are often harder to compute than their counterparts in \eqref{eq:2nd_sheet_asy_intro}.\footnote{Under suitable conditions such as non-chirality, there are universal contributions on the first sheet from a small neighborhood of $x=0$ or $u=0$, of the form $Z^{\ast}_{4d}\sim e^{\frac{16\pi^2}{3\beta}(c_{4d}-a_{4d})}$ \cite{Aharony:2013dha,DiPietro:2014bca} or $Z^{\ast}_{3d}\sim\frac{1}{\beta^{\frac{c_{2d}}{3}}}$ \cite{Aharony:2017adm}, but these contributions do not necessarily dominate the index (see \cite{ArabiArdehali:2015ybk,Hwang:2018riu,DiPietro:2016ond} for 4d examples and Section~\ref{subsec:SQED} for a 3d example).} The 3d computation seems to becomes particularly challenging due to the sum over $m$ if some near zones have non-vectorlike matter making them difficult to analyze directly.

Finally, in the 4d context the multiple scales difficulty was not severe, since subtraction methods allowed obtaining homogenized (single-scale) estimates on inner patches (see footnote~20 of \cite{Ardehali:2021irq} in particular). Cardy limit of the 4d index is thus roughly a two-scale problem. In the 3d context, besides the most natural scales of the \emph{near zones} and the \emph{outer patch} common between the two settings, as alluded to above there seem to be additional scales in the problem arising from the tree-level 3d CS terms (Gaussian zones), the 2d running FI parameters, and also the large-$m$ sectors. Cardy limit of the 3d index hence appears to be roughly a five-scale problem.

\subsection{Terminology}\label{subsec:terminology}

\paragraph{\small Chiral versus non-chiral.} We call a 3d gauge theory \emph{chiral} if it has non-vectorlike
matter content and/or nonzero tree-level gauge-gauge Chern-Simons (CS) coupling. Gauge theories with vectorlike matter content and zero tree-level gauge-gauge CS coupling are called \emph{non-chiral}. Although \emph{time-reversal invariance} can have more requirements than we demand of a theory to call it non-chiral, the two notions are essentially the same for our practical purposes in this work.

A non-chiral theory taken to the second sheet becomes chiral according to our definitions, because it is no longer vectorlike due to the $R$ twists around $S^1$. Alternatively, the $R$-twisted background breaks the time-reversal invariance.

\paragraph{\small Asymptotics.} On the 2nd sheet of the 3d index the asymptotics are of the form $e^{i\frac{D}{\beta}}$, with $D$ some real number depending on the theory. Although for $\beta\in\mathbb{R}$ (as we assume throughout our analysis) this is just a phase, we refer to this type of asymptotics as \emph{exponential growth}. This is because we expect from local holomorphy in $q$ that at least for suitably complexified $\beta$ (such that $e^{i\frac{D}{\beta}}$ amounts to growth) the asymptotic formula remains valid. 

When studying asymptotics of surface integrals via the method of stationary phase, we sometimes refer to the \emph{stationary-phase points} of the oscillating exponent as \emph{saddles} for brevity. This is not a major abuse of terminology, as the multidimensional stationary phase method relies on \emph{local} (but not global) factorization of the surface integrals into products of contour integrals near the stationary-phase points, and then applies the saddle point technique on the resulting small contours \cite{miller2006applied}.

\paragraph{\small Effective field theory.} We refer to the 2d $\mathcal{N}=(2,2)$ effective field theory arising from Kaluza-Klein (or harmonic) expansion of the 3d $\mathcal{N}=2$ multiplets around the circle $S^1_\beta$ as \emph{Kaluza-Klein (or KK) EFT}. In the literature, these types of EFT are sometimes called ``high-temperature'' or ``thermal'' EFT, to hint at the analogy with thermal physics, but since we do not consider thermal boundary conditions around the circle, we tend to avoid those terms.

The 2d multiplets whose mass is less than some (large enough) cutoff $\Lambda$ will be called \emph{light}, those with mass greater than $\frac{2\pi\epsilon}{\beta}$ for some (small enough) $\epsilon$ will be called \emph{heavy}, and the rest will be called \emph{middleweight}. With a slightly abusive extension of the terminology, we call a 3d multiplet light (or middleweight) if there is among its KK modes a 2d light (or middleweight) multiplet, and call it heavy otherwise.

We only keep the light 2d multiplets in the KK EFT in this work, and integrate out the rest. The effective action of the light 2d fields contains a tree-level piece coming from the 3d action, as well as corrections arising from integrating out the rest of the fields. The leading corrections are encoded\footnote{We are neglecting all actions that would vanish on the BPS locus of the 2d EFT on $S^2$.} in the one-loop effective twisted superpotential (which is analogous to the prepotential in the 4d $\mathcal{N}=2$ context) and the one-loop effective dilaton (which captures the supersymmetrized Einstein-Hilbert terms).

\paragraph{\small Multiscale analysis.} Postponing a systematic formulation to the future, we illustrate our multiscale definitions with a simple example here. Consider the function
\begin{equation}
    f_\text{toy}(z,\beta):=e^{-\frac{|z|}{\beta}}+(1-e^{-|z|}),\label{eq:ftoy}
\end{equation}
defined on the complex $z$-plane. We want to find uniform, single-scale approximations to $f_\text{toy}$ in the $\beta\to0$ limit. By \emph{uniform} we mean valid up to a certain error (which here we take to be $o(\beta^0)$) for all $z\,.$ By \emph{single-scale} we mean a function $h$ of only a single rescaled variable $\frac{z}{\beta^\alpha}$, with some real $\alpha$ which may be called the scaling exponent.\footnote{To make at least part of the formulation unambiguous, one must commit to a certain tolerable error (here $o(\beta^0)$), as well as to a permissible class of field renormalizations (here, linear with coefficients that are powers of $\beta$). In some contexts, it may be desirable to impose stronger notions than uniformity---for example suitably decaying error along non-compact directions---or to allow more flexible functional forms than $h(\frac{z}{\beta^\alpha})$---for example $\sum_j g_j(\beta)\, h_j(\frac{s_j(z)}{\beta^{\alpha_j}})$. Such generalizations are, in fact, necessary for some of the applications in the main text, but we leave their formalization to future work.

These forms of commitment in multiscale analysis are reminiscent of Galois’s insistence on expressing solutions in terms of radicals in his approach to polynomial equations. The remaining cutoff ambiguities in the choice of $\epsilon$s and $\Lambda$s, encoded often in renormalization groups (see e.g.~\cite{ArabiArdehali:2023bpq}), seem analogous to the permutation ambiguities in the choice of root orderings encoded in Galois groups. See also~\cite{Connes:2005zp,Ebrahimi-Fard:2006yct}, where renormalized perturbation theory---which from a Wilsonian perspective is seen as multiscale analysis of weakly coupled QFTs---is linked to Galois theory.}

The function $f_\text{toy}$ above has \emph{multiscale structure}. That is, it is a function of two rescaled variables: the fast variable $z/\beta$ and the slow variable $z$. In order to find uniform single-scale approximations to it, we decompose the $z$-plane into an \emph{outer patch} $|z|>\epsilon$ with some fixed $\epsilon,$ and a
\emph{near zone} $|z|<\Lambda\beta$ with some fixed $\Lambda$. This would leave an \emph{intermediate zone} $\beta\Lambda<|z|<\epsilon\,.$ We can do away with an intermediate zone in the present case (essentially because it does not contain any new scales) by letting $\epsilon$ and $\Lambda$ depend on $\beta$, for instance as $\beta\Lambda=\epsilon=\beta^\gamma$ with some 
\begin{equation}
    \gamma\in(0,1).
\end{equation}
We then speak of an \emph{extended near zone} $|z|<\beta^\gamma$ and an \emph{extended outer patch} $\beta^\gamma<|z|$.

On the extended near zone the 1st term dominates, and the leading asymptotics is
\begin{equation}
    f_\text{toy}(z,\beta)=e^{-|z|/\beta}+\mathcal{O}({\beta}^\gamma),\qquad\text{\small(extended near zone)}.\label{eq:ftoy_near}
\end{equation}
Hence $h(\cdot)=e^{-(\cdot)}$, $\alpha=1.$ On the (non-extended) near zone the error improves to $\mathcal{O}(\beta).$

On the extended outer patch the 2nd term dominates.
The leading asymptotics is
\begin{equation}
    f_\text{toy}(z,\beta)=1-e^{-|z|}+\mathcal{O}(e^{-\frac{1}{\beta^{1-\gamma}}}),\qquad\text{\small(extended outer patch).}\label{eq:ftoy_far}
\end{equation}
Hence $h(\cdot)=1-e^{-(\cdot)}$, $\alpha=0.$ On the (non-extended) outer patch the error is $\mathcal{O}(e^{-1/\beta})\,.$

In Section~\ref{sec:generalities}, the moduli space of holonomies $x\in(-\frac{1}{2},\frac{1}{2}]$ will be divided into an \emph{outer patch} where all Cartan charged fields (i.e. $\rho_\Phi\neq0$ and $\alpha_\pm$ weights) in the EFT are heavy, together with matter inner patches (or \emph{matter patches}, for short) where charged matter fields (i.e. $\rho_\Phi\neq0$ weights) become non-heavy, as well as \emph{gauge inner patches} where {only} W boson multiplets (i.e. $\alpha_\pm$ weights) become non-heavy. As we shall see, the contribution of the gauge inner patches is captured correctly if we treat them as if they are parts of the outer patch. This motivates defining
\begin{equation}
    \text{gauge patch}:=\text{outer patch}\cup(\cup\,\text{gauge inner patches})\,.
\end{equation}
Note that $(\text{gauge patch})=(\cup\,\text{matter patches})'\,.$ Once the flux sectors labeled by $m\in\mathbb{Z}$ are taken into account, since $x$ and $m$ naturally combine into the complex variable $u=x-im\beta/4\pi,$ the gauge patch naturally complexifies. The complexified gauge patch looks roughly like a punctured $x$-$m$ (or $u$-) strip, except that in the $m$ (or $\mathrm{Im}u$) direction it is discrete. Poisson resummation (or 2d duality) provides an alternative description of the complexified gauge patch as infinitely many \emph{continuous} punctured $u$-strips (or punctured $y$-planes)
\begin{equation}
    \text{complexified gauge patch}=\sum_w \text{punctured $u$-strip}=\sum_w \text{punctured $y$-plane}\,.
\end{equation}
When the context is clear, we often simply say \emph{gauge patch} instead of complexified gauge patch, and do not bother distinguishing between the complexified gauge patch and the puncture $u$-strip/$y$-plane.

\subsection{Relation to previous work and highlight of open problems}\label{subsec:previous_work}

Our work is a continuation of a long line of earlier papers on the Cardy limit of the 3d superconformal index \cite{Benini:2012ui,Aharony:2017adm,Jockers:2018sfl,Pasquetti:2019uop,Choi:2019zpz,Bobev:2019zmz,Nian:2019pxj,Choi:2019dfu,David:2020ems,GonzalezLezcano:2022hcf,Bobev:2022wem,BenettiGenolini:2023rkq,Amariti:2023ygn,Gukov:2023cog,Hayashi:2023txz,Bobev:2024mqw,Bobev:2025ltz,BenettiGenolini:2025jwe}. However, it is primarily a follow-up to the works of Aharony-Razamat-Willett \cite{Aharony:2017adm} where the connection with 2d EFT was established through the effective twisted superpotential and the effective dilaton, and Choi-Hwang \cite{Choi:2019dfu} where the exponential growth of 2nd-sheet indices was demonstrated.

\vspace{.4cm}
Compared to Aharony-Razamat-Willett \cite{Aharony:2017adm}, we have
\begin{itemize}
    \item achieved a clearer view of the \emph{multiscale structure of the BPS moduli-space}, and thereby succeeded in determining the dominant field configurations as $\beta\to0$ in various examples with periodic boundary conditions around $S^1_\beta$;
    \item recognized the role of \emph{Poisson resummation} in determining the asymptotics of 3d indices, and found its 2d EFT interpretation as duality on an appropriate patch of the BPS moduli-space (note that without Poisson resummation one can not get to surface integrals in chiral theories where Euler-Maclaurin does not apply);
    \item given precise \emph{arguments based on Lorentzian factorization for exponential suppression of the contribution of gauge-enhancing Bethe vacua} on the punctured moduli-space, at least for non-chiral theories (in \cite{Aharony:2017adm} instead expectations from dynamical SUSY breaking in the 2d EFT were appealed to, in order to motivate discarding of various gauge-enhancing Bethe vacua).
\end{itemize}

\vspace{.4cm}
Compared to Choi-Hwang \cite{Choi:2019dfu}, we have
\begin{itemize}
    \item developed tools for obtaining the \emph{power-law asymptotics typical on the 1st sheet} (in \cite{Choi:2019dfu} only the exponential asymptotics on the 2nd sheet was studied);
    \item given precise arguments, at least for non-chiral theories, to demonstrate the \emph{factorization of 2nd-sheet indices not only at the level of the integrand, but fully}, and used our factorization formula to argue for exponential suppression of the contribution of the gauge-enhancing Bethe vacua (note that our factorization formula \eqref{eq:fully_factorized_intro} is not easy to guess from just the factorization of the integrand; in particular we have an infinite sum over $w$ and both $\ell,r$ contours);
    \item clarified that without suitable contour deformations---or puncture excisions---and a subsequent \emph{Poisson resummation}, it is not possible to get surface integrals in general, as for instance Euler-Maclaurin does not necessarily apply in chiral cases; moreover, once arrived at surface integrals, the systematic method for asymptotic analysis (for $\beta\in\mathbb{R}$) is the method of multidimensional stationary phase.
\end{itemize}

Our multiscale analysis is a natural extension of the strategy that has been very effective in the context of the Cardy limit of the 4d superconformal index \cite{Ardehali:2021irq,ArabiArdehali:2023bpq}. While in the 4d context the underlying BPS moduli-spaces have been the real moduli-spaces of holonomies, in the 3d context here we encounter \emph{complex moduli-spaces} arising in winding sectors on the gauge patch. In \cite{Ardehali:2021irq} all multiplets that are not heavy were called light. Here, the new scale (of the \emph{Gaussian zones}) associated with the 3d tree-level CS couplings forced us to refine that definition by dividing the non-heavy multiplets into light and \emph{middleweight}.

Our linking of the Cardy limit of the 3d index to the 2d KK EFT through \emph{supersymmetric localization of EFT} on $S^2$ is a natural extension of the analysis of \cite{ArabiArdehali:2021nsx} where Cardy-like limits of the 4d index were linked to the 3d KK EFT via localization.

The appearance of \emph{Dotsenko-Fateev integrals} on the first sheet of 3d superconformal indices of non-chiral theories in the Cardy limit was first recognized in \cite{Pasquetti:2019uop}. We use this connection to obtain the subleading asymptotics on the first sheet in some examples.

Our factorization formula \eqref{eq:fully_factorized_intro} for the second-sheet index of non-chiral theories was anticipated in \cite{Beem:2012mb}, but the need for two contours $\ell,r$ as well as the $w$ sum over the weight lattice does not seem to have been envisioned there. Note that compared with the Higgs branch factorizations in \cite{Hwang:2012jh,Hwang:2015wna,Hwang:2017kmk,Crew:2020psc,Bullimore:2020jdq}, ours is a \emph{Coulomb branch factorization}.

Cardy limit of the 3d index has an interesting relation with the $b\to0$ limit of the partition function on the squashed three-sphere $S_b^3$ with the squashing parameter $b$ \cite{Choi:2019dfu,Gang:2019jut}. We do not explore this connection in the present work. It would be interesting to see what our factorization formula may imply in that context.

\subsection*{Open problems by topic}

Seven open problems are outlined in Section~\ref{sec:open_problems}. The more mathematically oriented readers may find problems \textbf{1}, \textbf{4}, \textbf{5}, \textbf{7} worth studying, while the more EFT oriented readers may find problems \textbf{2}, \textbf{3}, \textbf{6}, \textbf{7} interesting.

\paragraph{\small Complex analysis.} Problem \textbf{1} concerns extending the Lorentzian factorization formula \eqref{eq:fully_factorized_intro} to cases with nonzero tree-level gauge-gauge CS coupling.


\paragraph{\small Asymptotic analysis.} Problem \textbf{4} is to prove (or disprove) our conjecture that non-Gaussian parts of the intermediate zones can never dominate the index.

Problem \textbf{5} is obtaining a multiscale resolution of the estimate \eqref{eq:qPoch_master_estimate_body} for the $q$-Pochhammer symbol (similar to what \eqref{eq:ftoy_near}--\eqref{eq:ftoy_far} do for \eqref{eq:ftoy}).


Problem \textbf{7} concerns systematic treatment of the puncture (or complexified matter patch) contributions to the index, via GLSM/LG dualities, at least in some examples.

\paragraph{\small Anomalies in EFT.} To demonstrate non-perturbative screening of gauge-enhancing saddles outside punctures, we used the oddity of the integrand of the corresponding holomorphic block. Problem \textbf{2} is to find the EFT counterpart of the oddity as a (possibly mixed) $\mathbb{Z}_2$ anomaly.

\paragraph{\small Vortex-instantons in EFT.} Problem \textbf{3} is to relate patch competition on the 1st sheet (of at least non-chiral 3d gauge theories) to 2d gauge dynamics induced by vortex-instanton superpotentials \cite{Hori:2000kt}, analogously to the 4d$\,\to\,$3d story in \cite{ArabiArdehali:2019zac,ArabiArdehali:2024ysy}.

\paragraph{\small Dualities in EFT.} Problem \textbf{6} is to relate the Poisson resummation on the gauge inner patches to the proposals in \cite{Aharony:2016jki,Gu:2018fpm}.

\vspace{.2cm}
Also hints of previously unknown 2d $\mathcal{N}=(2,2)$ dualities arise in Section~\ref{sec:example},  as discussed below Eqs.~\eqref{eq:2d_duality?} and \eqref{eq:2d_duality2?}.

\section{Generalities on the index and its Cardy limit}\label{sec:generalities}

In this section, we first review the localization formula for the superconformal index of 3d $\mathcal{N}=2$ gauge theories. This will be the starting point of our asymptotic analysis in the rest of the paper. The formula, spelled out in Section~\ref{subsec:general_index}, expresses the index as a sum over (co-character valued) magnetic fluxes $m$ on $S^2$, and an integral over the moduli space of (Cartan torus valued) gauge holonomies $x$ around $S^1$.

Restricting to small fluxes (i.e. fluxes of order one, or  $m=\mathcal{O}(1/\beta^0)$), we show in Section~\ref{subsec:Cardy_estimates} that, in the Cardy limit, a single-scale (or practically useful) estimate valid uniformly over the moduli space of holonomies is not available for the integrand of the index. As usual in multiscale analysis, we hence decompose the moduli space into different patches, and use uniform estimates tailored to each patch. This multiscale approach sits naturally within the Wilsonian framework when comparing our asymptotic estimates with the 2d KK EFT governing each patch. We then proceed to argue, by recognizing that fluxes and holonomies naturally pair up in the combination $u:=x-im\beta/4\pi$, that the patches defined on the moduli space of holonomies naturally complexify into regions on the complex $y$-plane, where
\begin{equation}
    y:=e^{2\pi i u}=e^{2\pi i x+m\beta/2}.
\end{equation}
This ``holomorphy'' suggests, among other things, a hierarchy of scales in the magnetic flux, e.g. that fluxes of order $\frac{1}{\beta^0}$ or $\frac{1}{\beta}$ amount to different estimates for the integrand of the index. We will verify this from the 2d EFT point of view at the end of Section~\ref{subsec:Cardy_estimates}.

Note that since $m$ takes discrete values, our $y$ does not yet sweep a surface. In Section~\ref{subsec:surface_integral}, we show that actual surfaces emerge upon Poisson resummation of the integrand on the (complexified) \emph{patches that do not support massless charged chiral multiplets in the 2d EFT}. Since the patches supporting massless charged chirals must be excised from the $y$-plane, we end up with a \emph{punctured $y$-plane}.

Poisson resummation turns out to greatly simplify asymptotic analysis of the patches without massless charged chirals. This is because it replaces the sum over fluxes $m$ with a sum over windings $w$, and (as we shall see in Section~\ref{subsec:asymptotics}) only finitely many winding sectors host saddles in the Cardy limit.

We will also see in Section~\ref{subsec:asymptotics} that, for some theories and on some sheets of the index, it is not the punctured surfaces just described that dominate the asymptotics, but the punctures. In fact, the punctures themselves display multiscale behavior, and need to be further decomposed into subzones. The innermost such zone, with $u$ in a disk of radius $\sim\mathcal{O}(\beta)$ around the center of the puncture, will be called \emph{the near zone}. On the near zones, asymptotics of the integrand contains gamma functions (associated to the 2d light charged chirals) with poles inside the zone. Therefore Poisson resummation is not allowed on near zones (at least not immediately). Fortunately though---and this is what a useful multiscale decomposition must achieve---their asymptotic contribution is often simple enough to determine directly in our various examples, at least on the first sheet (except in Section~\ref{subsec:mirror_sym} where results from \cite{Benini:2012ui} will be used).

\subsection{The index as a sum-integral over fluxes and holonomies}\label{subsec:general_index}

The 3d $\mathcal{N}=2$ superconformal index is defined as \cite{Bhattacharya:2008zy,Imamura:2011su}
\begin{equation}
    Z(q):=\mathrm{Tr}^{}_{S^2}(-1)^{2j_3} q^{R/2+j_3}.\label{eq:index_def_main}
\end{equation}
Here $R$ is the $U(1)_R$ charge, the trace is over the Hilbert space on $S^2$, and $j_3$ is the spin quantum number associated with the $\mathrm{SO}(3)$ isometry of the $S^2$. When the theory is superconformal, we can alternatively trace over the space of local operators. We do not turn on fugacities for flavor symmetries, or FI terms for Abelian factors of the gauge group, although we expect our methods to straightforwardly generalize to accommodate their presence.

The superconformal index can be computed by a path integral on $S^2 \times S^1$, for which convenient Coulomb branch localization formulae are available. From this perspective, the problem is equivalent to studying a 2d theory with $\mathcal{N} = (2,2)$ supersymmetry with an infinite number of fields corresponding to the towers of KK modes around the circle. We normalize the radius of $S^2$ to be $1,$ and denote the circumference of $S^1$ by $\beta$. One may refer to $\beta$ as ``inverse temperature'', but we emphasize that we always consider supersymmetric boundary conditions around $S^1.$ The ``high-temperature'' limit we will ultimately be interested in corresponds to taking $\beta \to 0$. As is well known, the high-temperature asymptotics of the partition function will depend crucially on what boundary conditions one imposes for the fields as they go around the ``thermal'' circle. We begin the discussion by considering the periodic boundary condition, and discuss the $U(1)_R$ twisted boundary conditions afterwards.

The index of a 3d $\mathcal{N}=2$ gauge theory takes the form \cite{Kim:2009wb,Imamura:2011su,Kapustin:2011jm,Dimofte:2011py,Aharony:2013kma,Closset:2019hyt}:
\begin{equation}
    Z(q)=\frac{1}{|W|}\sum_{m \in \Gamma_{G}^{\vee}} \oint \prod_{a = 1}^{\text{rk}(G)}\frac{dz_a}{2\pi i z_a}\ e^{-S_{\text{CS}^+}}\times I^+_{\text{Chiral}}\times Z^{}_{\text{Gauge}}\,,\label{eq:index_factors}
\end{equation}
where the sum is over the GNO monopole sectors of the gauge group $G$. We often rewrite $\oint\prod_{a = 1}^{\text{rk}(G)}\frac{dz_a}{2\pi i z_a}$ as $\int\mathrm{d}^{r_G}x$, where $z_a=e^{2\pi ix_a},$ and take the fundamental domain $-1/2<x_a\le 1/2.$

The various pieces in \eqref{eq:index_factors} are as follows. The factor
$e^{-S_{\text{CS}^+}}$ encodes contributions from gauge-gauge and gauge-$U(1)_R$ Chern-Simons terms:
\begin{equation}
    e^{-S_{\text{CS}^+}} = (-1)^{k^{+}_{ab}m_a m_b}q^{\frac{k^+_{aR} m_a}{2}}z^{k^+_{ab}m_b}_{a},
\end{equation}
with the levels shifted to incorporate certain chiral multiplet contributions:
\begin{equation}
\label{eq:kgg_eff}
    k^+_{ab}:=k_{ab}+\sum_{\Phi}\sum_{\rho\in \mathcal{R}_\Phi}\rho_a\rho_b/2\,,
\end{equation}
\begin{equation}
k^+_{aR}:=k_{aR}+\sum_{\Phi}\sum_{\rho\in \mathcal{R}_\Phi}\rho_a(r_{\Phi}-1)/2\,.
\label{eq:kgR_eff}
\end{equation}

The second piece on the RHS of \eqref{eq:index_factors} is the one-loop determinant of the matter multiplets $\Phi$, and is expressed in terms of  $(a;q):=\prod_{n=0}^{\infty}(1-a\,q^n)$ as\footnote{The chiral multiplet contributions on the right-hand sides of \eqref{eq:kgg_eff} and \eqref{eq:kgR_eff} combine with $I^+_\text{Chiral}$ to form $Z_\text{Chiral}.$ We have kept them separate for analytic convenience later on.}
\begin{equation}
  I^+_{\text{Chiral}}=  \prod_{\Phi}\prod_{\rho\in \mathcal{R}_\Phi}\frac{(z^{-\rho}q^{-\rho_a m_a/2+1-r_\Phi/2}\,;q)}{(z^{\rho}q^{-\rho_a m_a/2+r_\Phi/2};q)}\,.
\end{equation}
Here $r_\Phi$ denotes the charge under $U(1)_R$, while $\mathcal{R}_\Phi$ denotes the representation of $\Phi$ under $G$, and $\rho$ labels the weights of this representation. We use the notation,
\begin{equation*}
    z^{\rho}=\prod_{a\in\{1,\dots,\mathrm{rk}\left(G\right)\}}z_a^{\rho_a}\,.
\end{equation*}

The third piece on the RHS of \eqref{eq:index_factors} is the one-loop determinant of the vector multiplet and reads
\begin{equation}
    Z_{\text{Gauge}} =q^{\frac{\alpha_{+j}m_j}{2}} \prod_{\alpha_+} (1-q^{-\frac{\alpha_{+j}m_j}{2}}z^{\pm\alpha_+})\,,
\end{equation}
with $\alpha_+$ denoting the positive roots of $G$. The $\pm$ notation means that we have left implicit a product over both choices of signs. 

Note that the zero-point  gauge- and $R$-charge pieces of $Z_\text{chiral}$ are incorporated into $e^{-S_{\text{CS}^+}}$ as seen in Eqs.~\eqref{eq:kgg_eff} and \eqref{eq:kgR_eff}---hence the $+$ superscripts. Rewritings of \eqref{eq:index_factors} are possible in terms of the gauge- and $R$-charges of the BPS monopoles
\begin{equation}
    c_j(m)=k_j(m)-\frac{1}{2}\sum_{\Phi}\sum_{\rho\in R_\Phi}\rho_j|\rho(m)|,\label{eq:c_j(m)}
\end{equation}
\begin{equation}
    R(m)=k_{gR}(m)-\frac{1}{2}\sum_{\Phi}(r_{\Phi}-1)\sum_{\rho\in R_\Phi}|\rho(m)|-\frac{1}{2}\sum_\alpha |\alpha(m)|,\label{eq:R(m)}
\end{equation}
as reviewed e.g. in \cite{ArabiArdehali:2024vli}, and also in terms of the other ``holomorphic'' combinations $k^-_{ab},\,k^-_{aR}$, but we stick to \eqref{eq:index_factors} in this work.\\

Now consider $U(1)_R$ twisted supersymmetric boundary conditions around $S^1$:
\begin{equation}
    \phi(x+\beta)=\, e^{i\pi\nu (F+R)}\, \phi(x).\label{eq:R_twisted_bc}
\end{equation}
These are conveniently encoded in the sheet of the index under consideration. Indeed, the index is generically not single-valued as a function of $q = e^{-\beta}$,  because $R/2+j_3$ in \eqref{eq:index_def_main} transforms non-trivially under $q \to q\, e^{2\pi i \nu}$ $(\nu \in \mathbb{Z})$ when $R/2+j_3$ is fractional. The main cases of physical interest correspond to the first and second sheet $\nu = 0,1$. The boundary condition is periodic on the first sheet ($\nu=0$) and amounts to an insertion of $(-1)^F$ in the trace.  On the second sheet ($\nu=1$) the boundary condition is minimally $R$-twisted and amounts to an insertion of $(-1)^{R}$. Expressions on a generic sheet are obtained by replacing $q \to q \,e^{2\pi i \nu},$  for the $(\nu+1)^{\text{th}}$ sheet. We also perform the change of variables $z_a \to z_a e^{i\pi\nu m_a}$ to compensate various phases.
Explicitly, the index on the $(\nu+1)^{\text{th}}$ sheet reads:
\begin{equation}
\begin{split}
\label{eq:indexsumint}
    Z_{\nu+1}(q):=Z(q\,e^{2\pi i\nu}) =\frac{1}{|W|}\sum_{m \in \Gamma_{G}^{\vee}} \oint \prod_{a = 1}^{\text{rk}(G)}\frac{dz_a}{2\pi i z_a}\,&e^{i\pi(\nu k^+_{gR}+(1+\nu) k^+_{gg})m}\ z^{k^+_a(m)}_{a}\\
    {\color{blue}q^{k^+_{gR} m/2}}\,{\color{blue}\prod_{\alpha_+} (q^{\frac{\alpha_+(m)}{4}}z^{\mp\frac{\alpha_+}{2}}-q^{-\frac{\alpha_+(m)}{4}}z^{\pm\frac{\alpha_+}{2}})}&\prod_{\Phi}\prod_{\rho\in \mathcal{R}_\Phi}\frac{(e^{-i\pi\nu r_{\Phi}}z^{-\rho}q^{-\rho(m)/2+1-r_\Phi/2}\,;q)}{(e^{i\pi\nu r_{\Phi}}z^{\rho}q^{-\rho(m)/2+r_\Phi/2};q)}\,.
    \end{split}
\end{equation}

Integrals of the form \eqref{eq:indexsumint} are sometimes called \emph{basic hypergeometric integrals} \cite{Krattenthaler:2011da,Gahramanov:2016wxi}. Our study concerns their \emph{classical limit} $q\to1$.

\subsection*{Holomorphic-antiholomorphic factorization of the integrand}

In terms of $u_j,y_j$ defined via
\begin{equation}
     u_j:=x_j-\frac{i\beta}{4\pi}m_j \Longrightarrow 2\pi i\,u_j=2\pi i\, x_j+\beta\, \frac{m_j}{2},\label{eq:uDef}
\end{equation}
\begin{equation}
    y_j:=e^{2\pi i u_j}=z\,q^{-m/2},\qquad \bar y_j=e^{-2\pi i \bar u_j}=z^{-1}q^{-m/2},\label{eq:yDef}
\end{equation}
we can write the various pieces in the integrand of \eqref{eq:indexsumint} in factorized form as (cf.~\cite{Choi:2019dfu}):
\begin{equation}
\label{eq:CS_fac}
    e^{-S_{\text{CS}^+}} = e^{-4\pi^2(\nu k^+_{jR}+(1+\nu) k^+_{jj})\frac{u_j-\bar{u}_j}{2\beta}}(y \bar{y})^{-k^+_{gR}/2}e^{-4\pi^2\,k^+_{ij}\frac{u_i u_j-\bar{u}_i\bar{u}_j}{2\beta}},
\end{equation}

\vspace{-.3cm}
\begin{equation}
\label{eq:Ga_fac}
    Z_{\text{Gauge}} = \prod_{\alpha_+} (y^{-\alpha_+
/2}-y^{\alpha_+
/2})(\bar{y}^{\, -\alpha_+\!/2}-\bar{y}^{\,\alpha_+\!/2})\,,
\end{equation}

\vspace{-.3cm}
\begin{equation}
\label{eq:Ch_fac}
  I^+_{\text{Chiral}}=  \prod_{\Phi}\prod_{\rho\in \mathcal{R}_\Phi}\frac{(\bar{y}^{\rho} e^{-i\pi\nu r_\Phi} q^{1-r_\Phi/2};q)}{(y^{\rho}e^{i\pi\nu r_{\Phi}}q^{r_\Phi/2};q)}\,.
\end{equation}

This factorization is the 3d lift of
(or KK product over)
the one for the integrand of the $S^2$ partition function of 2d $\mathcal{N}=(2,2)$ gauge theories \cite{Benini:2012ui,Gomis:2012wy}.

\subsection{Cardy limit and 2d EFT}\label{subsec:Cardy_estimates}

The Cardy limit of the CS and gauge pieces are straightforward. For the chiral multiplets piece, we use the small-$\beta$ asymptotic of the Pochhammer symbol
\begin{equation}
\label{eq:qPoch_master_estimate_body}
    (q^X;q)= \left(\frac{\sqrt{2\pi}}{\Gamma({X})}\cdot e^{\, X\,(\log  X-1)}\cdot e^{-\frac{1}{2}\log X}\right)\,
    \times \left( e^{-\frac{\mathrm{Li}_2(e^{-\beta X})}{\beta}}\, (1-e^{-\beta X})^{1/2} \right)\times \big(1+ \mathcal{O}(\beta)\big).
\end{equation}
To evaluate the RHS we must first shift $X$ vertically by an element of $\frac{2\pi i}{\beta}\mathbb{Z}$ to go to the principal strip $-\frac{\pi}{\beta}<\mathrm{Im}X\leq\frac{\pi}{\beta},$ and then use the principal sheet of all the multi-valued functions. See Appendix~\ref{app:estimates} for more details.

To adapt the above estimate to \eqref{eq:Ch_fac}, we should take $X$ to be
\begin{equation}
\begin{split}
    X_\text{num} = 1 -\frac{r_\Phi}{2}+\frac{2\pi i}{\beta}\big(\rho(\bar{u})+\frac{\nu}{2} r_\Phi\big)_\mathbb{Z}\,,\\
    X_\text{den} =\frac{r_\Phi}{2}-\frac{2\pi i}{\beta}\big(\rho(u)+\frac{\nu}{2} r_\Phi\big)_\mathbb{Z}\,,\label{eq:Xnum_Xden}
    \end{split}
\end{equation}
with the symbol $x^{}_\mathbb{Z}$ denoting $x-\mathrm{nint}(\mathrm{Re}\,x)$, where $\mathrm{nint}$ is the nearest integer function.\\


Putting the pieces together, we obtain our master estimate:
\begin{equation}
\begin{split}
    &Z_{\nu+1}(q)\approx\sum_{m \in \Gamma_{G}^{\vee}} \int \frac{\mathrm{d}^{r_G}(\frac{2\pi x}{{\color{green}\beta}})}{(2\pi)^{r_G}|W|}\ e^{W_0(u;\,\beta)-\widetilde{W_0(u;\,\beta)}-i\pi\big(\Omega_0(u;\,\beta)-\widetilde{\Omega_0(u;\,\beta)}\big)}\\
    &\hspace{-.4cm}\text{$\times{\color{olive}\left(\prod_{\alpha_+}\frac{-2\pi i \big(\alpha_+(u)\big)^{}_\mathbb{Z}}{\beta}\cdot\frac{2\pi i\big(\alpha_+(\bar u)\big)^{}_\mathbb{Z}}{\beta}\right)}$}\!\left(\prod_{\Phi}\prod_{\rho^\Phi} \frac{\Gamma\big(\frac{r_\Phi}{2}-\frac{2\pi i}{\beta}(\rho(u)+\nu\frac{r_\Phi}{2})^{}_\mathbb{Z}\big)}{\Gamma\big(1-\frac{r_\Phi}{2}+\frac{2\pi i}{\beta}(\rho(\bar{u})+\nu\frac{r_\Phi}{2})^{}_\mathbb{Z}\big)}\right),
    \end{split}\label{eq:index_Cardy_master}
\end{equation}
with relative $\mathcal{O}(\beta)$ error similarly to~\eqref{eq:qPoch_master_estimate_body}, and with
{\small\begin{equation}
    \begin{split}
        W_0(u;\beta)&=-\frac{4\pi^2}{\beta}\left({\frac{1}{2}k^+_{jl}\big(u_j u_l}+(1+\nu)\delta_{jl}u_j\big)+\frac{\nu}{2}k^+_{jR}u_j+\sum_{\Phi}\sum_{\rho\in \mathcal{R}_\Phi}\frac{1}{-4\pi^2}\mathrm{Li}_2\big(q^{\frac{r_\Phi}{2}}e^{2\pi i \rho(u)+i\pi\nu r_\Phi}\big)\,\right)\\
        &\qquad-\sum_{\Phi}\sum_{\rho^\Phi}\big(\,\frac{r_\Phi}{2}-\frac{2\pi i}{\beta}(\rho(u)+\nu\frac{r_\Phi}{2})^{}_\mathbb{Z}\big)\big(\log(\frac{r_\Phi}{2}-\frac{2\pi i}{\beta}(\rho(u)+\nu\frac{r_\Phi}{2})^{}_\mathbb{Z})-1\big)\\
        \Omega_0(u;\beta)&={{\color{blue}k^+_{jR}u_j}}+{\color{blue}\sum_{\alpha_+}\frac{1}{-i\pi}\log\big(e^{-i\pi  \alpha_+(u)}-e^{i\pi  \alpha_+(u)}\big)}{\color{olive}-\sum_{\alpha_+}\frac{1}{-i\pi }\log\frac{-2\pi i \big(\alpha_+(u)\big)_\mathbb{Z}}{\beta}}-{\color{green}\frac{r_G}{2\pi i}\log\beta}\\
        &+\sum_{\Phi}\sum_{\rho\in \mathcal{R}_\Phi}\frac{1}{2\pi i}\log\big(1-q^{\frac{r_\Phi}{2}}e^{2\pi i \rho(u)+i\pi\nu r_\Phi}\big)-\sum_{\Phi}\sum_{\rho^\Phi}\frac{1}{2\pi i}\log\big(\frac{r_\Phi}{2}-\frac{2\pi i}{\beta}(\rho(u)+\nu\frac{r_\Phi}{2})^{}_\mathbb{Z}\big)
        \end{split}\label{eq:W0_Omega0}
        \end{equation}
        \begin{equation}
        \begin{split}
        \widetilde{W_0(u;\beta)}&=-\frac{4\pi^2}{\beta}\left({\frac{1}{2}k^+_{jl}\big(\bar u_j \bar u_l}+(1+\nu)\delta_{jl}\bar u_j\big)+\frac{\nu}{2}k^+_{jR}\bar u_j+\sum_{\Phi}\sum_{\rho\in \mathcal{R}_\Phi}\frac{1}{-4\pi^2}\mathrm{Li}_2\big(q^{1-\frac{r_\Phi}{2}}e^{-2\pi i \rho(\bar u)-i\pi\nu r_\Phi}\big)\,\right)\\
        &\qquad-\sum_{\Phi}\sum_{\rho^\Phi}\big(1-\frac{r_\Phi}{2}+\frac{2\pi i}{\beta}(\rho(\bar u)+\nu\frac{r_\Phi}{2})^{}_\mathbb{Z}\big)\big(\log(1-\frac{r_\Phi}{2}+\frac{2\pi i}{\beta}(\rho(\bar u)+\nu\frac{r_\Phi}{2})^{}_\mathbb{Z})-1\big)\\
        \widetilde{\Omega_0(u;\beta)}&={{\color{blue}k^+_{jR}\bar u_j}}+{\color{blue}\sum_{\alpha_+}\frac{1}{i\pi}\log\big(e^{i\pi \alpha_+(\bar u)}-e^{-i\pi  \alpha_+(\bar u)}\big)}{\color{olive}-\sum_{\alpha_+}\frac{1}{i\pi }\log\frac{2\pi i\big(\alpha_+(\bar u)\big)_\mathbb{Z}}{\beta}}+{\color{green}\frac{r_G}{2\pi i}\log\beta}\\
        &\hspace{-.7cm}+\sum_{\Phi}\sum_{\rho\in \mathcal{R}_\Phi}\frac{1}{2\pi i}\log\big(1-q^{1-\frac{r_\Phi}{2}}e^{-2\pi i \rho(\bar u)-i\pi\nu r_\Phi}\big)-\sum_{\Phi}\sum_{\rho^\Phi}\frac{1}{2\pi i}\log\big(1-\frac{r_\Phi}{2}+\frac{2\pi i}{\beta}(\rho(\bar u)+\nu\frac{r_\Phi}{2})^{}_\mathbb{Z}\big)
    \end{split}\label{eq:W0_Omega0_tilde}
\end{equation}}
The color coding here is to help the reader recognize that: $i$) the blue terms come from the blue terms in \eqref{eq:indexsumint}, $ii$) the olive and green terms cancel those in \eqref{eq:index_Cardy_master}. The trivial ``multiplication and division'' by the olive and green terms is for later convenience in comparing with the 2d EFT.

A crucial point to notice is that the small-$\beta$ expansion of the RHS of \eqref{eq:qPoch_master_estimate_body} depends strongly on whether $\beta X$ is close to or far from ${2\pi i}\mathbb{Z}.$ In other words, the $\beta\to0$ limit introduces a scale hierarchy in the problem, and the asymptotics of $(q^X;q)$ depends on where the distance of $X$ to the ``singularities'' $\frac{2\pi i}{\beta}\mathbb{Z}$ sits in the hierarchy. We shall see momentarily that the singularity condition $X\in\frac{2\pi i}{\beta}\mathbb{Z}$ corresponds to the loci supporting massless charged chirals in the 2d Kaluza-Klein EFT. The integrand of the index \eqref{eq:indexsumint} has poles at these locations. In the rank-1 cases these poles are points, and we denote the closest point to them on the $x$ contour by $\{x_{p_1},x_{p_2},\dots\}.$ We see from \eqref{eq:Xnum_Xden} that the defining equation of $x_{p_j}$ is
\begin{equation}
    \rho(x_{p_j})+\frac{\nu}{2} r_\Phi\in\mathbb{Z}\,.\label{eq:xpj_def}
\end{equation}

There could also be loci where only W bosons become massless. These do not correspond to poles in the integrand of the index \eqref{eq:indexsumint}. In the rank-1 cases, where these are also points, we denote their locations by $\{x_{g_1},x_{g_2},\dots\}$. Their defining equation is
\begin{equation}
    \alpha_+(x_{g_j})\in\mathbb{Z}\,.\label{eq:xgj_def}
\end{equation}
We denote the locations of all $p_i,$ $g_i$ collectively by $\{x_{s_1},x_{s_2},\dots\}.$\\

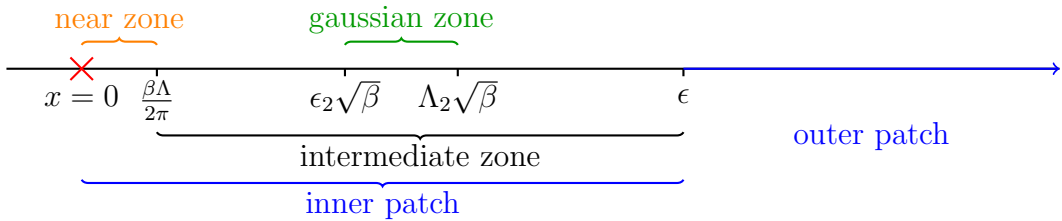
\begin{figure}[t]
    \centering
\begin{tikzpicture}[thick]
  \coordinate (X0) at (0,0);
  \coordinate (XLab) at (1,0);  
  \coordinate (XeSqrtB) at (3.5,0); 
  \coordinate (XLsSqrtB) at (5,0); 
  \coordinate (Xe) at (8,0);    
  \coordinate (Xend) at (13,0); 

  \draw[black] (-1,0) -- (Xend);

  \draw[thick] ($(X0) + (-0.15,0.15)$) -- ($(X0) + (0.15,-0.15)$);
  \draw[thick] ($(X0) + (-0.15,-0.15)$) -- ($(X0) + (0.15,0.15)$);
  \node[below=2pt] at (X0) {$x=0$};

  \foreach \pt/\lbl in {XLab/$\frac{\beta\Lambda}{2\pi}$, XeSqrtB/$\epsilon_2\sqrt{\beta}$, XLsSqrtB/$\Lambda_2\sqrt{\beta}$, Xe/$\overset{\,}{\epsilon}$} {
    \draw (\pt) -- ++(0,-0.1);
    \node[below] at (\pt) {\lbl};
  }

  \draw[decorate, decoration={brace, raise=9pt}]
    (X0) -- node[midway, above=11pt] {near zone} (XLab);

  \draw[decorate, decoration={brace, raise=11pt}]
    (XeSqrtB) -- node[midway, above=12pt] {gaussian zone} (XLsSqrtB);

  \draw[decorate, decoration={brace, mirror, raise=24pt}, black]
    (XLab) -- node[midway, below=25pt] {intermediate zone} (Xe);

  \draw[decorate, decoration={brace, mirror, raise=44pt}]
    (X0) -- node[midway, below=44pt] {inner patch} (Xe);

  \draw[->, thick] (Xe) -- node[midway, below=16pt] {outer patch} (Xend);

\end{tikzpicture}
\caption{Schematic illustration of the various patches and zones on the moduli space of gauge holonomies. As explained in Section~\ref{subsec:surface_integral}, nonzero magnetic flux $m$ naturally complexifies the above structure, so that e.g. the inner patch becomes a disk of radius $\epsilon$ around $u=0$.}
    \label{fig:patches_zones}
\end{figure}

To illustrate the scale hierarchy, let us consider a rank-1 situation with $\nu=0$, such that there is an $x_p=0$ (the first sheet of 3d $\mathcal{N}=2$ SQED or SQCD are examples). We define different patches and zones according to the distance to $x_p$ schematically as in Figure~\ref{fig:patches_zones}. Although in Figure~\ref{fig:patches_zones} we have focused on $x>0,$ there are similar patches and zones to the left of $x_p=0$.

\subsection*{Two-dimensional $\mathcal{N}=(2,2)$ EFT}

The circle reduction of  3d $\mathcal{N}=2$ gauge theories gives rise to  2d $\mathcal{N}=(2,2)$ EFTs.
The 3d $\mathcal{N}=2$ multiplets are the vector and chiral multiplets. We will focus on the bosonic content of the multiplets. The vector contains a gauge field $A_\mu$ and a real scalar $\sigma$. Upon reduction, the third component of $A_\mu$ yields another real scalar $x$, which combines with the reduction of $\sigma$ to form a 2d complex scalar. 
The first two components of $A_\mu$ with the aforementioned complex scalar form a 2d $\mathcal{N}=(2,2)$ vector multiplet.
The 3d chiral multiplet simply reduces to a 2d chiral multiplet. See \cite{Aharony:2017adm,Closset:2019hyt} for background.


\subsubsection{Heavy multiplets and the outer patch}

Consider the case where $\mathrm{Im}(\beta X)/2\pi$ is at least $\epsilon$ away from $\mathbb{Z},$ with $\epsilon$ a small constant (say $\epsilon=0.01$). This corresponds to the outer patch of the chiral multiplet. Physically, this is the region of the BPS moduli space where the 2d twisted mass of the lightest KK mode of the chiral multiplet exceeds the cut-off $2\pi\epsilon/\beta$: in other words
\begin{equation}
    \frac{2\pi}{\beta}\big(\rho(x)+\frac{\nu}{2} r_\Phi\big)^{}_\mathbb{Z}\ge\frac{2\pi\epsilon}{\beta}.
\end{equation}
We refer to such 3d chiral multiplets (whose KK modes are \emph{all} heavy in the 2d sense) as \emph{heavy} multiplets. Similarly, we can define heavy vector multiplets as those satisfying
\begin{equation}
    \frac{2\pi}{\beta}\big(\alpha_+(x)\big)^{}_\mathbb{Z}\ge\frac{2\pi\epsilon}{\beta}.
\end{equation}
The corresponding region on the moduli space is the outer patch of the vector multiplet.

We refer to the intersection of the outer patches of all multiplets in the theory as \emph{the outer patch}. While so far this is defined as a subset of the moduli space of holonomies, we will see below that nonzero magnetic flux naturally complexifies the outer patch to a punctured $u$-strip, with the punctures around $x_{s_j}$. (Since $x_{g_j}$ do not correspond to poles, later we will fully shrink the $g_j$ punctures and keep only the $p_j$ punctures.)

For heavy chiral multiplets, the complicated multiscale estimate \eqref{eq:index_Cardy_master} simplifies as follows. We have $\mathrm{Im} X\ge\frac{2\pi\epsilon}{\beta}$, so we can use the Stirling approximation 
\begin{equation}
    \Gamma(x)\xrightarrow{x\to\infty}\sqrt{2\pi}\,x^{\,x-\frac{1}{2}}\,e^{-x}\,,\label{eq:Stirling}
\end{equation}
to simplify \eqref{eq:qPoch_master_estimate_body} as
\begin{equation}
\label{eq:outer_estimate}
    (q^X;q) \approx e^{-\frac{\mathrm{Li}_2(e^{-\beta X})}{\beta}}\, (1-e^{-\beta X})^{1/2}.
\end{equation}
Using $\mathrm{Li}'_2(y)=-\frac{\log(1-y)}{y}$, which implies the dilogarithm Taylor expansion
\begin{equation}
    \frac{\mathrm{Li_2}(q^n y)}{\beta} \approx \frac{\mathrm{Li_2}( y)}{\beta}+n\log(1-y),
\end{equation}
the estimate \eqref{eq:outer_estimate} further simplifies for $X$ of the form \eqref{eq:Xnum_Xden} as
\begin{equation}
\begin{split}
    \frac{1}{(y^\rho e^{i\pi\nu r_\Phi}q^{r_\Phi/2};q)}&\approx e^{\frac{\mathrm{Li}_2(y^\rho\,e^{i\pi\nu r_\Phi})}{\beta}}\, (1-y^\rho\,e^{i\pi\nu r_\Phi})^{\frac{r_\Phi-1}{2}},\\
    (\bar y^\rho e^{-i\pi\nu r_\Phi}q^{1-r_\Phi/2};q)&\approx e^{-\frac{\mathrm{Li}_2(\bar y^\rho\,e^{-i\pi\nu r_\Phi})}{\beta}}\, (1-\bar y^\rho\,e^{-i\pi\nu r_\Phi})^{\frac{r_\Phi-1}{2}}.
    \end{split}\label{eq:outer_est_Poch_main}
\end{equation}
Denoting the set of all heavy multiplets by $H$, the contribution of the heavy multiplets to the integrand of the index \eqref{eq:index_Cardy_master} can thus be estimated as
\begin{equation}
    e^{W^H(u;\,\beta)\,-\,\overline{W^H(u;\,\beta)}\,-\,i\pi\left(\Omega^H(u)\,-\,\overline{\Omega^H(u)}\,\right)},\label{eq:heavy_integrand}
\end{equation}
where the twisted superpotential $W^H$ and the dilaton $\Omega^H$ are\footnote{To compare with other conventions, note that in comparison with Closset~et~al~\cite{Closset:2017zgf} we~have
\begin{equation*}
    W^H_\text{here}=-\frac{4\pi^2}{\beta}W^H_\text{there},\qquad \Omega^H_\text{here}=\Omega^H_\text{there},
\end{equation*}
while in comparison with Benini-Cremonesi \cite{Benini:2012ui} we have $W_\text{here}=4\pi i\,W_\text{there}\,.$}
\begin{equation}
        W^H(u;\beta)=\frac{1}{\beta}
        \sum_{\Phi}\sum_{\rho^{}_\Phi\in H}\mathrm{Li}_2\big(e^{2\pi i \rho^{}_\Phi(u)+i\pi\nu r_\Phi}\big)\,,\label{eq:heavy_W}
        \end{equation}
        \begin{equation}
        \Omega^H(u)=
        -{\color{blue}\sum_{\alpha_+\in H}\frac{1}{i\pi}\log\big(e^{-i\pi \alpha_+(u)}-e^{i\pi \alpha_+(u)}\big)}-{\color{red}\sum_{\Phi}\sum_{\rho^{}_\Phi\in H}\frac{r_\Phi-1}{2\pi i}\log\big(1-e^{2\pi i \rho^{}_\Phi(u)+i\pi\nu r_\Phi}\big)}\,.
    \label{eq:heavy_Omega}
\end{equation}
Note that $W^H\!,\, \Omega^H$ arise from the $\beta\to0$ limit of $W_0,\Omega_0$ in \eqref{eq:W0_Omega0} respectively, except that the red piece of $\Omega^H$ comes from $W_0,\widetilde{W_0}$.\\

For heavy vector multiplets, we do not make any further simplifications in \eqref{eq:index_Cardy_master}, besides enacting the trivial cancelations of the olive-colored terms.\\

Note that the 3d chiral multiplets (heavy or not) make contributions to $S_{\text{CS}^+}$ as well. For technical convenience, we have decided to treat these separately, but, as our EFT discussion below should make clear, those $S_{\text{CS}^+}$ pieces coming from the heavy chiral multiplets conceptually belong to $W^H,\Omega^H$ instead.

\subsubsection{Inner patches and their multiscale resolution into subzones}

\subsubsection*{Light multiplets and near zones}

If the lightest KK mode of a 3d multiplet has twisted mass less than some cutoff $\Lambda$:
\begin{equation}
    \frac{2\pi}{\beta}\big(\rho(x)+\frac{\nu}{2} r_\Phi\big)^{}_\mathbb{Z}<\Lambda,\qquad\text{or}\qquad \frac{2\pi}{\beta}\big(\alpha_+(x)\big)^{}_\mathbb{Z}<\Lambda,
\end{equation}
we call it \emph{light}. Note that the zero modes of the Cartan photons have zero twisted mass, so we consider them also light, irrespective of $x$. Note also that the light 3d multiplets have only one KK mode in their tower that is light in the 2d sense.

On the near zone of a chiral multiplet $X_\text{num},X_\text{den}$ defined above are $\mathcal{O}(1)$. Therefore we have $(1-e^{-\beta X})\approx \beta X$, and also (see~Eq.~\eqref{eq:Li2_est_app})
\begin{equation}
    e^{\frac{1}{\beta}\mathrm{Li}_2(e^{-\beta X})}= e^{\frac{\pi^2}{6\beta}+X\,(\log X-1)+X\log\beta -\beta\frac{X^2}{4}+\mathcal{O}(\beta^2)}.
\end{equation}
This simplifies \eqref{eq:qPoch_master_estimate_body} as
\begin{equation}
    ( q^X;q)\xrightarrow{\beta\to0}e^{-\frac{\pi^2}{6\beta}}\,\beta^{\frac{1}{2}-{X}}\,\frac{\sqrt{2\pi}}{\Gamma({X})}\, \big(1+o(\beta^0)\big).\label{eq:qPochNearEst_main}
\end{equation}

These relations simplify the complicated multiscale estimate \eqref{eq:index_Cardy_master} for the light multiplets as follows. We first change variables
\begin{equation}
    u= x_s+u',\quad x=x_s+x',
\end{equation}
where $x_s$ is the location of the center of the zone. We also define
\begin{equation}
    u'_{2d}:=\frac{2\pi u'}{\beta},\quad x'_{2d}:=\frac{2\pi x'}{\beta},\quad x_{2d}:=\frac{2\pi x}{\beta},
\end{equation}
so that now $u'_{2d}=x'_{2d}-\frac{i}{2}m.$ (For ranks higher than $1,$ where the singularity in the zone may be extended, one may want to rescale in the directions perpendicular to the strata of the singularity, cf.~\cite{ArabiArdehali:2015ybk,Ardehali:2021irq}.) The simplified contribution of the light multiplets to the integrand of the index \eqref{eq:index_Cardy_master} then reads
\begin{equation}
    \begin{split}
        & e^{W^L(u'_{2d};\,\beta)-\overline{W^L(u'_{2d};\,\beta)}-i\pi \bigl(\Omega^L(\beta)-\overline{\Omega^L(\beta)}\bigr)}\\
        &\qquad\left(\prod_{\alpha_+\in L}\big(\alpha_+(u'_{2d})\big)\big(\alpha_+(\bar{u'}_{2d})\big)\right)
    \left(\prod_{\Phi}\prod_{\rho^\Phi\in L} \frac{\Gamma\big(\frac{r_\Phi}{2}-i\rho(u'_{2d})\big)}{\Gamma\big(1-\frac{r_\Phi}{2}+i\rho(\bar{u'}_{2d})\big)}\right),
    \end{split}\label{eq:near_zone_integrand}
\end{equation}
where the set of all light multiplets on the near zone is denoted $L$, and
\begin{equation}
    W^L(u'_{2d};\beta)=\sum_{\rho\in L}-i\big(\rho(u'_{2d})\big)\log\beta\,,
\end{equation}
\begin{equation}
    \Omega^L(\beta)=-\frac{1}{2\pi i}\big(\mathrm{dim}G_L+\sum_{\Phi}\sum_{\rho^\Phi\in L}({\color{red}r_\Phi-1})\big)\log\beta=\frac{1}{2\pi i}\big(\frac{c^{}_L}{3}\big)\log\beta\,.\label{eq:light_Omega}
\end{equation}
Here $\mathrm{dim}G_L=r_G+2\sum_{\alpha_+\in L}1,$ is the dimension of the effective gauge group in the near zone, and
\begin{equation}
    c^{}_L:=-\,3\,\mathrm{Tr}^{}_LR=-3\big(\mathrm{dim}G_L+\sum_{\Phi}\sum_{\rho^\Phi\in L}({\color{red}r_\Phi-1})\big)\,,
\end{equation}
is the central charge of the light 2d fields.

Note that $W^L\!,\, \Omega^L$ arise from the $\beta\to0$ limit of $W_0,\Omega_0$, respectively, {except} that the red piece of $\Omega^L$ comes from $W_0,\widetilde{W_0}$.

\subsubsection*{Middleweight multiplets and intermediate zones}

If
\begin{equation}
    \Lambda\le\frac{2\pi}{\beta}\big(\rho(x)+\frac{\nu}{2} r_\Phi\big)^{}_\mathbb{Z}<\frac{2\pi\epsilon}{\beta},\qquad\text{or}\qquad \Lambda\le\frac{2\pi}{\beta}\big(\alpha_+(x)\big)^{}_\mathbb{Z}<\frac{2\pi\epsilon}{\beta},
\end{equation}
for some fixed but large $\Lambda$ (say $\Lambda=100$), we refer to the multiplet as \emph{middleweight}.
The middleweight multiplets reside on the \emph{intermediate zones} of the moduli space. An \emph{inner patch} is comprised of a near zone and its surrounding intermediate zone(s).

Consider an intermediate zone $x\in(\frac{\beta\Lambda}{2\pi},\epsilon)$ as in Figure~\ref{fig:patches_zones}. Near the left end $x\sim\frac{\beta\Lambda}{2\pi}$ of the zone, the useful estimate is \eqref{eq:qPochNearEst_main}, while near the right end $x\sim\epsilon$ the useful estimate is \eqref{eq:outer_estimate}. It is therefore impossible to treat the intermediate zones uniformly (at least with the methods at our disposal). This motivates further decomposition of the intermediate zones.

We may consider $O(\beta^{1/n})$ subzones, where $n\in\mathbb{N}_{>1}\,$, inside the intermediate zone $x\in(\frac{\beta\Lambda}{2\pi},\epsilon)$. These can be more precisely defined as open subsets $(\epsilon_n\beta^{1/n},\Lambda_n\beta^{1/n}).$ Over such subzones, the lightest 2d mode has twisted mass $\propto \beta^{-1/n}\to\infty$ as $\beta\to0$. Therefore, it ought to be integrated out. In other words, on such subzones we can treat middleweight multiplets similarly to heavy multiplets, and thus use for them \eqref{eq:heavy_integrand}--\eqref{eq:heavy_Omega}. Importantly, however, the expressions for $W_H,\Omega_H$ can be further simplified in these subzones so that their $\beta$ scaling becomes manifest.

In the present work, we study only the subzone corresponding to $n=2$. We refer to it as the Gaussian zone, since $e^{\frac{u^2}{\beta}}$ is $\mathcal{O}(\beta^0)$ there. Our neglect of other parts of the intermediate zones is based on the observation that no new physics is associated with them, as discussed in Section~\ref{sec:intro}. It will also be corroborated by our analysis of various rank-$1$ examples in Section~\ref{sec:example}.

Denote the set of middleweight multiplets on the Gaussian zone by $M$. Their asymptotic contribution to the index
\begin{equation}
    e^{W^M(u_\text{gz};\,\beta)\,-\,\overline{W^M(u_\text{gz};\,\beta)}\,-\,i\pi\left(\Omega^M(u_\text{gz};\,\beta)\,-\,\overline{\Omega^M(u_\text{gz};\,\beta)}\,\right)},\label{eq:middleweight_integrand}
\end{equation}
can be found by simplifying $W^H,\Omega^H$ (using \eqref{eq:Li2_est_app} in particular) as
\begin{equation}
    \begin{split}
        W^M(u_\text{gz};\beta)&\approx\sum_{\Phi}\sum_{\rho\in M}\left(
        -\frac{2\pi i\rho(u_\text{gz})}{\sqrt{\beta}}\Big(\log\big(\!-\!2\pi i\sqrt{\beta}\,\rho(u_\text{gz})\big) \, -1\Big)+\pi^2\big(\rho(u_\text{gz})\big)^2\right)\,,\\
        \Omega^M(u_\text{gz};\beta)&\approx -{\color{blue}\sum_{\alpha_+\in M}\frac{1}{i\pi}\log\big(\!-\!{\sqrt{\beta}\,2\pi i \,\alpha_+(u_\text{gz})}\big)}-\sum_{\Phi}\sum_{\rho\in M}\frac{{\color{red}r_\Phi}-1}{2\pi i}\log\big(\!-\!\sqrt{\beta}\,2\pi i \,\rho(u_\text{gz})\big)\,.
    \end{split}\label{eq:intermediate_W_Omega}
\end{equation}
Here we have dropped a real constant in $W^M$ because it cancels against a similar term in $\overline{W^M}$, and have defined $u_\text{gz}$ via 
\begin{equation}
    \sqrt{\beta}\,\rho(u_\text{gz})=\big(\rho(u)+\frac{\nu}{2} r_\Phi\big)^{}_\mathbb{Z}\,,\label{eq:ugz_def}
\end{equation}
for the chiral multiplet Gaussian zones, and $\sqrt{\beta}\,\alpha_+(u_\text{gz})=\big(\alpha_+(u)\big)^{}_\mathbb{Z}$ for vector multiplet Gaussian zones. This is essentially $u_\text{gz}=u/\sqrt{\beta}=u_{2d}\sqrt{\beta}/2\pi$, but with an appropriate shift before scaling.

\subsubsection{Summary and comparison with EFT}

In terms of
\begin{equation}
    x_{2d}=\frac{2\pi x}{\beta}\,, \qquad u'_{2d}=x'_{2d}-\frac{i}{2}m\,,
\end{equation}
we can put together our results for various patches and zones to obtain (neglecting the intermediate zones except for their Gaussian zones):
\begin{equation}
    \boxed{\begin{split}
        Z_{\nu+1}(q)\approx{\left(\frac{1}{2\pi}\right)^{r_G}} \sum_{m \in \Gamma_{G}^{\vee}}& \int_{(\frac{2\pi}{\beta})^{r_G}\,\mathfrak{h}_\text{cl}}\  \frac{\mathrm{d}^{r_G}x^{}_{2d}}{|W|}\ e^{W(u'_{2d};\,\beta)-\overline{W(u'_{2d};\,\beta)}\,-i\pi \bigl(\Omega(u'_{2d};\,\beta)-\overline{\Omega(u'_{2d};\,\beta)}\bigr)}\\
        &\left(\prod_{\alpha_+\in L}\big(\alpha_+(u'_{2d})\big)\big(\alpha_+(\bar{u'}_{2d})\big)\right)
    \left(\prod_{\Phi}\prod_{\rho^\Phi\in L} \frac{\Gamma\big(\frac{r_\Phi}{2}-i\rho(u'_{2d})\big)}{\Gamma\big(1-\frac{r_\Phi}{2}+i\rho(\bar{u'}_{2d})\big)}\right).
    \end{split}}\label{eq:master_EFT}
\end{equation}
Here 
\begin{equation}
    \begin{split}
        W&=W_{\text{CS}^+}+W^L+W^M+W^H,\\
        \Omega&=\Omega_{\text{CS}^+}+\Omega^L+\Omega^M+\Omega^H,
    \end{split}
\end{equation}
with the (shifted) CS level contributions reading
\begin{equation}
\begin{split}
    W_{\text{CS}^+}&=-\frac{\beta}{2} k^+_{jl}{u'_{2d}}^{}_{j}\,{u'_{2d}}^{}_{\,l}-\pi\big((1+\nu) k^+_{jj}+2k^+_{jl}x^s_l-\nu k^+_{jR}\big){u'_{2d}}^{}_{j}\,,\\
    \Omega_{\text{CS}^+}&=\frac{\beta}{2\pi}k^+_{jR}{u'_{2d}}^{}_{j}\,.
\end{split}\label{eq:WCS_OmegaCS}
\end{equation}
The first line above is obtained by using $u=x_s+u'$, and neglecting the real constants arising in $W_{\text{CS}^+}$ since they cancel against $\overline{W_{\text{CS}^+}}$.

In the 2d EFT, the piece of $W_{\text{CS}^+}$ that is linear in $u'_{2d}$ amounts to an effective theta-angle:
\begin{equation}
    \vartheta^{2d}_j\longleftrightarrow\pi\big((1+\nu) k^+_{jj}+2k^+_{jl}x^s_l-\nu k^+_{jR}\big),
\end{equation}
while the part of $\Omega_{\text{CS}^+}$ that is linear in $u'_{2d}$ is a linear dilaton
\begin{equation}
    \text{2d linear dilaton coefficient}\longleftrightarrow\frac{\beta}{2\pi} k^+_{jR}\,.
\end{equation}
The quadratic-in-$u'_{2d}$ piece of $W_{\text{CS}^+}$ is an effective mass for the 2d twisted chiral multiplets associated with $u'_{2d\ j}$:
\begin{equation}
    m^{2d}\longleftrightarrow\frac{\beta}{2}k^+_{jl}\, {u'_{2d}}^{}_{j}\,{u'_{2d}}^{}_{\,l}\,.
\end{equation}
On the near zones where $u'_{2d\, j},\,u'_{2d\, k}$ are of order 1, this mass term is negligible of course as $\beta\to0.$ Here we have obtained these relations from asymptotic analysis of the index. Direct 2d EFT derivations of them can be done through e.g. the results in Sections~2,4 and Appendix~C of~\cite{Closset:2017zgf}.

The EFT explanation of $W^H,\Omega^H$ is clear: they (together with the chiral multiplet pieces in $W_{\text{CS}^+},\,\Omega_{\text{CS}^+}$) are the contributions to the effective twisted superpotential and effective dilaton arising from integrating out the whole KK towers of 2d modes of a heavy 3d multiplet \cite{Nekrasov:2014xaa}.\footnote{To compare with \cite{Nekrasov:2014xaa} note that $u^\text{here}_{2d}=i\sigma^\text{there}$. See also Eq.~(2.17) in~\cite{Closset:2017zgf}.} We will illustrate the EFT derivation of $W^H$ momentarily. That of $\Omega^H$ is similar, and that of $W^M,\,\Omega^M$ is the same as they are only simplifications of $W^H,\,\Omega^H$ adapted to the Gaussian zones.

The EFT explanation of $W^L,\Omega^L$ is that they arise from the contributions of all the KK towers of 2d modes of a light 3d multiplet, \emph{except} the light 2d multiplet in the tower which we do not integrate out:
\begin{equation}
    \begin{split}
        W_{\rho^{}_L}^L&\longleftrightarrow W_{\rho^{}_L}^H-W_{{\rho^{}_L}\ 2d}\,,\\
        \Omega^L&\longleftrightarrow\Omega^H-\Omega_{2d}\,,
    \end{split}
\end{equation}
where $W_{\rho^{}_L}^H,\Omega^H$ stand for the would-be contributions of the whole towers, while
\begin{equation}
\begin{split}
    W_{\rho^{}_L\ 2d}&=-i\,\rho^{}_L({u'_{2d}})\Big(\log\big(\!-\!i\rho^{}_L({u'_{2d}})\big) -1\Big),\\
    \Omega_{2d}&=-\frac{1}{2\pi i}\sum_{\alpha^{}_L\in L}\log\big(\!-\!i\alpha^{}_L({u'_{2d}})\big)-\sum_{\rho^{}_L\in L}\frac{r-1}{2\pi i}\log\big(\!-\!i\rho^{}_L({u'_{2d}})\big),
\end{split}\label{eq:W2d_Omega2d}
\end{equation}
are the would-be contributions of the 2d light multiplet, which are subtracted since we are not integrating out the light 2d multiplet.

Actually, the preceding paragraph applies for the 3d chiral multiplets and W bosons, for which $W^H,\Omega^H$ are defined, not the Cartan photons. We  justify the same origin for the Cartan photon multiplets piece of $\Omega^L$ (even though $\Omega^H$ is not defined for the Cartan photon multiplets), separately by taking the formal limit $\alpha_+\to0$. Alternatively, summation over all but one mode of the KK towers can be done directly to derive the Cartan photon multiplets piece of $\Omega^L$.\\

Let us now briefly demonstrate the EFT derivation of $W^H.$ It arises from summing the one-loop effective twisted superpotentials arising from integrating out all the 2d KK modes of a 3d chiral multiplet. We see from \eqref{eq:W2d_Omega2d} and $u'_{2d}=\frac{2\pi(u-x_s)}{\beta},$ that the contribution of a weight $\rho$ of a heavy 3d chiral multiplet to $W^H$ is
\begin{equation}
    \sum_{n\in\mathbb{Z}}-\frac{2\pi i\big(\rho(u)+\nu \frac{r_\Phi}{2}+n\big)}{\beta}\,\bigg(\log\Big(\!-\frac{2\pi i\big(\rho(u)+\nu \frac{r_\Phi}{2}+n\big)}{\beta}\Big)-1\bigg).\label{eq:KKsum_W}
\end{equation}
There is a large-$n$ divergence in this formula, which maps to the zero sector of the Poisson re-summed expression. Dropping that zero sector defines a regularization scheme. In that scheme the only term in \eqref{eq:KKsum_W} giving a nonzero answer is
\begin{equation}
    W^H_\rho(u)=\sum_{n\in\mathbb{Z}}-\frac{2\pi i\big(\rho(u)+\nu \frac{r_\Phi}{2}+n\big)}{\beta}\log{\big(\rho(u)+\nu \frac{r_\Phi}{2}+n\big)}.\label{eq:KKsum_W_simp}
\end{equation}
This, first of all, manifests the crucial $\frac{1}{\beta}$ factor. Next, its evaluation using the Poisson resummation scheme just described yields $\frac{1}{\beta}\mathrm{Li}_2\big(e^{2\pi i \rho^{}_\Phi(u)+i\pi\nu r_\Phi}\big)$, compatibly with \eqref{eq:heavy_W}, together with a piece\footnote{There is another choice of scheme involved in regularizing sums of the form $\sum_{n\in\mathbb{Z}}f(x+n)$, when $f$ has branch cuts as in our case. That is in evaluating the Fourier transform $\hat{f}$ of $f,$ which enters the Poisson resummation formula as $\sum_{n\in\mathbb{Z}}f(x+n)=\sum_{k\in\mathbb{Z}}e^{2\pi ixk} \hat{f}(k).$ This Fourier transform involves a contour integral, and the position of the contour with respect to the branch cut may be ambiguous. We resolve this ambiguity by demanding that the end result be compatible with the regularization scheme used in \cite{Closset:2017zgf}. We leave a more thorough analysis of these very subtle issues to future work.}
that accounts for the chiral multiplet contributions to $W_{\text{CS}^+}$.\\

The main point of the present subsection is that the expression in \eqref{eq:master_EFT} matches the localization formula for the $S^2$ partition function of the 2d EFT \cite{Benini:2012ui,Doroud:2012xw,Benini:2016qnm}, patch by patch (at least on the patches without gauge enhancement as we discuss shortly, and neglecting the non-Gaussian parts of the intermediate zones as we do in this work). In other words, the index is well approximated in the Cardy limit by the sum over the $S^2$ partition functions of all patches. This is the partition function manifestation of the fact that the 3d theory is well approximated by the direct sum of the 2d EFTs associated to each patch. (One must be careful also with the range of fluxes, since the scale hierarchy is really in the $u$ variable, and not $x.$ We shall discuss this momentarily.)

We have to be cautious with the gauge inner patches, since on those we can not dualize the 2d vector multiplet to a twisted chiral multiplet which would feature in a twisted superpotential. The EFT interpretation of the asymptotic analysis on these patches hence requires extra care. We postpone this to the future. See Problem~\textbf{6}.\\

In the outer patch, we work with $u$ instead of $u'_{2d}$, i.e. we do not introduce shifts in holonomies and also do not rescale. We thus write
\begin{equation}
    \begin{split}
        W_{\text{CS}^+}=-\frac{4\pi^2}{\beta}\Big(\frac{1}{2}k^+_{jl}\big(u_ju_l+(1+\nu)\delta_{jl}u_j\big)+\frac{\nu}{2}k^+_{jR}u_j\Big),\qquad&\Omega_{\text{CS}^+}=k^+_{jR}u_j\,,\\
        W^L=0,\qquad\qquad\qquad\qquad\quad&\Omega^L =-\frac{\mathrm{Tr}^{}_{L_\text{o}}\!R}{2\pi i}\log\beta\,.
    \end{split}\label{eq:W_Omega_simpl_out}
\end{equation}
The set of light multiplets on the outer patch, $L_\text{o}$, consists of Cartan photon multiplets and zero weights (i.e. $\mathcal{R}_\Phi\ni\rho=0$) of chiral multiplets with $\frac{\nu r_\Phi}{2}\in\mathbb{Z}$.
None of these contributes to $W^L,$ but they do contribute to $\Omega^L,$ as seen in the above equation. For convenience, we denote the contribution of the zero weights of the chiral multiplets with $\frac{\nu r_\Phi}{2}\in\mathbb{Z}$ to \eqref{eq:master_EFT} by $Z_{S^2}^{\rho=0,\frac{\nu r_\Phi}{2}\in\mathbb{Z}}$:
\begin{equation}
    Z_{S^2}^{\rho=0,\frac{\nu r_\Phi}{2}\in\mathbb{Z}}:=\prod_{\Phi|\frac{\nu r_\Phi}{2}\in\mathbb{Z}}\ \prod_{\rho^\Phi=0} \frac{\Gamma\big(\frac{r_\Phi}{2}\big)}{\Gamma\big(1-\frac{r_\Phi}{2}\big)}.\label{eq:Z_rho=0}
\end{equation}
It is also more convenient on the outer patch to use the 3d measure:
\begin{equation}
    \Big(\frac{1}{2\pi}\Big)^{r_G}\int'_{(\frac{2\pi}{\beta})^{r_G}\,\mathfrak{h}_\text{cl}}\mathrm{d}^{r_G}x_{2d}\ \xrightarrow{\ \ \ }\, \frac{1}{\beta^{r_G}}\int'_{\mathfrak{h}_\text{cl}}\mathrm{d}^{r_G}x\,.\qquad\label{eq:2d_vs_3d_measure}
\end{equation}

As we shall see below, the gauge inner patches are negligible both on the first and second sheet. Adding these negligible contributions to the outer patch completes it into the gauge patch. Therefore, the outer patch contribution with the gauge inner patches $g_j$ fully shrunk, correctly gives the contribution from the gauge patch.

\subsection*{Holomorphy of the integrand and expectations for the large-flux EFT}

Eq.~\eqref{eq:master_EFT}  manifests a sort of holomorphy in $u'_{2d}=x'_{2d}-im/2$. This is not full holomorphy yet, since the real part of $u'_{2d}$ is continuous but its imaginary part is discrete. In other words, we do not have a surface yet. We will overcome this limitation shortly on the gauge patch, where Poisson resummation yields a punctured $y$-plane.

Postponing the issue of holomorphy at the level of the domain of integration to the next subsection, we now discuss an important implication of holomorphy at the level of the integrand.

Many of the asymptotic estimates used so far for large $x'_{2d}=2\pi x'/\beta,$ are equally valid for large $m.$ For example, the Stirling approximation \eqref{eq:Stirling} for $\Gamma(\frac{r}{2}-i\rho(u'_{2d}))$, is valid not only for large imaginary part of the argument of the gamma function, but also for large real part. This means that our various EFT simplifications must extend from the $x_{2d}$ line to the whole complex $u_{2d}$  strip. In particular, we expect that just as large $x'_{2d}$ Higgses the gauge group and makes the chiral multiplets charged under the corresponding Cartan direction massive, so would large $m.$ We now demonstrate this for charged chiral multiplets, and leave a similar demonstration for the W bosons to the interested reader.

Recall that the scalar in a chiral multiplet couples to the gauge multiplet schematically as
\begin{equation}
    \label{eq:ScalarQED}
    \left|\left(\partial_i-ieA_i\right)\phi\right|^2+e^2\sigma^2\left|\phi\right|^2
\end{equation}
The BPS locus in the localization procedure \cite{Closset:2019hyt} for theories on $ S_t^1\times S^2$ is
\begin{equation}
    \label{eq:scalarBPSlocus}
A=\frac{2\pi x}{\beta}\,\mathrm dt+\frac{m}{2}\left(1-\cos{\left(\theta\right)}\right)\mathrm d\varphi\,,\quad \sigma=-\frac{m}{2}\,,
\end{equation}
with $m$ the magnetic flux.
Therefore the mass-like term in the tree-level analysis following the action \eqref{eq:ScalarQED} for the light 2d mode of the chiral multiplet is ($e\longleftrightarrow\rho$)
\begin{equation}
    M^2_{\text{tree}}=\rho\left(x'_{2d}\right)^2+\frac{\rho(m)^2}{4}.
\end{equation}
This shows that the matter patches are indeed complexified, from $\frac{2\pi\epsilon}{\beta}$ intervals around $x'_{2d}=0$ on the real $x_{2d}$ line, into disks of that size on the complex $u_{2d}$ strip.

The preceding two paragraphs demonstrate, through asymptotic analysis and EFT, that the matter patches are complexified into (disk shaped) punctures. One may still wonder though, why the poles of the index at $|\mathrm{Im}(u)|>\epsilon,$ which are away from the punctures just described, do not cause problems (and do not necessitate their own neighborhoods to be punctured for instance). The reason is that on the actual BPS locus (i.e. $\mathrm{Re}u\in(-\frac{1}{2},\frac{1}{2}]$, $\mathrm{Im}u\in\frac{\beta}{4\pi}\mathbb{Z}$), the potential worry that we might be only $\mathcal{O}(\beta)$ away from poles is relieved for $|\mathrm{Im}(u)|=\frac{\beta|m|}{4\pi}>\epsilon$ by the exponential suppression of the functions involved. In short, on the BPS locus we are safely away from the poles at large $\mathrm{Im}u$. The poles still leave an imprint on the BPS locus at large $\mathrm{Im}u$ in the form of branch cuts of the integrand, but these cuts are too mild a singularity to cause problems for our Poisson resummation.

\subsection{Poisson$\,$resummation on$\,$gauge$\,$patch$\,${\it\&} emergence of punctured surfaces}\label{subsec:surface_integral}

It is hard to directly extract the high-temperature asymptotics of the 3d index from its sum-integral expressions discussed above. The presence of the sum over the monopole fluxes makes the analysis particularly challenging. We now rewrite the gauge-patch contribution to the index into a form that is more directly amenable to standard methods of asymptotic analysis. As we shall soon see, one may effectively turn the sum over fluxes into an integral, which will combine with the integral enforcing the projection on the gauge-invariant sector into a complex surface integral, amenable to the multi-dimensional stationary phase technique. Another sum (over winding sectors of the 2d dual photon multiplet) arises through the Poisson resummation, but this happens to be easy to deal with, since only finitely many terms in that sum (i.e. finitely many winding sectors) host stationary-phase loci and dominate the asymptotics.

To achieve our desired rewriting, we use the following distributional identity: 
\begin{equation}
    \sum_{m\in\mathbb Z} f(m) =\int_{\mathbb R} \mathrm dm \,f(m)\left( \sum_{w\in\mathbb{Z}}\delta \left(m- w\right)\ \right) =\int_{\mathbb R}  \mathrm dm\, f(m)\left(\sum_{w\in\mathbb{Z}}\exp\left( 2\pi i m w\right)\right) \,.\label{eq:Poisson_resum}
\end{equation}
The first equality is trivial, provided one can interchange the sum and integral, and the second equality follows from performing the Fourier transform of $\sum_w\delta(m - w)$. Alternatively, this is just the Poisson resummation formula. It is valid if $f$ is bounded, but not necessarily if it has poles.

In the present context, we also have an integration over the gauge holonomies $x$. The total integration measure can be written in terms of the holomorphic combinations $2\pi i u = \frac{\beta m}{2} + 2\pi i x$ or $y=e^{2\pi i u}$ as 
\begin{equation}
\label{eq:measure}
    \int_{\mathbb{R}} \mathrm dm\int_{-\frac{1}{2}}^{\frac{1}{2}} \mathrm dx=\int_\mathbb{C}\frac{i\,\mathrm{d}y\wedge\mathrm{d}\bar{y}}{2\pi\beta|y|^2}=\int_S\frac{2\pi i\,\mathrm{d}u\wedge\mathrm{d}\bar{u}}{\beta}\,,
\end{equation}
with $S$ the $u$-strip. In practice, it may be useful to replace $i\,\mathrm{d}y\wedge\mathrm{d}\bar{y}$ with its polar version $2r\mathrm{d}r\mathrm{d}\theta.$ The Poisson phase factors in \eqref{eq:Poisson_resum} can be written in terms of $u$ as 
\begin{equation}
    e^{2\pi i m w} = e^{\frac{-4\pi^2}{\beta}  w (u-\bar{u})}.
\end{equation}

A crucial point here is that Poisson resummation can not be applied---at least immediately---on patches supporting 2d light charged chirals, because of the corresponding gamma function poles in the integrand. We thus exclude such patches from our resummation procedure (and analyze their asymptotic contributions through multiscale decomposition into near zones and Gaussian zones instead). We do not need to exclude the gauge inner patches, because they do not host poles.

Another point to notice is that now that we are focusing on the complement of the matter patches, we are excising $\epsilon$ neighborhoods of $x_p$ where the chiral multiplet singularities are. In terms of $x,m$, this means leaving out $\mathcal{O}(\epsilon)$ neighborhoods of $x_p$, as well as $\mathcal{O}(\epsilon/\beta)$ neighborhoods of $m=0$ in that vicinity. Adapting Poisson resummation to such excision scenarios is achieved quite simply, by inserting suitable cut-off functions
on both sides of \eqref{eq:Poisson_resum}, to screen the punctures around $x_p$. Since implementation of such cut-offs is rather trivial, we will suppress them from here on. The surface integral on the RHS of \eqref{eq:measure} should thus be over the gauge patch, but not the matter patches. The domain of integration is hence not the whole complex $y$-plane (or the complex $u$-strip), but rather the $y$-plane (or the $u$-strip) punctured around the charged chiral singularities at $e^{2\pi ix_p}$ (or $x_p$). We thus write our gauge-patch Poisson resummation formula as
\begin{equation}
    \sum_m' \int'\mathrm{d}x=\sum_w\int_{\mathbb{C}_p}\frac{i\,\mathrm{d}y\wedge\mathrm{d}\bar{y}}{2\pi\beta|y|^2}\,e^{-\frac{4\pi^2}{\beta}w(u-\bar u)}=\sum_w\int_{S_p}\frac{2\pi i\,\mathrm{d}u\wedge\mathrm{d}\bar{u}}{\beta}\,e^{-\frac{4\pi^2}{\beta}w(u-\bar u)}.\label{eq:punctured_PR}
\end{equation}
Here $\mathbb{C}_p:=\mathbb{C}\setminus\{p_j\}$ (resp $S_p:=S\setminus\{p_j\}$) denotes the $y$-plane (resp $u$-strip), with $\mathcal{O}(\epsilon)$-sized punctures $p_j$ around the charged chiral singularities removed.

The above expressions give the explicit transformation for one factor $U(1) \subset U(1)^{\text{rk}(G)}$. For groups of higher rank than one, the Poisson resummation turns the sum over the (magnetic) lattice of co-weights $\Lambda_m$ into a sum over the (electric) weight lattice $\Lambda_w$.
This allows us to write the gauge-patch piece of \eqref{eq:master_EFT} as\footnote{The factor of $1/\beta^{r_G}$ arises from the measure replacement \eqref{eq:2d_vs_3d_measure}. Note that it cancels against the Cartan photon multiplets contribution to $\Omega-\bar\Omega$; see \eqref{eq:W_Omega_simpl_out}.}
\begin{equation}
   \frac{Z_{S^2}^{\rho=0,\frac{\nu r_\Phi}{2}\in\mathbb{Z}}}{\beta^{r_G}|W|\mathrm{vol}\Lambda_m}\sum_{w\in\Lambda_w}\int_{S^{r_G}\setminus\{s_j\}}\left(\prod_j\frac{2\pi i\,\mathrm{d}u_j\wedge\mathrm{d}\bar{u}_j}{\beta}\right)\,e^{-\frac{4\pi^2}{\beta}w^a(u_a-\bar u_a)+W-\overline{W}-i\pi\big(\Omega-\overline{\Omega}\big)},
\end{equation}
where $s_j$ denote $\mathcal{O}(\epsilon)$ neighborhoods of the singular strata in the multi-strip $S^{r_G}$ covered by $u_j$. 
The Poisson (or Dirac comb) exponent can be incorporated into $W$ by defining
\begin{equation}
    W^{(w)}=W-\frac{4\pi^2}{\beta} w^a u_a\,.
\end{equation}

We thus get
\begin{equation}
\label{eq:surf_int_out}
    \boxed{Z_{\text{gp}} = \frac{Z_{S^2}^{\rho=0,\frac{\nu r_\Phi}{2}\in\mathbb{Z}}}{\beta^{r_G}|W|\mathrm{vol}\Lambda_m}
    \sum_{w \in \Lambda_w} \int_{\mathbb{C}_p^{\text{rk}(G)}}\left(\prod_{a = 1}^{\text{rk}(G)}\frac{i\mathrm{d}y_a\wedge\mathrm{d}\bar{y}_a}{2\pi\beta|y_a|^2}\right)\ e^{\,W^{(w)}-\overline{W^{(w)}}-i\pi\big(\Omega-\overline{\Omega}\big)},}
\end{equation}
with the subscript gp standing for the gauge patch $S^{r_G}\setminus\{s_j\}.$

\subsection*{Singularities, branch cuts, and monodromies}

As $\beta\to0,$ the points on the BPS moduli space that get dangerously close to singularities can be characterized as\footnote{Recall that the gauge enhancement loci $\{u\ |\  e^{{2\pi i}\,\alpha(u)}=1\}$ do not introduce singular behavior in the integrand or the asymptotics.}
\begin{equation}
    \mathcal{S}:=\{u\ |\ e^{{2\pi i}(\rho(u)+\frac{\nu}{2} r_\Phi)}=1\}.
\end{equation}
These are located at $u=x_{p_j},$ as seen from \eqref{eq:xpj_def}, which is where charged chirals become light in the 2d EFT. These singularities sit at the end of cuts as we explain next.

We have radial cuts on the $y_i$-plane arising from $W^H$, emanating from the matter singularities and going to infinity if $\rho_i>0$, and to zero if $\rho_i<0$. These cuts are formed by sequences of poles of the integrand of the 3d index condensing in the Cardy limit.

Let us now discuss the jumps across the cuts. First, we consider the monodromies in $W.$ Upon crossing the cuts on the $y_i$-plane, we have a shift in $W^H$ arising from 
\begin{equation}
    \mathrm{Li}_2(z)\xrightarrow{z\,\to\, e^{2\pi i}z\,}\begin{cases}
        \mathrm{Li}_2(z)+2\pi i\log|z| \qquad &\text{$|z|>1$,}\\
        \mathrm{Li}_2(z) &|z|\le1.
    \end{cases}
\end{equation}
A full $2\pi$ rotation on the $y_i$-plane, which makes the argument of a term $\frac{1}{\beta}\mathrm{Li}_2(y^\rho\, e^{i\pi\nu r})$ in $W^H$ cross $\rho_i$-many cuts of the type just described, yields
\begin{equation}
    \Delta_i W^H=2\pi i\rho_i\,\frac{\rho(m)}{2}\,\frac{1+\mathrm{sign}(\rho(m))}{2}=-\frac{4\pi^2 i}{\beta}\rho^{}_i\rho^{}_j\,\mathrm{Im}u_j\Big(\frac{1-\mathrm{sign}(\mathrm{Im}\rho(u))}{2}\Big).
\end{equation}
Importantly, this is equivalent to a change of $w_j$ as
\begin{equation}
    \Delta_i w_j=\rho_i\rho_j\,\Big(\frac{1+\mathrm{sign}(\rho(m))}{2}\Big)=\rho_i\rho_j\,\Big(\frac{1-\mathrm{sign}(\mathrm{Im}\rho(u))}{2}\Big).\label{eq:matter_delta_k}
\end{equation}

The monodromy arising from $W_{\text{CS}^+}$ under $u_i\to u_i+1$ is
\begin{equation}
    \Delta_i W_{\text{CS}^+}=-\frac{4\pi^2}{\beta}k^{+}_{ij}u_j,
\end{equation}
modulo real constants that cancel against those from $\overline{W}$. The corresponding cut appears to be from zero to infinity on the $y_i$ plane, along an arbitrarily fixed ray. Again, this monodromy is equivalent to changing the winding sector:
\begin{equation}
    \Delta_i w_j=k^+_{ij}.\label{eq:CS_delta_k}
\end{equation}

All other monodromies, in particular those arising from $\Omega^H$, cancel between the holomorphic and anti-holomorphic pieces.

Since we can think of the sum over $w$ alternatively as a sum over infinitely many sheets of the $y_i$-plane, we see that the shifts \eqref{eq:matter_delta_k} and \eqref{eq:CS_delta_k} can be thought of as moving on a higher sheet of the $y_i$-plane, such that the integrand remains continuous over the multi-sheeted domain. In this work, we do not adopt this perspective and instead consider one $y_i$-plane for each $w$ sector. In other words, we always take $\frac{1}{2}<\mathrm{Re}u_j\leq\frac{1}{2}$, and sum over the winding sectors. However, the smooth multi-sheeted perspective guarantees one thing for us: that even the apparent branch cut discontinuities do not pose an obstacle to our Poisson resummation.

\subsubsection{EFT interpretation as dualizing and summing over windings}

On the (complexified) outer patch, our Poisson resummation has a clear interpretation in the 2d EFT. It corresponds to dualizing the $\mathcal{N}=(2,2)$ Cartan photon gauge multiplets to their field strength (flux-constrained) twisted chiral multiplet, and then trading the latter for a Lagrange multiplier (periodic, but unconstrained) twisted chiral superfield which has winding sectors.

As mentioned above, we do not Poisson re-sum on the patches $p^{}_i$ supporting massless charged chirals.

There remain the gauge inner patches $g_j$. On these patches, our Poisson resummation seems related to the proposals in \cite{Aharony:2016jki,Gu:2018fpm}. This point deserves further investigation. See Problem~\textbf{6}.

\subsection*{Inside the punctures}


A puncture consists of an intermediate zone and a near zone. As in the rest of this paper, let us consider only the Gaussian zone of the intermediate zone. On the Gaussian zone we can Poisson resum, because we would be far away from the singularities.

The near zone is where the real difficulty with Poisson resummation arises. In cases where the near-zone EFT is non-chiral, its asymptotic contribution is actually simple enough to be determined directly (essentially thanks to the multiscale decomposition having isolated the relevant scale in the problem). This is how various near-zone contributions will be estimated in the next subsection, as well as on the 1st sheet of our non-chiral examples in Section~\ref{sec:example}.

If the near-zone EFT is chiral, a new scale ($x=\mathcal{O}(\frac{1}{\log\frac{1}{\beta}})$) associated with the 2d running FI parameter arises, which prevents a direct evaluation of the asymptotics. It may be possible to further decompose the near zone in such cases to bring the asymptotic contributions under control, but we do not attempt that in this work. In preparation for future continuations of our project, however, we note that duality suggests an alternative way forward: one may be able to appeal to the GLSM/Landau-Ginzburg dualities, to rewrite the contributions of the near zones in a Landau-Ginzburg form that is amenable to standard tools of asymptotic analysis. This seems to be related to how the contribution of a near zone in the $U(1)_{1/2}+\Phi_{+1}$ example around Eq.~\eqref{eq:Iin_LambdaToInfty} is computed in \cite{Benini:2012ui}. At a computational level, the problem is approached in \cite{Benini:2012ui} by Poisson resummation after residue evaluation.

\subsection{Asymptotics}\label{subsec:asymptotics}

We divide the cases of interest into two classes, with each requiring a separate strategy.

The first class of problems, which we call \emph{the exponential class}, has
\begin{equation}
    W^{(w)}-\overline{W}^{(w)}\propto \frac{f(u)-f(\bar u)}{\beta},
\end{equation}
on the punctured $u$-strip (or the complexified outer patch) with a non-linear $f(u)$. The appropriate strategy for the asymptotic analysis of these problems is the multidimensional stationary phase method, reviewed in Appendix~\ref{app:AA}.


The stationary phase method allows writing the asymptotics of \eqref{eq:surf_int_out} as a sum over contributions from small neighborhoods of the stationary-phase points of $W^{(w)}-\overline{W^{(w)}}.$ The latter are easily shown to coincide with the critical points of $W^{(w)}$, which are in turn encoded in a Bethe-type equation for $W$.

Of these critical points, some lie inside gauge inner patches.
These can be shown, by a careful analysis via the stationary phase method order by order in $\beta$, to give contributions that are suppressed by powers of $\beta$, due to completely destructive interferences. (Presumably, an all-order analysis would demonstrate suppression to all orders.) This \emph{screening} of the gauge inner patches can actually be demonstrated beyond all orders, via holomorphic blocks, as we shall explain in Section~\ref{sec:factorization&screening}.

There could also be critical points corresponding to $\mathrm{Im}u\to\pm\infty.$ These are subject to screening in the non-abelian cases as well, so that the actual asymptotics of the index is governed by the non-gauge-enhancing Bethe roots at finite $u.$

There remain the punctures. We will argue below that in cases where the index has exponential growth, the exponential growth is captured correctly by pretending that the outer-patch estimates extend to the punctures as well. If the naive extension yields power-law (rather than exponential) growth, one has to compute the puncture contributions more carefully, which we do through multiscale decomposition into near zones and Gaussian zones.

\vspace{.3cm}
The second class of problems, which we call \emph{the power-law class},
has
\begin{equation}
    W^{(w)}-\overline{W}^{(w)}=\mathcal{O}(\beta^0),
\end{equation}
in a certain winding sector $w_0$, and has $W^{(w)}\propto u/\beta$ in all other sectors such that they lack stationary-phase points and are suppressed. We propose that the appropriate strategy for the asymptotic analysis of these problems is multiscale decomposition in the $w_0$ sector, into a (complexified) outer patch, Gaussian zones, and near zones. The rest of the winding sectors are suppressed by at least additional powers of $\beta$ (and in fact beyond all orders as we argue in Section~\ref{sec:factorization&screening}).

\vspace{.3cm}
Second-sheet indices typically belong to the exponential class, while first-sheet indices of non-chiral theories belong to the power-law class.

\subsubsection{Exponential growth: typical on the 2nd sheet}\label{subsec:exponential}


We begin with the contribution of the punctured $u$-strip, and discuss the puncture contributions afterwards.

\subsection*{Punctured strip}

We recite the expression for the contribution of the punctured $u_a$-strips (or the complexified gauge patch) to the index from \eqref{eq:surf_int_out}:
\begin{equation}
\label{eq:surf_int_out_u}
    Z_{\text{gp}} \approx \frac{Z_{S^2}^{\rho=0,\frac{\nu r_\Phi}{2}\in\mathbb{Z}}}{|W|\,\mathrm{vol}\Lambda_m}\sum_{w \in \Lambda_w} \int_{\text{gp}}\frac{\prod_{a = 1}^{\text{rk}(G)}2\pi i\,\mathrm{d}u_a\wedge\mathrm{d}\bar{u}_a}{\beta^{\,2\,\mathrm{rk}(G)}}\ e^{\,W^{(w)}-\overline{W^{(w)}}-i\pi(\Omega-\overline{\Omega})}.
\end{equation}

We are assuming that the problem belongs to the exponential class, so that
\begin{equation}
    W^{(w)}-\overline{W^{(w)}}=\frac{f(u)-f(\bar u)-4\pi^2 w(u-\bar u)}{\beta},
\end{equation}
with $f(u)$ a non-linear function. The method of stationary phase instructs us to look for  points $u^\ast$ where
\begin{equation}
    \partial_{u_x}(f(u)-f(\bar u)-4\pi^2 w(u-\bar u))=0,\qquad \partial_{u_y}(f(u)-f(\bar u)-4\pi^2 w(u-\bar u))=0,
\end{equation}
where $u_x,u_y$ are defined through $u=u_x+i\,u_y\,.$ It is easily seen from $\partial_{u_x}=\partial_u+\partial_{\bar u}$ and $\partial_{u_y}=i(\partial_u-\partial_{\bar u})$ that the above two equations combine to give
\begin{equation}
    \frac{\partial}{\partial u_a}f(u)=4\pi^2 w^a\Longrightarrow \frac{\partial}{\partial u_a}W^{(w)}(u)=0\Longrightarrow \frac{\partial}{\partial u_a}W(u)=\frac{4\pi^2 w^a}{\beta}\,.\label{eq:nonExpBetheEq}
\end{equation}
The winding sectors are often made implicit by rewriting the latter equation as
\begin{equation}
    \exp\Big(\frac{\beta}{2\pi i}\,\partial_{u_a}\!W\Big)=1\,.\label{eq:ExpBetheEq}
\end{equation}
This is referred to as the Bethe-type equation of the outer-patch\footnote{Its interpretation on gauge inner patches seems more subtle. See Problem~\textbf{6}.} 2d $\mathcal{N}=(2,2)$ EFT.

Since the derivatives of the poly-logarithms give logarithms, the Bethe equation will simply give a polynomial equation in the $y_a$ variables, which will have finitely many solutions.\\

If there are no stationary-phase points outside the punctures, Eq.~\eqref{eq:no_stationary_asy_d} implies that the punctured $u$-plane gives an asymptotic contribution to the index that is
\begin{equation}
    \mathcal{O}(\beta^{\mathrm{Tr}_{L_\text{o}}\!R\,-r_G+\frac{1}{2}}).\label{eq:no_stationary_asy_gp}
\end{equation}
This is a combination of the explicit factor of $\beta^{r_G+\frac{1}{2}}$ arising from the measure \eqref{eq:no_stationary_asy_d}, the explicit factor $\beta^{-2r_G}$ in \eqref{eq:surf_int_out_u}, and the contribution $\beta^{\mathrm{Tr}^{}_{\!L^{}_\text{o}}\!\!R}$ of the light multiplets $L_\text{o}$ on the outer patch to $\Omega^L$.

When there \emph{are} points of stationary phase inside the punctured $u$-plane, we can use \eqref{eq:stationary_asy}. Denoting them by $u^\ast$, and assuming that they are isolated, we get the sum over the leading contributions from each stationary-phase point as:
\begin{equation}
    Z(\beta) \approx Z_{S^2}^{\rho=0,\frac{\nu r_\Phi}{2}\in\mathbb{Z}}\sum_{u^\ast}\frac{1}{|W_{u^\ast}|}\,\frac{e^{\,W^{(w^\ast)}-\overline{W^{(w^\ast)}}-i\pi(\Omega-\overline{\Omega})}}{\big|\mathrm{det}_{a,b}\,\partial^2W/\partial u_a\partial u_b\big|\, \cdot\,\beta^{r_G}}\Bigg|_{u=u^\ast}.
\end{equation}
We have not included the sum over $w^\ast$, since $w^\ast$ and $u^\ast$ are tied via \eqref{eq:nonExpBetheEq}. To deduce the factor $1/|W_{u^\ast}|$, where $|W_{u^\ast}|$ is the order of the Weyl group preserved at $u^\ast$, we argue as follows. If $u^\ast$ breaks $W$ to $W_{u^\ast}$, then there are as many as $|W|/|W_{u^\ast}|$ degenerate Bethe roots. We identify such roots on a Weyl orbit as one. Summing over the orbit multiplies the $1/|W|$ in \eqref{eq:surf_int_out_u} to give $1/|W_{u^\ast}|$.

The above result can be rewritten in the form:
\begin{equation}
    \boxed{Z(\beta) \approx Z_{S^2}^{\rho=0,\frac{\nu r_\Phi}{2}\in\mathbb{Z}}\sum_{u^\ast\in S_p}\frac{1}{|W_{u^\ast}|}\,\frac{\mathcal{F}}{\mathcal{H}^{1/2}\,\overline{\mathcal{H}^{1/2}}\ \ \overline{\mathcal{F}}}\Bigg|_{u=u^\ast}\Big(1+\mathcal{O}(\beta)\Big),}
    \label{eq:stationary_asy_main}
\end{equation}
with
\begin{equation}
    \mathcal{F}(u^\ast;\beta):=e^{W^{(w^\ast)}(u^\ast;\,\beta)}\overset{\eqref{eq:nonExpBetheEq}}{=}e^{W-u_a\,\partial_{u_a}W}\bigg|_{u=u^\ast}\!\!,\quad\mathcal{H}(u;\beta):=\beta^{r_G}e^{2\pi i\,\Omega}\,\mathrm{det}_{a,b}\,\frac{\partial^2\big(\frac{\beta\, W}{4\pi^2}\big)}{\partial u_a\partial u_b}.
\end{equation}
We have replaced $W^{(w)}$ with $W$ in $\mathcal{H}$, because the second derivatives kill the difference, which is linear in $u$. A simple calculation also shows that $\mathcal{H}\propto \beta^{-\mathrm{Tr}R|_{\rho=0,\frac{\nu r_\Phi}{2}\in\mathbb{Z}}}$, so $\mathcal{H}$ is independent of $\beta$ in the absence of Cartan-neutral chirals ($\rho=0$) with $\frac{\nu r_\Phi}{2}\in\mathbb{Z}$.

\vspace{.3cm}
In various examples below, we will find that the set of Bethe roots $u^\ast$ has more than one element. We expect that each (Weyl orbit) of them corresponds to a vacuum of the 2d $\mathcal{N}=(2,2)$ EFT; except when the root is screened!


\paragraph{\small Perturbative screening of saddles.} It may happen that $\mathcal{H}(u^\ast;\beta)\to\infty.$ Then the leading term of the $u^\ast$ contribution shown in \eqref{eq:stationary_asy_main} would vanish. Absent additional information, we expect the next-to-leading term of the $u^\ast$ contribution to be nonzero, and hence the suppression due to $\mathcal{H}(u^\ast;\beta)=\infty$ be of order $\beta$. We refer to this scenario as a \emph{perturbative} (in the high-temperature EFT) or power-law screening. More of the subleading terms of the $u^\ast$ contribution may be zero due to various cancellations, and we would still call this power-law screening if the leading $u^\ast$ contribution is of the form
\begin{equation}
    \frac{e^{W^{(w^\ast)}(u^\ast;\,\beta)-\overline{W^{(w^\ast)}(u^\ast;\,\beta)}}}{\beta^{\mathrm{Tr}R|_{\rho=0,\frac{\nu r_\Phi}{2}\in\mathbb{Z}}}}\times \beta^n,
\end{equation}
with a non-zero coefficient, and some $n>0$.

\paragraph{\small Non-perturbative screening through the lens of factorization.} In non-abelian examples below we see that for gauge-enhancing saddles (those fixed by a nontrivial subgroup of the Weyl group) that do not support light charged chirals, not only $\mathcal{H}(u^\ast;\beta)\to\infty$, but also that the cancellations persist to all orders. In fact, such saddles are completely suppressed. We refer to this scenario as \emph{non-perturbative} or exponential screening. Instead of a tedious order-by-order demonstration of such cancellations, we appeal to the powerful tool of Lorentzian factorization to study this phenomenon. We shall see that the blocks arising from the factorization have an odd integrand (and a contour that can be deformed to remain intact) under the said subgroup of the Weyl group, and thus vanish.

Another class of Bethe roots subject to screening are those with infinite or vanishing $y_a$. We will see in the next section an abelian example (the electric side of the elementary mirror symmetry on the first sheet) where such a saddle exists and is not screened, and a non-abelian example (SU(2) SQCD with $N_f=3$ on the second sheet) where the saddle appears to be non-perturbatively screened.

\subsection*{Punctures}

When the outer-patch Bethe equation suggests that there is no Bethe root inside a puncture, we expect that the contribution of the puncture is negligible. Arguments analogous to those around Eq.~(2.56) of \cite{Ardehali:2021irq} might help establish this rigorously, but we leave that for future work.

If the outer-patch Bethe equation suggests a Bethe root inside a puncture, we expect, as in \cite{Ardehali:2021irq}, that simply extending the outer-patch result to the puncture correctly captures the exponential asymptotics. We moreover conjecture that the power-law and the $O(\beta^0)$ asymptotics are captured correctly by the near zone or the Gaussian zone in such cases. (Our main example is the first sheet of the electric side of the duality appetizer.) Demonstrating these appears to be more difficult than the analogous claims in~\cite{Ardehali:2021irq}; stronger control over the near zones is needed (possibly through GLSM/LG dualities, see Problem~\textbf{7}), which we leave for the future.




\subsubsection{Power-law growth: typical on the 1st sheet}\label{subsec:power_law}

The simplest scenario for power-law growth of the 3d index is when the outer-patch effective twisted superpotential vanishes. This happens in particular on the 1st sheet of non-chiral theories. The fluctuations hence spread over the {outer patch}, which has size $\propto\frac{1}{\beta^{\,r^{}_G}}$, yielding power-law growth. In our examples below, we will encounter this simple scenario on the 1st sheet of SQED in Section~\ref{subsec:SQED}. 

Interestingly, we will also encounter more sophisticated scenarios on the 1st sheet of various other theories. For SQED$_{N_f=3}$ in Section~\ref{subsec:SQED_Nf3} and SU(2) SQCD$_{N_f=3}$ in Section~\ref{subsec:SQCD}, we will find a second scenario where fluctuations are enhanced on near zones. So even though the near zones have parametrically smaller size compared to the outer patch, they end up dominating over and setting the leading power-law asymptotic in those cases.

There is a third scenario, in chiral cases, where the outer-patch twisted superpotential does not identically vanish, but since it ends up localizing the fluctuations to neighborhoods where the superpotential goes to zero, we get sub-exponential growth. For the $U(1)_{1/2}$ theory with a charge-1 chiral multiplet, also known as the electric side of the elementary mirror symmetry, we will find in Section~\ref{subsec:mirror_sym} a non-vanishing outer-patch twisted superpotential on the 1st sheet (as the theory is chiral due to non-vectorlike matter content as well as tree-level CS coupling). The non-zero superpotential yields runaway behavior (on magnetic flux sectors with $m\sim\frac{1}{\beta}$) reminiscent of Liouville theory. The dominant contribution comes from asymptotic \emph{large-flux sectors} (with $m\sim\frac{1}{\beta}\log\frac{1}{\beta}$) in that case, and since the superpotential asymptotes to zero at large fluxes, we get power-law growth. The SU(2)$_{-2}$ theory with a chiral multiplet in the triplet representation, aka the electric side of the duality appetizer, discussed in Section~\ref{subsec:appetizer} also has a non-vanishing outer-patch twisted superpotential on the 1st sheet (as the theory is chiral due to non-zero tree-level CS coupling). The twisted superpotential in this case localizes the fluctuations to the intermediate zones between the outer patch and the two disjoint near zones (related by center symmetry). More precisely, the \emph{Gaussian zones} dominate, and since the outer-patch superpotential goes to zero as we approach the intermediate zones, we again get power-law asymptotics.


Since power-law growth is typical on the 1st sheet, we focus on the 1st sheet in this subsection. However, our methods can be applied to similar situations even if they arise on higher sheets.

\subsection*{Punctured strip}

As explained above,  theories with non-vectorlike matter typically have a nontrivial outer-patch effective twisted superpotential even on the 1st sheet. So the method of analysis of the punctured $u$-strip is similar to the typical 2nd sheet case discussed already. We will thus focus on vectorlike theories here. By definition, the matter content of these theories is such that whenever there is matter associated to a weight $\rho$ of $G$, there is also matter associated to the weight $-\rho$.

First, assume that there is no tree-level CS term. Then on the gauge patch of a theory with vectorlike matter content, we would not have any exponential ($e^{1/\beta}$ type) growth in the integrand of the index. This is because when the heavy multiplets come in pairs of opposite weights $\rho$, $-\rho$ under the gauge group, we can benefit from the simplification:
\begin{equation}
    W^H_\rho+W^H_{-\rho}=\frac{1}{\beta}\big(\mathrm{Li}_2(e^{2\pi i\rho(u)})+\mathrm{Li}_2(e^{-2\pi i\rho(u)})\big)=\frac{1}{\beta}\Big(\!-\frac{\pi^2}{6}+2\pi^2\big(\rho(\tilde{u})\pm\frac{1}{2}\big)^2\Big).\label{eq:WH_pair}
\end{equation}
Here $\rho(\tilde u)= \rho(u)\ \mathrm{mod}\,\mathbb{Z}$ such that $\mathrm{Re}\rho(\tilde{u})\in(-\frac{1}{2},\frac{1}{2}].$ The ambiguous sign on the RHS can be shown to be $-\mathrm{sign}\,\mathrm{Re}\,\rho(\tilde{u}),$ although that is not important for our purposes.
What is important, is that the $u$-independent terms on the RHS cancel against those in $\overline{W}^H_\rho+\overline{W}^H_{-\rho}$, the quadratic part in $u$ cancels against the quadratic chiral multiplet contributions to $W_{\text{CS}^+},$ and the linear part in $u$ combines with the linear chiral multiplet contributions to $W_{\text{CS}^+}$ to yield a trivial phase.\footnote{To demonstrate the latter simplification we must use $\rho_j^2\pm\rho_j=\rho_j(\rho_j\pm1)\in2\mathbb{Z}\,.$} 

In such cases, we see from \eqref{eq:surf_int_out} that the $w=0$ sector contributes
\begin{equation}
    \begin{split}
    \beta^{\mathrm{Tr}^{}_{L_\text{o}}\!R-2r^{}_G}\frac{Z_{S^2}^{\rho=0}}{\mathrm{vol}\Lambda_m}
    \int_{\mathbb{C}_p^{\text{rk}(G)}}\left(\prod_{a = 1}^{r^{}_G}\frac{i\mathrm{d}y_a\wedge\mathrm{d}\bar{y}_a}{2\pi |y_a|^2|W|}\right)\ &e^{-i\pi k^+_{jR}(u_j-\bar u_j)}\\
    &\times\frac{\prod_{\alpha_+}|y^{\,\frac{\alpha_+}{2}}-y^{\,-\frac{\alpha_+}{2}}|^2}{\prod_{\Phi,\,\rho^{}_\Phi\neq0}|1-y^{\rho^{}_\Phi}|^{1-r^{}_\Phi}},
    \end{split}\label{eq:1st_power_gen}
\end{equation}
to the index. This is
\begin{equation}
    \mathcal{O}(\beta^{\mathrm{Tr}^{}_{L_\text{o}}\!R-2r^{}_G}),\label{eq:outer_1st_sheet}
\end{equation}
and arises as a combination of the explicit factor of $1/\beta^{2r_G}$ in \eqref{eq:surf_int_out} and $\Omega^{L}$ as in \eqref{eq:W_Omega_simpl_out}.

It may be that the integral over the $y$-plane(s) is divergent in the limit $\epsilon\to0$ of shrinking a puncture supporting light charged matter, due to the factors in the denominator of the second line in \eqref{eq:1st_power_gen}. For simplicity, let us focus on rank one. Then a simple power-counting, keeping track of the zeros of the denominator and numerator of the second line in \eqref{eq:1st_power_gen}, shows that we have such divergence if
\begin{equation}
    \mathrm{Tr}_{L^c_p}\,R+1\leq-1\,.\label{eq:outer_div_criter_rank1}
\end{equation}
Here $\mathrm{Tr}_{L^c_p}\,R$ indicates the sum over the $R$-charges of those charged (hence the subscript $c$) fermions, including the gaugini, that becomes light inside the corresponding puncture---leading to zeros in the numerator or denominator of \eqref{eq:1st_power_gen}. The $+1$ on the LHS corresponds to the radial coordinate $r$ arising from the measure $r\mathrm{d}r\mathrm{d}\theta$, while the inequality demands that the integrand diverges as or faster than $1/r$ as $r\to0.$ 
Such divergences of the outer-patch integrals as $\epsilon\to0$ would signify that the corresponding intermediate zone, whose contribution is more and more included as the puncture shrinks, has a significantly larger contribution, and dominates over the outer patch. (The intermediate zone in turn may be itself dominated by the near zone inside it, in the $\Lambda\to\infty$ limit.) 

The trace in relation \eqref{eq:outer_div_criter_rank1} seems related to the $R$-charge of instanton-vortex configurations in the near zone EFT. This relation might allow generalizing \eqref{eq:outer_div_criter_rank1} to higher rank, and giving it physical meaning. We leave that to future work.

With a nonzero bare CS level, the outer patch effective twisted superpotential would be nonzero. The most general 1st-sheet outer-patch effective twisted superpotential for theories with vectorlike matter takes the form
\begin{equation}
\label{eq:genericfirstouter}
    {W}^{(w)}(u;\beta) =-\frac{4\pi^2}{2\beta}\big( k_{ij}u_i u_j + k_{jj} u_j+2w_j u_j\big) ,
\end{equation}
with $k_{ij}$ the bare CS level, and $w_j$ the Dirac comb  integer (or the winding). It is clear that the corresponding saddles $u^\ast$ are real-valued, and so are the critical values of ${W}^{(w)}$. These real-valued critical values cancel against the critical value of $\overline{\mathcal{W}}$. We thus get no exponential growth. Instead we expect generically a power-law behavior in the Cardy limit, with the dominant contribution coming from the region
around $u^\ast$.
Typically, the stationary points of the twisted superpotential fall inside punctures, in which case the gauge patch itself would be further suppressed (than $\mathcal{O}(\beta^{\mathrm{Tr}_{L_\text{o}}\!R-2r_G})$) due to its lack of stationary-phase points. See Section~\ref{subsec:appetizer} for an example.

\subsection*{Punctures}

Punctures (complexified inner patches) consist of near zones and intermediate zones.

On the 1st sheet, a distinguished near zone is the one around $x_j=0$. For $|m_j|=\mathcal{O}(\beta^0)<\Lambda_m,$ it has the property that all the 3d multiplets are light. Therefore $W^H=\Omega^H=0.$ Moreover, since $x_j=\mathcal{O}(\beta),$ we have
\begin{equation}
\begin{split}
    W_{CS}&=-\pi k^+_{jj}{u^{}_{2d}}_j+\mathcal{O}(\beta),\\
    \Omega_{CS}&=\mathcal{O}(\beta).
\end{split}
\end{equation}
We thus get the simplification
\begin{equation}
\begin{split}
    Z_{\text{nz}_0}&\approx\left(\frac{1}{2\pi}\right)^{r_G}\ \frac{1}{\beta^{\frac{\ \text{\footnotesize$c$}_{\text{\tiny$L$}_{p}}}{\text{\footnotesize3}\ }}}\ \sum_{|m| <\Lambda_m} \int_{-\Lambda}^{\Lambda} \frac{\mathrm{d}^{r_G}x^{}_{2d}}{|W|}\ e^{-\pi k^+_{jj}({u_{2d}}^{}_j-\bar{u}_{2d_j})}\beta^{-i\sum_{\rho}\rho(u_{2d}+{\bar{u}_{2d}})}\\
    &\qquad\left(\prod_{\alpha_+}\big(\alpha_+(u_{2d})\big)\big(\alpha_+(\bar{u}_{2d})\big)\right)\left(\prod_{\Phi}\prod_{\rho^\Phi} \frac{\Gamma\big(\frac{r_\Phi}{2}-i\rho(u_{2d})\big)}{\Gamma\big(1-\frac{r_\Phi}{2}+i\rho(\bar{u}_{2d})\big)}\right),
    \end{split}\label{eq:in0_1st_sheet}
\end{equation}
with $L_p$ the set of light fields on the near zone. Recall that $u_{2d}=x_{2d}-i\frac{m}{2}$. The contribution of a general near zone is given by \eqref{eq:master_EFT}.

It may be that a near zone contribution is divergent as $\Lambda_m,\Lambda\to\infty$. Assume vectorlike matter, so that the $\beta^{-i\sum_\rho\rho(u_{2d}+\bar{u}_{2d})}$ factor in the integrand on the 1st line of \eqref{eq:in0_1st_sheet} drops out. Then we can use the Stirling approximation together with the integral test for convergence of series, to obtain the large $u$ behavior of the integrand:
\begin{equation}
\label{eq:powers}
    Z_{\text{nz}_0} \sim \frac{1}{\beta^{\frac{\ \text{\footnotesize$c$}_{\text{\tiny$L$}_{p}}}{\text{\footnotesize3}\ }}}
   \int^{\Lambda}\mathrm{d}^{r_G}u_{2d}\ \mathrm{d}^{r_G}\bar{u}_{2d}\ \, e^{-\pi k_{jj}({u_{2d}}^{}_j-\bar{u}_{2d_j})}\, \frac{\prod_{\alpha_+}|\alpha_+(u_{2d})|^2}{\ \prod_{\Phi}\prod_{\rho\in R^+_\Phi}|\rho(u_{2d})^{1-r_\Phi}|^2}\,,
\end{equation}
We assume for simplicity that moreover $k_{jj}\in2\mathbb{Z},$ so that the factor $e^{-\pi k_{jj}({u_{2d}}^{}_j-\bar{u}_{2d_j})}$ in the above integral becomes a trivial phase before Poisson resummation and can be dropped. Even then, finding a general formula for the asymptotic of the integral in \eqref{eq:powers} seems difficult, possibly requiring tools from real algebraic geometry and hyperplane arrangements. So again we restrict attention to rank one. Convergence or divergence of the integral then follows from a simple power counting. In particular we get divergence if
\begin{equation}
    \mathrm{Tr}_{L^c_p}\,R+1\ge-1\,,\label{eq:inner_div_criter_rank1}
\end{equation}
where $\mathrm{Tr}_{L^c_p}\,R$ indicates the sum over the $R$-charges of those charged fermions, including the gaugini, that are light on the near zone. Such divergence would signify that the surrounding intermediate zone (whose contribution is more and more included as $\Lambda_m,\Lambda\to\infty$) has a significantly larger contribution, dominating over the near zone.

As discussed below Eq.~\eqref{eq:outer_div_criter_rank1}, we expect the higher rank generalization of \eqref{eq:inner_div_criter_rank1} to be related to the $R$-charges of instanton-vortices.

Now consider the intermediate zones.
Focusing on the Gaussian zone of the intermediate zone, using $u=\sqrt{\beta}u_\text{gz}$ (or more generally \eqref{eq:ugz_def}), we have from 
\eqref{eq:1st_power_gen} at rank one the simplification:
\begin{equation}
\begin{split}
    \beta^{\mathrm{Tr}^{}_{L_\text{o}}\!R-1}\ Z_{S^2}^{\rho=0}
    \int_{\text{g.z}}\ \frac{2\pi i\,\mathrm{d}u_\text{gz}\wedge\mathrm{d}\bar{u}_\text{gz}}{ |W|}\ &e^{-{4\pi^2}(\frac{1}{2}k_{gg}(u_\text{gz}^2-\bar u_\text{gz}^2+\frac{u^{}_\text{gz}-\bar u^{}_\text{gz}}{\sqrt{\beta}}))}\\
    &\hspace{-1cm}\times{|2\pi \sqrt{\beta}\alpha_+(u_\text{gz})|^2}{\prod_{\Phi,\,\rho^{}_\Phi\neq0}|2\pi\sqrt{\beta}\rho(u_\text{gz})|^{r^{}_\Phi-1}}.\label{eq:gaussian_rank1_gen_int}
    \end{split}
\end{equation}
where the $\beta^{\mathrm{Tr}^{}_{L_\text{o}}\!R-1}$ in the prefactor arises as a combination of the $\beta^{\mathrm{Tr}^{}_{L_\text{o}}\!R-2}$ prefactor in \eqref{eq:1st_power_gen}, and a $\beta$ factor arising from rescaling the measure: ${\mathrm{d}u\wedge\mathrm{d}\bar u}= {\beta}\ {\mathrm{d}u_\text{gz}\wedge\mathrm{d}\bar u_\text{gz}}$.
We assume for simplicity that moreover $k_{gg}\in2\mathbb{Z},$ so that the factor $e^{-2\pi^2 k_{gg}\frac{({u_\text{gz}}-\bar{u}_\text{gz})}{\sqrt{\beta}}}$ in the above integral becomes a trivial phase before Poisson resummation and can be dropped. Then the Gaussian zone contribution becomes
\begin{equation}
    \mathcal{O}(\beta^{\mathrm{Tr}^{}_{L_\text{o}}\!R-1}\times\beta^{\frac{\mathrm{Tr}_M R}{2}}),\label{eq:gaussian_rank1_gen}
\end{equation}
with the factor $\beta^{\frac{\mathrm{Tr}_M R}{2}}$ arising from the middleweight multiplet contributions on the second line of \eqref{eq:gaussian_rank1_gen_int}. 

\subsubsection{Summary for rank-one gauge theories}\label{subsec:asy_summary}

Assume first that we have a non-linear effective twisted superpotential $W(u)$ on the punctured $u$-strip. This is generically the case \textbf{\small on the second sheet, or on the first sheet of chiral theories}. 
There are often Bethe roots---i.e. solutions to \eqref{eq:ExpBetheEq}---in the finite punctured $u$-strip. By \eqref{eq:stationary_asy_main}, these roots amount to exponential growth of the index. In non-abelian cases, the Bethe roots at gauge-enhancement points (if there are no light charged chirals there), as well as those at infinite flux $\mathrm{Im}u=\pm\infty$, are screened. The perturbative screening was discussed in this section, and the non-perturbative (or beyond-all-order) versions will be discussed in Section~\ref{sec:factorization&screening}. When the finite punctured $u$-strip does not host Bethe roots, one has to see whether the punctures yield exponential growth. If not, we end up with power-law growth. The electric side of the elementary mirror symmetry and the duality appetizer exemplify this latter power-law growth scenario on their 1st sheet, as seen in the next section. These are both chiral theories with nonlinear $W(u)$ on the 1st sheet, but lack Bethe roots in the finite punctured $u$-strip. Their punctures do not contribute exponentially either---in the duality appetizer case the asymptotics is dominated by
\begin{equation}
    \begin{cases}
    \text{Gaussian zones with growth $\quad\beta^{\,\mathrm{Tr}^{}_{L_\text{o}}\!R-1+\frac{1}{2}\mathrm{Tr}^{}_{\!M} \!R}$},
    \end{cases}\label{eq:gaussian_rank1_gen_rep}
\end{equation}
as follows from the general rank-one formula \eqref{eq:gaussian_rank1_gen} valid on the first sheet of vectorlike theories with $k_{gg}\in2\mathbb{Z}$.
So both examples end up having power-law growth on the 1st sheet, despite their non-linear twisted superpotential.

Next assume the effective twisted superpotential $W(u)$ on the punctured $u$-strip is linear. This is the case on the \textbf{\small first sheet of non-chiral theories}. The index has then power-law growth. Assume that there is a sector $w$ where $W^{(w)}(u)$ is independent of $u.$ This stationary-phase sector contributes $\mathcal{O}(\beta^{\mathrm{Tr}^{}_{L_\text{o}}\!R-2})$ as in \eqref{eq:outer_1st_sheet}, and dominates over the rest of the winding sectors. The exponent of the power-law growth of the index is determined in a competition between this punctured strip and various near zones:
\begin{equation}
    \begin{cases}
        \text{punctured plane with growth $\quad\beta^{\mathrm{Tr}^{}_{L_\text{o}}\!R-2}$},\\
        \text{near zones with growth $\qquad \beta^{\mathrm{Tr}^{}_{L_p}R}$}.
    \end{cases}\label{eq:cases_pp_nz}
\end{equation}

In the special case where there is a puncture with $\mathrm{Tr}^{}_{{L^c_p}}\,R=-2$, both the punctured plane and the near zone contributions diverge logarithmically in the limit of removing the cut-offs (around the puncture or around the near zone, respectively; see \eqref{eq:outer_div_criter_rank1}, \eqref{eq:inner_div_criter_rank1}). This marginal case corresponds to when the punctured plane and the near zone contribute similarly:
    $\beta^{\mathrm{Tr}^{}_{L_\text{o}}\!R-2}= \beta^{\mathrm{Tr}^{}_{L_p}R}\,.$
To see this, note that the set $L_p$ of light multiplets on the near zone of a puncture $p$, is the union of the set $L_\text{o}$ of light multiplets on the outer patch, and the set $L^c_p$ of the charged multiplets that becomes light only inside the puncture: $L_p=L_\text{o}\,\cup\, L^c_p\,.$ Therefore $\mathrm{Tr}^{}_{L_p}R=\mathrm{Tr}^{}_{L_\text{o}}\!R+\mathrm{Tr}^{}_{L^c_p}R\xrightarrow{\mathrm{Tr}^{}_{{L^c_p}}\,R=-2} \mathrm{Tr}^{}_{L_\text{o}}\!\!R-2\,.$ We expect the logarithmic divergences (of the punctured strip and the near zone) cancel each other, and the index has then growth $\beta^{\mathrm{Tr}^{}_{L_\text{o}}\!R-2}$.


\section{Examples}\label{sec:example}

Having described our method for analyzing the Cardy limit of the index, we now apply it to various examples.

We have three principal aims in this section. First, to show that the tools developed in the previous section are \emph{almost} mature enough to establish matching of the Cardy-like asymptotics of dual indices both on the 1st and 2nd sheet, in various simple examples. We say almost because, as we shall see, the gauge-enhanced and infinite-flux saddles of non-abelian gauge theories need extra care, which we will provide in the next section through Lorentzian factorization. Second, to demonstrate that the competition between the punctured plane and punctures can be non-trivial, especially on the 1st sheet of vectorlike theories where the punctures may dominate the asymptotics, as seen in Table~\ref{tab:patches_intro}. Third, to illustrate how once the dominant ``scale'' (or patch) in the problem is identified, subleading asymptotics can often be obtained (up to relative $o(\beta)$ error) without much technical difficulty.

\subsection{Abelian examples}\label{subsec:abelian}

\addtocontents{toc}{\protect\setcounter{tocdepth}{2}}

We begin by considering 3d dualities involving on one side an ``electric'' theory consisting of a $U(1)$ gauge group with some charged matter, and on the other side a ``magnetic'' theory consisting of only chiral superfields (possibly with a non-trivial superpotential). Since the duality implies equality of the superconformal indices, we should also expect agreement of the Cardy-like asymptotics. The asymptotic analysis on the magnetic side is completely straightforward since there are no complicated sums or integrals. This will allow us to validate our methods for analyzing the more complicated electric indices. 

\subsubsection{SQED}\label{subsec:SQED}

Our first example consists of SQED with $k=0$, $k_{gR}=0$, and with one flavor, by which we mean a pair of chiral superfields with gauge charges $\pm1$ and with $U(1)_R$ charge $r_\Phi = 1/3$. This is IR dual to the XYZ model, which consists of three chiral superfields $X,Y,Z$ and a superpotential $W = XYZ$. Consistently with the superpotential, the $U(1)_R$ charge of these three chirals is $r^{}_{X,Y,Z} = 2/3$. By virtue of their IR duality, these two theories share the same superconformal index: 
\begin{equation}
    \sum_{m \in \mathbb{Z}} \oint \frac{dz}{2\pi i z}(-z)^m \frac{(q^{5/6}(z^{-1} q^{-m/2})^{\pm};q)}{(q^{1/6}(z q^{-m/2})^{\pm};q)} = \frac{(q^{2/3};q)^3}{(q^{1/3};q)^3},\label{eq:SQED=XYZ}
\end{equation}
where we have specialized to the first sheet to make the expressions more readable. We will use this identity as a check on our methods.

\begin{center}
\subsection*{\small{$\boxed{\text{Second sheet:}}$}}
\end{center}
\vspace{-.3cm}

\noindent We begin by analyzing the magnetic side without gauge group. The outer patch estimates \eqref{eq:heavy_integrand}--\eqref{eq:heavy_Omega} give:
\begin{equation}
    Z^{\text{mag}}(q\,e^{2\pi i}) = \frac{(q^{2/3}e^{4\pi i/3};q)^3}{(q^{1/3}e^{2\pi i/3};q)^3}=C\, e^{\frac{i D}{\beta}}\big(1+O(\beta)\big),\label{eq:XYZasy}
\end{equation}
with $C=\frac{1}{(1-e^{\frac{2\pi i}{3}})(1-e^{\frac{-2\pi i}{3}})}=\frac{1}{\sqrt{3}} \approx 0.58$ and $D=3\text{Im}\big(\mathrm{Li}_2(e^{2\pi i/3})-\mathrm{Li}_2(e^{4\pi i/3})\big)\approx 4.06$.\\

Now we consider the more interesting gauge theory side.

\paragraph{\small Punctured strip.}\hspace{-.3cm}The punctured $u$-strip contributes to the second-sheet index according to \eqref{eq:surf_int_out_u} as
\begin{equation}
    \sum_{w\in\mathbb{Z}}\int_{S\setminus \{p^{}_{1},p^{}_{2}\}}\frac{2\pi i\,\mathrm{d}u\wedge\mathrm{d}\bar u}{{\beta^2}}\,  e^{W^{(w)}-\overline{W^{(w)}}-i\pi(\Omega-\overline{\Omega})},
\end{equation}
where $S\setminus \{p^{}_{1},p^{}_{2}\}$ is the $u$-strip punctured around $x_{p_1}=-\frac{1}{6}$ and $x_{p_2}=\frac{1}{6}$, while
\begin{equation}
\begin{split}
    {W}&=\frac{1}{\beta}\left(-{2\pi^2}(u^2+2u)+\mathrm{Li}_2(e^{2\pi i u+i\pi/3})+\mathrm{Li}_2(e^{-2\pi i u+i\pi/3})\right),\\
    \beta \, e^{2\pi i\Omega}&={(1-e^{+ 2\pi i u+i\pi/3})^{2/3}(1-e^{- 2\pi i u+i\pi/3})^{2/3}}\,.
\end{split}
\end{equation}
Solving the Bethe equation \eqref{eq:ExpBetheEq} we find a single saddle at $u^\ast = 0$. The corresponding winding sector is $w=-1$ as can be checked from \eqref{eq:nonExpBetheEq}. 
Using \eqref{eq:stationary_asy_main} we then find:
\begin{equation}
    Z_{S\setminus \{p^{}_{1},\,p^{}_{2}\}}^\text{el}(q\,e^{2\pi i}) \approx C \exp{\frac{iD}{\beta}},
\end{equation}
where $C \approx 0.58,\ D \approx 4.06$. This matches the asymptotics on the dual side.

\paragraph{\small Punctures.}\hspace{-.3cm}Since the inner patches around $u=\pm\frac{1}{6}$ do not host Bethe vacua in the present case, we know their contribution is exponentially suppressed and hence do not consider them in detail.

\begin{center}
\subsection*{\small{$\boxed{\text{First sheet:}}$}}
\end{center}
\vspace{-.3cm}

\noindent For the side of the duality with trivial gauge group, we use the near-zone estimate \eqref{eq:qPochNearEst_main} to obtain
\begin{equation}
    Z^\text{mag}(q) = \frac{\Gamma^3(1/3)}{\Gamma^3(2/3)}\frac{1}{\beta}\,\big(1+\mathcal{O}(\beta)\big).\label{eq:SQED_1st_asy}
\end{equation}

\vspace{.3cm}
We now reproduce the above asymptotics from the gauge theory side.

\paragraph{\small Punctured strip.}\hspace{-.3cm}The $w\neq0$ terms do not have stationary-phase points. Therefore we focus on the $w=0$ sector.

The punctured $u$-strip maps to the punctured $y$-plane $\mathbb{C}\setminus p$, where $p$ is a puncture of size $\epsilon$ around $e^{2\pi i x_p}=1$. It contributes according to \eqref{eq:1st_power_gen} as:
\begin{equation}
    Z^\text{el}_{\mathbb{C}\setminus p}(q) \approx \frac{1}{\beta}\int_{\mathbb{C}\setminus p} \frac{\mathrm{d}^2y}{\pi|y|^{4/3}}\, \frac{1}{|y-1|^{4/3}}\,.
\end{equation}
 We have used $\mathrm{d}^2 y=\frac{i\,\mathrm{d}y\wedge \mathrm{d}\bar y}{2}$.
The resulting integral is of Dotsenko-Fateev form \cite{Dotsenko:1984nm}, and it has a finite value as $\epsilon\to0$, readily found in the literature: 
\begin{equation}
    R(a,b):=\int\mathrm{d}^2y\, |y|^{2a}\, |y-1|^{2b}=-\sin(\pi b)\frac{\Gamma(1+a)\Gamma(1+b)}{\Gamma(2+a+b)}\frac{\Gamma(-1-a-b)\Gamma(1+b)}{\Gamma(-a)},
\end{equation}
with $a=b=-2/3.$ We hence get
\begin{equation}
   Z^\text{el}_{\mathbb{C}\setminus p}(q)\xrightarrow{\epsilon\to0}\frac{\tilde{C}}{\beta}\,\big(1+\mathcal{O}(\beta)\big),\label{eq:Iout_epsTo0}
\end{equation}
with $\tilde{C}=(\sin(2\pi/3)/\pi)(\Gamma^4(1/3)/\Gamma^2(2/3))\approx 7.74$. This exactly reproduces the asymptotics \eqref{eq:SQED_1st_asy} of the dual index as one can check with Gamma function identities.

\paragraph{\small Puncture.}\hspace{-.3cm}On the 1st sheet the $u$-strip has a puncture around $u=0.$ We know from Section~\ref{subsec:power_law} that the near zone inside that puncture contributes $\propto\beta^{\mathrm{Tr}^{}_{L_p}R}$, which is $\mathcal{O}(\beta^{1-2\times\frac{2}{3}})=\mathcal{O}(\beta^{-1/3})$ and therefore negligible in the present case. It remains to compute the contribution of the intermediate zone inside the puncture. At finite $\epsilon,$ we know from the convergence of $Z^\text{el}_{\mathbb{C}\setminus p}(q)$ as $\epsilon\to0$, that the intermediate zone makes a contribution, from its outer end, that completes the asymptotics of $Z^\text{el}_{\mathbb{C}\setminus p}(q)$ to \eqref{eq:Iout_epsTo0} (i.e. it cancels the $\epsilon$ dependent piece of $Z^\text{el}_{\mathbb{C}\setminus p}(q)$). The only other part of the intermediate zone that we consider in this work is the Gaussian zone. Its contribution can be evaluated via \eqref{eq:gaussian_rank1_gen} as $\mathcal{O}(\beta^{\frac{2\times(\frac{1}{3}-1)}{2}})=\mathcal{O}(\beta^{-2/3})$. This is negligible compared to \eqref{eq:Iout_epsTo0}, as expected, since we have no tree-level CS coupling in this case, and hence no new scale associated with the Gaussian zone.

\subsubsection{$U(1)_{1/2}+\Phi_{+1}$ (elementary mirror symmetry)}\label{subsec:mirror_sym}

Next consider the $U(1)$ gauge theory with a single chiral multiplet of gauge charge $+1$ and $U(1)_R$ charge $r_\Phi = \frac{1}{3}$.
We take $k_{gR}=1/3,$ so that $k^+_{gR}=0.$ There is also a CS term with level $k = 1/2$ for the $U(1)$ gauge group to cancel the global gauge anomaly. This is dual to the theory of a single chiral superfield, which is identified with the gauge-invariant BPS monopole operator of the gauge theory and thus has $R$-charge
\begin{equation}
    R(m=+1) \overset{\eqref{eq:R(m)}}{=} \frac{2}{3}\,.
\end{equation}
The indices take the following form, again specialized to the first sheet for clarity:
\begin{equation}
    \sum_{m \in \mathbb{Z}} \oint \frac{dz}{2\pi i z}(-z)^m \frac{(q^{5/6}(z^{-1} q^{-m/2});q)}{(q^{1/6}(z q^{-m/2});q)} = \frac{(q^{2/3};q)}{(q^{1/3};q)}.
\end{equation}

\begin{center}
\subsection*{\small{$\boxed{\text{Second sheet:}}$}}
\end{center}
\vspace{-.3cm}

\noindent Once again, the asymptotic analysis on the magnetic side is straightforward. We find \begin{equation}
    Z^\text{mag}(q\,e^{2\pi i}) = \frac{(q^{2/3}e^{4\pi i/3};q)}{(q^{1/3}e^{2\pi i/3};q)} \approx \frac{e^{-\frac{\text{Li}_2(e^{4\pi i/3})-\text{Li}_2(e^{2\pi i/3})}{\beta}}}{(1-e^{4\pi i /3})^{1/6}(1-e^{2\pi i /3})^{1/6}} = C \exp(\frac{i D}{\beta}),
\end{equation}
where $C \approx 0.83,\ D \approx 1.35$.\\

Our task is now to reproduce the above asymptotics from the gauge theory side.

\paragraph{\small Punctured strip.}\hspace{-.3cm}On the punctured $u$-strip $S\setminus p,$ where $p$ is an $\epsilon$ sized puncture around $x_p=-1/6,$ we have
\begin{equation}
    {W}(u) = \frac{-2\pi^2(u^2+2u) + \text{Li}_2(e^{2\pi iu+i\pi/3})}{\beta},\qquad \beta\ e^{2\pi i\,\Omega(u)}={(1-e^{2\pi i\,u+i\pi/3})^{2/3}}\,.
\end{equation}
The only saddle is at $u^\ast= \frac{1}{2\pi i}\,\log \frac{1}{1+e^{i\pi/3}}$. The corresponding winding sector is $w=-1$ as can be checked from \eqref{eq:nonExpBetheEq}.
Using \eqref{eq:stationary_asy_main} we get:
\begin{equation}
    Z_{S\setminus p}^\text{el}(q\,e^{2\pi i}) \approx C \exp(\frac{i D}{\beta}),
\end{equation}
where $C = 3^{-1/6}\approx 0.83,\  D \approx 1.35$. This indeed agrees with the dual side.

\paragraph{\small Puncture.}\hspace{-.3cm}Since the puncture around $u=-\frac{1}{6}$ does not host Bethe vacua in the present case, we know its contribution is exponentially suppressed and hence do not consider it in detail.

\begin{center}
\subsection*{\small{$\boxed{\text{First sheet:}}$}}
\end{center}
\vspace{-.3cm}

\noindent On the magnetic side, we directly obtain the asymptotic scaling as
\begin{equation}
    Z^\text{mag}(q) = \frac{\Gamma(1/3)}{\Gamma(2/3)}\frac{1}{\beta^{1/3}}\,\big(1+\mathcal{O}(\beta)\big).\label{eq:electric_ems_asy}
\end{equation}

\vspace{.3cm}
We want to reproduce the above asymptotics from the gauge theory.

\paragraph{\small Punctured strip.}\hspace{-.3cm}The punctured $u$-strip maps to the punctured $y$-plane $\mathbb{C}\setminus p$, with $p$ a puncture of size $\epsilon$ around $e^{2\pi i x_p}=1.$ The contribution to the index is as in \eqref{eq:surf_int_out}:
\begin{equation}
    Z^\text{el}_{\mathbb{C}\setminus p}(q) \approx \sum_{w\in\mathbb{Z}}\int_{\mathbb{C}\setminus p} \frac{\mathrm d^2y}{\pi\beta|y|^2}\frac{1}{|1-y|^{{2/3}}}\ e^{{W}^{(w)}-\overline{W^{(w)}}},
\end{equation}
with the superpotential
\begin{equation}
    {W}^{(w)} = -\frac{4\pi^2}{\beta}\frac{(u^2+u)}{2} + \frac{\text{Li}_2\big(e^{2\pi iu}\big)}{\beta} -\frac{4\pi^2}{\beta}  w\, u\,.\label{eq:Ww:ems}
\end{equation}
The corresponding Bethe vacuum equation reads 
\begin{equation}
    \frac{1}{1-y}=\frac{1}{1-e^{2\pi i\,u}} = 0\,,\label{eq:Bethe_ems}
\end{equation}
which has a solution for $y\to\infty$ (or $\mathrm{Im}u \to -\infty$). The corresponding winding sector is $w=0$ as can be checked from \eqref{eq:nonExpBetheEq}.

To probe the region at infinity, we use the dilogarithm identity
\begin{equation}
\label{eq:reflectionish}
    \text{Li}_2(y) + \text{Li}_2(\frac{1}{y}) = -\pi^2/6 - \frac{(\log(-y))^2}{2}\,,
\end{equation}
with $y = e^{2\pi i\,u}$. This allows us to rewrite the superpotential as
\begin{equation}
    {W}^{(w)} = -\frac{\text{Li}_2(e^{-{2\pi i\,u}})}{\beta}-\frac{4\pi^2}{\beta}w\,u\,.
\end{equation}
Hence
\begin{equation}
    Z^\text{el}_{\mathbb{C}\setminus p}(q) \approx \sum_{w\in\mathbb{Z}}\frac{1}{\beta}\int_{\mathbb{C}\setminus p} \frac{\mathrm d^2y}{\pi|y|^2}\ \frac{1}{|1-y|^{{2/3}}}\ e^{\frac{1}{\beta}(-\text{Li}_2(y^{-1})+\text{Li}_2(\bar{y}^{-1})-4\pi^2\,w(u-\bar{u}))}.\label{eq:EMS_surface_simp}
\end{equation}
We moreover change variables $y\to s^{-1}$, so that the stationary-phase point of the $w=0$ term (corresponding to the Bethe vacuum as in \eqref{eq:Bethe_ems}) lies at $s=0$ rather than $y=\infty$. We get
\begin{equation}
    Z^\text{el}_{\mathbb{C}\setminus p}(q) \approx \sum_{w\in\mathbb{Z}}\frac{1}{\beta}\int_{\mathbb{C}\setminus \frac{1}{p}} \frac{\mathrm d^2s}{\pi|s|^2}\ \frac{1}{|1-s^{-1}|^{{2/3}}}\ e^{\frac{1}{\beta}(-\text{Li}_2(s)+\text{Li}_2(\bar{s})-4\pi^2\,w(u-\bar{u}))}.\label{eq:EMS_surface_simp_w}
\end{equation}
The only stationary-phase point corresponds to $w=0$ and $s=0$. Therefore the $w\neq0$ terms, as well as the region outside a small $\varepsilon$-neighborhood of $s=0$ for $w=0$, give a contribution
\begin{equation}
    \sim\frac{1}{\beta}\times\mathcal{O}(\beta)=\mathcal{O}(\beta^0),
\end{equation}
to $Z^\text{el}_{\mathbb{C}\setminus p}(q).$

We now focus on the contribution from a disk of radius $\varepsilon$ near $s=0$, in the $w=0$ sector. Writing $s=r\,e^{i\theta},$ we use convergent small-$r$ expansions to write this contribution as 
\begin{equation}
\begin{split}
    Z^\text{el}_{|s|<\varepsilon}(q)&\approx\frac{1}{\beta}\int\!\!\!\!\int_{|s|<\varepsilon}\ \frac{\mathrm d^2s}{\pi |s|^2}\ \big(|s|^{2/3}(1+\mathcal{O}(r))\big)\ e^{\frac{1}{\beta}(-s+\bar{s}+\mathcal{O}(r^2))}\\
    &=\frac{1}{\beta}\int_{0}^{\varepsilon}\int_0^{2\pi}\frac{r\mathrm{d}r\mathrm{d}\theta}{\pi r^2}\big(r^{2/3}(1+\mathcal{O}(r))\big)\ e^{-\frac{r}{\beta}(2i\sin\theta+\mathcal{O}(r))}\\
    &\overset{r=\tilde r\beta}{=}\frac{1}{\beta^{1/3}}\int_{0}^{\varepsilon/\beta}\int_0^{2\pi}\frac{\tilde r\mathrm{d}\tilde r\mathrm{d}\theta}{\pi \tilde r^2}\big(\tilde r^{2/3}(1+\mathcal{O}(\beta\,\tilde r))\big)\ e^{-{\tilde r}(2i\sin\theta+\mathcal{O}(\beta\,\tilde r))}\\
    &= \frac{1}{\beta^{1/3}}\int_{0}^{\infty}\int_0^{2\pi}\frac{\mathrm{d}\tilde r\mathrm{d}\theta}{\pi \tilde r^{1/3}}\ e^{-\tilde{r}(2i\sin\theta)}\big(1+\mathcal{O}(\beta)\big)\\&=\frac{3\times 2^{1/3}\times\Gamma(7/6)}{\sqrt{\pi}}\frac{1}{\beta^{1/3}}\big(1+\mathcal{O}(\beta)\big)\,.
    \end{split}
\end{equation}
Lagrange's duplication formula,
\begin{equation}
    \label{eq:LagrangeDuplication}
    \Gamma\left(z\right)\Gamma\left(z+\frac{1}{2}\right)=2^{1-2z}\sqrt{\pi} \,\Gamma\left(2z\right),
\end{equation}
together with  $\Gamma(z+1)=z\,\Gamma(z)$, with $z=2/3$, establish that this matches \eqref{eq:electric_ems_asy}.

We note that the dominant contribution to $Z^\text{el}_{|s|<\varepsilon}(q)$ comes from $\tilde r$ of order one, which means $y\sim\mathcal{O}(1/\beta),$ or since $|y|=e^{m\beta/2},$ from
\begin{equation}
    m\sim\mathcal{O}\big(\frac{1}{\beta}\log\frac{1}{\beta}\big)\,.
\end{equation}
This should be compared with Eqs.~(3.7) and (3.11) in \cite{Aganagic:2001uw}, where the calculations are performed in the ``Higgs phase'', with a nonzero FI parameter moving the vacuum to the zero-flux sector. See also \cite{Chen:2013pha}.

\paragraph{\small Puncture.}\hspace{-.3cm}We have a puncture around $u=0.$ First, consider the near zone $p^{}_\beta$ inside the puncture. We have according to \eqref{eq:in0_1st_sheet}
\begin{equation}
    Z_{p^{}_\beta}^\text{el}(q)\approx\frac{\beta^R}{2\pi}\sum_{|m|<\Lambda_m}e^{i\pi m}\int_{-\Lambda}^\Lambda\mathrm{d}x_{2d}\ e^{-2ix_{2d}\log\beta}\ \frac{\Gamma(\frac{R}{2}-ix_{2d}-\frac{m}{2})}{\Gamma(1-\frac{R}{2}+ix_{2d}-\frac{m}{2})},\label{eq:ZpEMS1st}
\end{equation}
with $R=1/3$. The integral can be subjected to the stationary phase method, as in Appendix~\ref{app:stationary_phase}. Since there are no stationary-phase loci, we find from \eqref{eq:no_stationary_asy} (upon a $\beta\to1/\log(1/\beta)$ replacement) that the integral is $\mathcal{O}(1/\log(1/\beta)),$ and since the sum in \eqref{eq:ZpEMS1st} has only finitely many terms we get $Z^\text{el}_{p^{}_\beta}=\mathcal{O}\big(\beta^R/\log(1/\beta)\big)$.\footnote{If $\Lambda,\Lambda_m$ were much larger than $1/\log(1/\beta)$, Eq.~(5.18) of \cite{Benini:2012ui} would imply a different asymptotics for $Z^\text{el}_{p^{}_\beta}$, but here we are interested in fixed $\Lambda,\Lambda_m$ as $\beta\to0.$ AA thanks M.~Ansari and S.~Fallahdoost for help with correcting an erroneous reliance on Eq.~(5.18) of \cite{Benini:2012ui} in the first preprint of this paper.}

It remains to compute the contribution of the intermediate zone inside the puncture. The inner end of the intermediate zone must makes a contribution comparable to $Z_{p^{}_\beta}^\text{el}(q)$, while its outer end must be negligible compared with $Z_{p^{}_\beta}^\text{el}(q)$ since it is far from the latter's stationary-phase point (which is at $u\to\infty$). The only other part of the intermediate zone that we care about in this work is the Gaussian zone. We have a contribution $\mathcal{O}(\beta^{\mathrm{Tr}_MR/2})=\mathcal{O}(\beta^{-1/3})$ there from the middleweight chiral multiplet, which is comparable to that of the punctured strip. However, since we do not have stationary-phase points inside the Gaussian zone, we get suppression by an additional power of $\beta.$
The Gaussian zone is hence negligible compared to the $\mathcal{O}(\beta^{-1/3})$ contribution of the punctured strip.

\subsubsection*{Possibility of exponential growth on the first sheet}

Non-chiral theories have nonzero outer-patch effective twisted superpotential even on the first sheet. We saw this in the preceding example in Eq.~\eqref{eq:Ww:ems}. In this special case, the critical value of $W^{(w)}-\overline{W^{(w)}}$ was zero and we obtained power-law growth for the first-sheet index. That the index does not have exponential growth on the first sheet in this example was of course guaranteed by the dual description.

However, it is not difficult to find sufficiently chiral examples where the critical value of $W^{(w)}-\overline{W^{(w)}}$ is nonzero, leading to exponential growth for the first-sheet index. We have checked that $U(1)_{5/2}+\Phi_{+1}$ is such an example.

Since first-sheet indices have a $q$ expansion with integer coefficients, their exponential asymptotics is not of the form $e^{i\frac{\#}{\beta}}$ as for typical second-sheet indices, but rather goes like $\cos\frac{\#}{\beta}.$ This of course still yields growth once $\beta$ is complexified.

\subsubsection{SQED $N_f = 3$}\label{subsec:SQED_Nf3}

For our final abelian example, we consider the case of SQED coupled to 3 flavors with $R=\frac{1}{3}$. This is dual to a theory of 9 chiral superfields with $R=\frac{2}{3}$, giving the index identity
\begin{equation}
    \sum_{m \in \mathbb{Z}} \oint \frac{dz}{2\pi i z}(-z)^{3m} \left(\frac{(q^{5/6}(z^{-1} q^{-m/2})^{\pm};q)}{(q^{1/6}(z q^{-m/2})^{\pm};q)}\right)^3 =\frac{(q^{2/3};q)^9}{(q^{1/3};q)^9}\,.
\end{equation}

We begin again with analyzing the second sheet and then discuss the more subtle first sheet.

\begin{center}
\subsection*{\small{$\boxed{\text{Second sheet:}}$}}
\end{center}
\vspace{-.3cm}

\noindent On the magnetic side, we have the following Cardy-like behavior:
\begin{equation}
    Z^\text{mag}(q\,e^{2\pi i}) \approx C e^{\frac{iD}{\beta}},\label{eq:Nf3_mag_2nd}
\end{equation}
where $C \approx 0.19$, $D \approx 12.18$.\\

We now reproduce the above result from the gauge theory.

\paragraph{\small Punctured strip.}\hspace{-.3cm}SQED with $N_f=3$ on the second sheet has its $u$-strip punctured around $x_{p_1}=-\frac{1}{6}$ and $x_{p_2}=\frac{1}{6}$. On the punctured $u$-strip:
\begin{equation}
    \hspace{-.4cm}{W}(u) = \frac{-6\pi^2(u^2+2u) + 3\text{Li}_2(e^{2\pi i\,u+i\frac{\pi}{3}}) + 3\text{Li}_2(e^{-2\pi i\,u+i\frac{\pi}{3}})}{\beta},\quad \beta\ e^{2\pi i\, \Omega(u)}={(1-e^{\pm 2\pi i\,u+i\frac{\pi}{3}})^2}\,.
\end{equation}
There is a single Bethe vacuum at $u^\ast = 0$ in the $w=-3$ sector. From \eqref{eq:stationary_asy_main} we get
\begin{equation}
    Z_{S\setminus\{p_1,\,p_2\}}^\text{el}(q\,e^{2\pi i}) \approx C e^{\frac{iD}{\beta}}\,,
\end{equation}
with $C \approx 0.19$, $D \approx 12.18$, giving a match with the magnetic side \eqref{eq:Nf3_mag_2nd}.

\paragraph{\small Punctures.}\hspace{-.3cm}Since the inner patches near $u=\pm\frac{1}{6}$ do not host Bethe vacua in the present case, we know their contribution is exponentially suppressed and hence do not consider them in detail.

\begin{center}
\subsection*{\small{$\boxed{\text{First sheet:}}$}}
\end{center}
\vspace{-.3cm}

\noindent It is straightforward to obtain the asymptotics of the matter theory. We obtain
\begin{equation}
    Z^\text{mag}(q) = \frac{C^9}{\beta^3}\big(1+\mathcal{O}(\beta)\big),
\end{equation}
with $C= \frac{\Gamma(1/3)}{\Gamma(2/3)}$.\\

Our task is to reproduce the above asymptotics on the gauge theory side.

\paragraph{\small Punctured strip.}\hspace{-.3cm}Similarly to the SQED case above, and in accordance with the general discussion in Section~\ref{subsec:power_law}---in particular \eqref{eq:outer_1st_sheet}---on the 1st sheet we find
\begin{equation}
    Z_{S\setminus p}^\text{el}(q)=\mathcal{O}(\beta^{-1}),
\end{equation}
where $p$ is the $\epsilon$ puncture around $u=0$. Looking more carefully at the $\mathcal{O}(\beta^0)$ piece of $Z_{S\setminus p}^\text{el}(q)$ reveals that it is, unlike the SQED case above, divergent in the limit of shrinking the puncture (see Eq.~\eqref{eq:outer_1st_SQCD} for a similar analysis). This signals that the punctured strip is dominated by the contribution from the intermediate zone (which is being more and more included as the puncture shrinks). As we shall see, in this case the intermediate zone is itself dominated by the near zone.

\paragraph{\small Puncture.}\hspace{-.3cm}We begin with the near zone $p^{}_\beta$ inside the puncture. We have according to \eqref{eq:in0_1st_sheet}:
\begin{equation}
    Z_{p^{}_\beta}^\text{el}(q)\approx\frac{\beta^{6R-5}}{2\pi}\sum_{|m|<\Lambda_m}\int_{-\Lambda}^\Lambda\mathrm{d}x_{2d}\ \left(\frac{\Gamma(\frac{R}{2}-ix_{2d}-\frac{m}{2})}{\Gamma(1-\frac{R}{2}+ix_{2d}-\frac{m}{2})}\frac{\Gamma(\frac{R}{2}+ix_{2d}+\frac{m}{2})}{\Gamma(1-\frac{R}{2}-ix_{2d}+\frac{m}{2})}\right)^3,\label{eq:Iin_SQED_Nf3}
\end{equation}
where $R=1/3$, and $p^{}_\beta$ is the puncture of size $\Lambda\beta/2\pi$ around $u=0$.
We rewrite the above result as
\begin{equation}
    Z_{ p^{}_\beta}^\text{el}(q)\approx\frac{1}{\beta^3}Z^{\text{$N_{\!f}\!=\!3$ SQED$_2$}}_{S^2}(\Lambda,\Lambda_m),\label{eq:near_Nf3}
\end{equation}
with $\Lambda,\Lambda_m$ the cut-offs on the Coulomb branch scalar and the magnetic flux, while 
\begin{equation}
    Z^{\text{$N_{\!f}\!=\!3$ SQED$_2$}}_{S^2}(\Lambda,\Lambda_m):=\sum_{|m|<\Lambda_m}\int_{-\Lambda}^\Lambda\frac{\mathrm{d}x_{2d}}{2\pi}\ \left(\frac{\Gamma(\frac{R}{2}-ix_{2d}-\frac{m}{2})}{\Gamma(1-\frac{R}{2}+ix_{2d}-\frac{m}{2})}\frac{\Gamma(\frac{R}{2}+ix_{2d}+\frac{m}{2})}{\Gamma(1-\frac{R}{2}-ix_{2d}+\frac{m}{2})}\right)^3.
\end{equation}

Now we first note that $Z^{\text{$N_{\!f}\!=\!3$ SQED$_2$}}_{S^2}(\Lambda,\Lambda_m)$ has a finite limit as $\Lambda,\Lambda_m\to\infty$. This can be established by applying the integral test to \eqref{eq:Iin_SQED_Nf3}, which amounts to analyzing
\begin{equation}
    \int_\mathbb{C}\mathrm{d}^2u_{2d}\  \left\vert\frac{\Gamma(1/6-iu_{2d})}{\Gamma(5/6-iu_{2d})}\right\vert ^6.
\end{equation}
The convergence of this is easy to demonstrate using Stirling's approximation:
\begin{equation}
     \left\vert\frac{\Gamma(1/6-iu_{2d})}{\Gamma(5/6-iu_{2d})}\right\vert ^6 \sim |u_{2d}|^{-4},
\end{equation}
so the integral is clearly convergent. A numerical evaluation with Mathematica by picking $\Lambda=50$ and $\Lambda_m=100$ yields
\begin{equation}
    Z^{\text{$N_{\!f}\!=\!3$ SQED$_2$}}_{S^2}:=\lim_{\Lambda,\Lambda_m\to\infty}Z^{\text{$N_{\!f}\!=\!3$ SQED$_2$}}_{S^2}(\Lambda,\Lambda_m)\approx\left(\frac{\Gamma(1/3)}{\Gamma(2/3)}\right)^9,\label{eq:2d_duality?}
\end{equation}
up to a (presumably numerical) error of less than $2\%$.
This establishes the match with the magnetic side, up to the said numerical error.

The identity in \eqref{eq:2d_duality?} is suggestive of a 2d $\mathcal{N}=(2,2)$ duality. We leave further investigation of this point to future work.

Finally, we treat the Gaussian zone inside the puncture. Its contribution is found from \eqref{eq:gaussian_rank1_gen} to be $\mathcal{O}(\beta^{\frac{6\times(\frac{1}{3}-1)}{2}})=\mathcal{O}(\beta^{-2})$. This is negligible compared to \eqref{eq:near_Nf3}, as expected, since we have no tree-level CS coupling in this case, and hence no new scale associated with the Gaussian zone.

\addtocontents{toc}{\protect\setcounter{tocdepth}{3}}

\subsection{Non-abelian examples and screened saddles}\label{subsec:nonabelian}


Having seen that our methods correctly reproduce the expected Cardy-like asymptotics for a few Abelian theories, we turn our attention to non-Abelian examples. Here, we will be confronted with the issue of apparent saddles at gauge-enhancing points on the moduli space, where a subgroup of the Weyl group acts trivially. We will find that if the theory
\begin{itemize}
    \item[$i$)] is non-chiral (i.e. zero CS level $k_{ab}$ and vectorlike matter content),
    \item[$ii$)] has no light charged chirals on the gauge enhancing patch (as generically happens on the second sheet),
\end{itemize}
the gauge-enhancing saddle is fully screened. In this section, we demonstrate suppression of such saddles by powers of $\beta$, which we refer to as perturbative screening. In the next section, we show how holomorphic block decompositions allow demonstrating full (or beyond-all-order) suppression of such saddles, which we will refer to as non-perturbative screening.


\addtocontents{toc}{\protect\setcounter{tocdepth}{2}}

\subsubsection{SU(2) SQCD $N_f=3$}\label{subsec:SQCD}

Our first non-abelian example is $\mathrm{SU}(2)$ SQCD with $N_f = 3$. Again, by flavor we mean a pair of chiral multiplets in the fundamental and anti-fundamental representations. These two representations are equivalent for $\mathrm{SU}(2)$, so there is a power of $(3+3=)$$6$ in the index below. The first sheet index for this theory can be written as 
\begin{equation}
Z^\text{el}(q) = \sum_{m\in\mathbb{Z}}\oint \frac{dz}{2\pi i z}  \left(z^{m}\frac{(q^{5/6}(q^{m/2}z)^{\pm1};q)}{(q^{1/6}(q^{-m/2}z)^{\pm1};q)}\right)^6 \frac{(1-q^{-m} z^{\pm2})}{2q^{-m}}.\label{eq:SQCDindex}
\end{equation}
This theory is Seiberg dual to a theory consisting of $15$ chiral multiplets, whose index takes the form 
\begin{equation}
Z^\text{mag}(q) = \frac{(q^{2/3};q)^{15}}{(q^{1/3};q)^{15}}.
\end{equation}

\vspace{.3cm}
We begin again with the second sheet. In the non-abelian case, however, even the second sheet has subtle aspects related to screenings of various saddles.

\begin{center}
\subsection*{\footnotesize{$\boxed{\text{Second sheet:}}$}}
\end{center}
\vspace{-.3cm}

\noindent As is by now usual, we begin by analyzing the simple, magnetic side asymptotics. One directly finds 
\begin{equation}
    Z^\text{mag}(q\,e^{2\pi i}) \to C e^{\frac{iD}{\beta}},\label{eq:sqcd_mag_2nd_asy}
\end{equation}
where $C = \frac{1}{9\sqrt{3}}$ and $D = 15 \left(\text{Li}_2(e^{2\pi i/3}) - \text{Li}_2(e^{4\pi i/3})\right) \approx 20.30 $.\\

We want to reproduce the above asymptotics from the gauge theory side.

\paragraph{\small Punctured strip.}\hspace{-.3cm}On the second sheet of SU$(2)$ SQCD with $N_f=3$, away from the two $\epsilon$ punctures at $x_{p_1}=\frac{1}{6},$ $x_{p_2}=-\frac{1}{6}$, we have
\begin{equation}
    \hspace{-.6cm}W(u)=\frac{-12\pi^2(u^2+2u)+6\big(\mathrm{Li}_2(e^{2\pi iu+i\frac{\pi}{3}})+\mathrm{Li}_2(e^{-2\pi iu+i\frac{\pi}{3}})\big)}{\beta}\,,\quad \beta\ e^{2\pi i\Omega(u)}=\frac{(1-e^{\pm 2\pi iu + i\pi/3})^{4}}{(e^{-2\pi iu}-e^{2\pi iu})^2}\,.\label{eq:W_Omega_SQCD}
\end{equation}
The Bethe equation \eqref{eq:ExpBetheEq} reads:
\begin{equation}
    ye^{i\pi  w/3}=\frac{1-y e^{i\pi/3}}{1-y^{-1}e^{i\pi/3}}\Rightarrow y=2,\,1,\,\frac{1}{2},\,0,\,-1,\,\infty,
\end{equation}
with the solutions corresponding to 
\begin{equation}
    w=-7,\ \,-6,\ \,-5,\ \,-4\text{ {\it\&}$\,-10$},\ \,-3\text{ {\it\&}$\,-9$},\ \,-2\text{ {\it\&}$\,-8$},\label{eq:SQCD_stationary_windings}
\end{equation}
respectively, according to \eqref{eq:nonExpBetheEq}. 

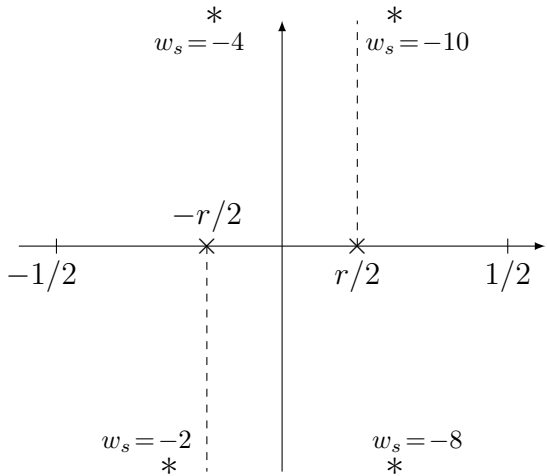
\begin{figure}
    \centering
    \begin{tikzpicture}
    \draw[-latex] (-3.5,0) -- (3.5,0) node[right] {};
    \draw[-latex] (0,-3) -- (0,3) node[above] {};

    \foreach \x in {-3, 3} {
        \draw (\x, 0.1) -- (\x, -0.1);
    }
    \node at (-3.2, -0.4) {$-1/2$}; 
    \node at (3, -0.4) {$1/2$};

    \node at (-1, 0) { $\times$};
    \node at (1, 0) { $\times$};
    \node at (-1, 0.4) {$-r/2$}; 
    \node at (1, -0.4) {$r/2$};

    \node at (-1.5, -3) {\large *};
    \node at (-1.8, -2.6) {\footnotesize$w_s\!=\!-2$};

    \node at (1.5, -3) {\large *};
    \node at (1.8, -2.6) {\footnotesize$w_s\!=\!-8$};

    \node at (-.9, 3) {\large *};
    \node at (-1.1, 2.7) {\footnotesize$w_s\!=\!-4$};

    \node at (1.5, 3) {\large *};
    \node at (1.8, 2.7) {\footnotesize$w_s\!=\!-10$};

    \draw[dashed] (-1,0) -- (-1,-3);
    \draw[dashed] (1,0) -- (1,3);

\end{tikzpicture}
\caption{The $u$-strip with the saddles at $\mathrm{Im}u=\pm\infty$ on the second sheet of $\mathrm{SU}(2)$ SQCD$_{N_f=3}$ marked by $\ast$. The corresponding winding sector is denoted $w_s$. The cuts on the $u$-strip arising from the dilogarithms in \eqref{eq:W_Omega_SQCD} are indicated by dashed lines.}
\label{fig:infinity_saddles}
\end{figure}

Two comments are in order regarding the windings in \eqref{eq:SQCD_stationary_windings}. First, the degeneracy $-3\text{ {\it\&}$\,-9$}$ has a very natural interpretation: $w=-3\text{ (or$\,-9$)}$ corresponds to $u^\ast=-\frac{1}{2}\text{ (or$\,\frac{1}{2}$)},$ each of which is on the boundary of the $u$-strip; instead of thinking of each as ``half a stationary-phase point'' (because they lie at the boundary), we combine them into a single ``full stationary-phase point''. Second, the degeneracy between $-4$ and $-10$, and between $-8$ and $-2$, has a natural interpretation: $y=0,\,\infty$ corresponds to $\mathrm{Im}u=+\infty,\,-\infty$, but the latter information is insufficient for locating a saddle on the $u$-strip due to the branch cuts; one has to also ``pick a side'' of the branch cuts, and then $w=-4$ and $w=-2$ correspond to the left side while $w=-10$ and $w=-8$ to the right. See Figure~\ref{fig:infinity_saddles}. In a sense, the saddles at $\mathrm{Im}u^\ast=\pm\infty$ also lie at ``the boundary'' of the $u$-strip, so it is natural for them to come in pairs to complete them into ``full stationary-phase points''.

Modulo the Weyl group action $y\to\frac{1}{y}$ (and modulo the pairings described in the previous paragraph), we have a total of four saddles at $y=1,\ 1/2,\ 0,\ -1.$ The critical point at $y=1/2$ gives according to \eqref{eq:stationary_asy_main}:
\begin{equation}
    Z^\text{el}_{S\setminus\{p_1,\,p_2\}}(q\,e^{2\pi i}) \to C e^{\frac{iD}{\beta}},\label{eq:SQCD_2nd_sheet_asy}
\end{equation}
with $C = \frac{1}{9\sqrt{3}}$ and $D = 15 \left(\text{Li}_2(e^{2\pi i/3}) - \text{Li}_2(e^{4\pi i/3})\right) \approx 20.30 $, matching our expectation \eqref{eq:sqcd_mag_2nd_asy} from duality.

It remains to discuss the two gauge-enhancing saddles at $y=1,-1,$ and the infinite-flux saddle at $y=0.$ Evaluation of $\mathcal{F}$ on these saddles naively suggests an $e^{\frac{i \text{\#}}{\beta}}$ growth, with $\text{\#}=\frac{6}{5}D,$ $-\frac{4}{5}D$, $0$, for $y=1,$ $-1$, $0$, respectively. However, on all three of these saddles $\mathcal{H}$ diverges, and therefore equation \eqref{eq:stationary_asy_main} gives zero. The divergence of $\mathcal{H}$ on the saddle does not by itself imply that the $\mathcal{O}(\beta)$ term in \eqref{eq:stationary_asy_main} is also zero. In other words, so far we only know that our saddle is screened by a factor of $\beta.$

Subleading corrections to \eqref{eq:stationary_asy_main} can be computed, and it turns out that the $\mathcal{O}(\beta)$ term in \eqref{eq:stationary_asy_main} is also zero. More and more accurate estimates would presumably show suppression/screening by higher and higher powers of $\beta.$ Although we do not attempt this, we expect \emph{perturbative} screening of this form to be possible to demonstrate to all orders in $\beta$. In Section~\ref{sec:factorization&screening} we will use techniques that are \emph{non-perturbative} in the Kaluza-Klein EFT, to demonstrate all-order (resp. exponential) screening of the saddles at $y=0$ (resp. $y=\pm1$).

\paragraph{\small Punctures.}\hspace{-.3cm}Since the punctures around $u=\pm\frac{1}{6}$ do not host Bethe vacua in the present case, we know their contribution is exponentially suppressed and hence do not consider them in detail.

\vspace{.2cm}
\begin{center}
\subsection*{\footnotesize{$\boxed{\text{First sheet:}}$}}
\end{center}
\vspace{-.3cm}

\noindent As in the previous cases, the asymptotic analysis on the magnetic side is immediate. We obtain
\begin{equation}
    \frac{(q^{2/3};q)^{15}}{(q^{1/3};q)^{15}}=\frac{C}{\beta^5}(1+O(\beta)),\label{eq:sqcd_mag_1st_asy}
\end{equation}
with $C=(\Gamma(1/3)/\Gamma(2/3))^{15} \approx 27 835.2$.\\

We now reproduce the above asymptotics from the gauge theory side.
\paragraph{\small Punctured strip.}\hspace{-.3cm}
On the first sheet the $u$-strip has a puncture around $x_p=0$. Only the $w=0$ sector has stationary phase. It gives according to \eqref{eq:1st_power_gen}:
\begin{equation}
    Z^\text{el}_{\mathbb{C}\setminus p}(q) \approx\int_{\mathbb{C}\setminus p} \frac{\mathrm{d}^2y}{2\pi\beta|y|^8}\, \frac{(1-y^{2})(1-\bar{y}^{2})}{|1-y^{-1}|^8}\, =\frac{1}{\beta}\int_{\mathbb{C}\setminus p} \frac{\mathrm{d}^2y}{2\pi}\, \frac{|y+1|^2}{|y-1|^6},\label{eq:outer_1st_SQCD}
\end{equation}
where $p$ is an $\epsilon$ puncture around $y=1.$ The $\mathcal{O}(1/\beta)$ scaling is in accordance with \eqref{eq:outer_1st_sheet}.
In the limit $\epsilon\to0$, the integral yields $R(-3,1)$, which is divergent, the divergence coming from near $y=1$. 
This is indicative that the dominant contribution to the index comes from the puncture, which is being more and more included in \eqref{eq:outer_1st_SQCD} as the puncture shrinks.

\paragraph{\small Puncture.}\hspace{-.3cm}We begin with the near zone $p^{}_\beta$ of the puncture. As explained in Section~\ref{subsec:power_law}---and in particular \eqref{eq:in0_1st_sheet}---we get ($\mathrm{Tr}R=3+6\times2(\frac{1}{3}-1)=-5$)
\begin{equation}
    Z_{ p^{}_\beta}^\text{el}(q)\approx\frac{1}{\beta^5}Z^{\text{$N_{\!f}\!=\!3$ SQCD$_2$}}_{S^2}(\Lambda,\Lambda_m),\label{eq:sqcd_1st_asy_nz}
\end{equation}
where
\begin{equation}
    Z^{\text{$N_{\!f}\!=\!3$ SQCD$_2$}}_{S^2}(\Lambda,\Lambda_m)=\frac{1}{2\pi}\sum_{|m|<\Lambda_m}\int_{-\Lambda}^\Lambda\frac{\mathrm{d}x_{2d}}{2}\,4|u_{2d}|^2\, \left\vert\frac{\Gamma(1/6-i u_{2d})}{\Gamma(5/6-iu_{2d})}\right\vert ^{12}.
\end{equation}
The ratio of $\Gamma$ functions is well behaved at small $u_{2d}$, and using the Stirling approximation, we can easily see that the integrand behaves well at large $u_{2d}$:
\begin{equation}
    \left\vert\frac{\Gamma(1/6-iu_{2d})}{\Gamma(5/6-iu_{2d})}\right\vert ^{12} \sim |u_{2d}|^{-8} \,.
\end{equation}
The integral is clearly convergent. A numerical evaluation with Mathematica by picking $\Lambda=50$ and $\Lambda_m=100$ yields
\begin{equation}
    Z^{\text{\,SU(2)\,SQCD$_{N_f=3}$}}_{S^2}:=\lim_{\Lambda,\Lambda_m\to\infty}Z^{\text{SU(2)\,SQCD$_{N_f=3}$}}_{S^2}(\Lambda,\Lambda_m)\approx\left(\frac{\Gamma(1/3)}{\Gamma(2/3)}\right)^{15},\label{eq:2d_duality2?}
\end{equation}
up to an error of less than $10^{-9}\%$. This establishes the matching of \eqref{eq:sqcd_1st_asy_nz} with the magnetic side \eqref{eq:sqcd_mag_1st_asy}, up to the said numerical error.

The identity in \eqref{eq:2d_duality2?} is suggestive of a 2d $\mathcal{N}=(2,2)$ duality, but we are not aware of such a duality being mentioned in the literature.


Finally, for the Gaussian zone of the puncture we get $O(\beta^{\frac{6\times2\times(\frac{1}{3}-1)+2}{2}})=\mathcal{O}(\beta^{-3})$ from \eqref{eq:gaussian_rank1_gen}. This is negligible as
expected, since we have no tree-level CS coupling in this case, and hence no new scale associated with the Gaussian zone.

\subsubsection{SU(2)$_{-2}\,+\Phi_{\Box\!\Box}$ (duality appetizer)}\label{subsec:appetizer}

Next consider the case of $\mathrm{SU}(2)$ Chern-Simons theory at level $-2$,
coupled to an adjoint chiral multiplet $\Phi$. As first pointed out in \cite{Jafferis:2011ns}, this is dual to a single free chiral superfield $X \sim \text{Tr}\Phi^2$.\footnote{The topological sector involved in the duality is irrelevant for us, as we have turned off the relevant fugacity. Note also that our normalization of the CS level is different from that of \cite{Jafferis:2011ns}.} Taking the $R$ charge of $\Phi$ to be $1/4$, we must assign $R$ charge $1/2$ to $X$. This leads to the following superconformal index identity:
\begin{equation}
    \sum_{m \in \mathbb{Z}} \oint \frac{dz}{2\pi i z}z^{2m}\frac{(q^{7/8}(q^{m/2}z)^{\pm2};q)(q^{7/8};q)}{(q^{1/8}(q^{-m/2}z)^{\pm2};q)(q^{1/8};q)}\frac{(1-q^{-m} z^{\pm2})}{2q^{-m}}=\frac{(q^{3/4};q)}{(q^{1/4};q)}.
\end{equation}

We begin again with the second sheet and then treat the first sheet.

\begin{center}
\subsection*{\small$\boxed{\text{Second sheet:}}$}
\end{center}
\vspace{-.3cm}

\noindent The result from the free chiral side is immediate; we obtain
\begin{equation}
    Z^\text{mag}(q\,e^{2\pi i}) \approx C e^{iD/\beta},\label{eq:appetizer_mag_2nd_asy}
\end{equation}
with $D \approx 1.83,\ C\approx 0.841$.\\

We now reproduce the above asymptotics from the gauge theory.
\paragraph{\small Punctured strip.}\hspace{-.3cm}On the electric side, outside the punctures around $x_p=\pm\frac{1}{16}$ we have
\begin{equation}
\begin{split}
    W(u)&=\frac{-4\pi^2(u^2+2u)+\text{Li}_2(e^{-4\pi iu + i\frac{\pi}{4}})+\text{Li}_2(e^{i\frac{\pi}{4}})+\text{Li}_2(e^{4\pi iu + i\frac{\pi}{4}})}{\beta}\,,\\
    \beta\ e^{2\pi i\Omega(u)}&=\frac{(1-e^{ 4\pi iu + i\pi/4})^{3/4}(1-e^{ i\pi/4})^{3/4}(1-e^{- 4\pi iu + i\pi/4})^{3/4}}{(e^{-2\pi iu}-e^{2\pi iu})^2}\,.
    \end{split}
\end{equation}

There are six Bethe roots: a pair at $u=0,1/2$ related via center symmetry and each fixed by the Weyl group, as well as a quadruple $\{\pm u^\ast,\pm u^\ast\mp\frac{1}{2}\}$, with $u^\ast\approx 0.1145 + 0.0817 i$ related via center symmetry and the Weyl group action.
The roots at $u=0,1/2$ are perturbatively screened since their $\mathcal{H}$ diverges. We expect that these saddles are in fact non-perturbatively screened, although our argument in the next section does not apply to cases with nonzero CS level.
The remaining Bethe roots give according to \eqref{eq:stationary_asy_main}:
\begin{equation}
    Z^\text{el}_{S\setminus\{p_1,\,p_2\}}(q\,e^{2\pi i}) \approx C e^{iD/\beta},
\end{equation}
with $D \approx 1.83,\ C \approx 0.841$, and with $p_{1,2}$ the $\epsilon$ punctures around $x_p=\pm\frac{1}{16}.$ This is in agreement with the asymptotics \eqref{eq:appetizer_mag_2nd_asy}
of the magnetic dual.

\paragraph{\small Punctures.}\hspace{-.3cm}Since the punctures near $u=\pm\frac{1}{16}$ do not host Bethe vacua in the present case, we know that their contribution is exponentially suppressed and hence do not consider them in detail.

\begin{center}
\subsection*{\small{$\boxed{\text{First sheet:}}$}}
\end{center}
\vspace{-.3cm}

\noindent The magnetic side gives
\begin{equation}
    Z^\text{mag}(q) \approx \frac{C}{\beta^{1/2}},\label{eq:mag_appetizer_asy}
\end{equation}
with $C=\frac{\Gamma(1/4)}{\Gamma(3/4)}$.\\
\color{black}

We now reproduce the above asymptotics from the gauge theory.
\paragraph{\small Punctured strip.}\hspace{-.3cm}On the first sheet we have two punctures this time: one around $y_{p_1}=1$ and the other around $y_{p_2}=-1$. The contribution of the punctured $y$-plane is
\begin{equation}
    \begin{split}
    \label{eq:surfaceSU2}
    Z_{\mathbb{C}\setminus\{p_1,\,p_2\}}^\text{el}(q)\approx\beta^{-3/4}\,\frac{\Gamma(1/8)}{\Gamma(7/8)}\,&\frac{1}{4\pi\beta}\sum_w\int_{\mathbb{C}\setminus\{p_1,\,p_2\}}\mathrm{d}^2y\  \frac{e^{-\frac{4\pi^2 (u^2+u)}{\beta}}}{(q^{1/8}y^{\pm2};q)}\frac{(1-y^2)}{y^2}\\
    &\times e^{\frac{4\pi^2 (\bar{u}^2+\bar u)}{\beta}}(q^{7/8}\bar{y}^{\pm2};q)\frac{1-\bar{y}^2}{\bar{y}^2}e^{-\frac{4\pi^2 w \left(u-\bar{u}\right)}{\beta}}.
    \end{split}
\end{equation} 
Away from the punctures around $u=0,\pm1/2$, we have the superpotential
\begin{equation}
W^{(w)} =-\frac{4\pi^2}{\beta}\big(-u^2+w\,u\big)\,.
\end{equation}
This yields two vacua: one at $u=0$, $w=0$, and the other its center symmetry image at $u=\frac{1}{2}$, $w=1$ (which pairs up with $u=-\frac{1}{2}$, $w=-1$ to form a ``full stationary-phase point''). However, these points are inside the punctures, and are therefore excised from the integration over the punctured strip. Instead, for integrals with stationary phases outside the integration region, we find the result from \eqref{eq:no_stationary_asy_gp} to be
\begin{equation}
    \mathcal{O}(\beta^{\overbrace{\text{\scriptsize${1}/{4}-1+1$}}^{\mathrm{Tr}_{L_\text{\tiny o}}\!R}-1+\frac{1}{2}})=\mathcal{O}({\beta^{-1/4}})\,.\label{eq:outer_appetizer_1st_sheet}
\end{equation}
The result is sub-leading when compared to the desired asymptotic contribution \eqref{eq:mag_appetizer_asy} that we expect from the magnetic side of the duality.

\paragraph{\small Punctures: near zones.}\hspace{-.3cm}On the 1st sheet we have punctures around $u=0,\pm1/2.$ We know from Section~\ref{subsec:power_law} that near zones of a non-chiral theory on the 1st sheet contribute $\propto\beta^{\mathrm{Tr}^{}_{L_p}R}$, which is $\mathcal{O}(\beta^{3-3\times\frac{3}{4}})=\mathcal{O}(\beta^{3/4})$ in the present case. (The fact that we have a nonzero tree-level CS coupling does not change the result, because the CS contributions are negligible on the near zones; in particular, the quadratic-in-$u'_{2d}$ term of $W_{\text{CS}^+}$ is $\mathcal{O}(\beta)$, as seen from e.g.~\eqref{eq:WCS_OmegaCS}.) Since $\mathrm{Tr}_{L_p^c}R=2-2\times\frac{3}{4}=\frac{1}{2},$ we know from \eqref{eq:inner_div_criter_rank1} that these contributions diverge in the limit $\Lambda\to\infty$ though, signaling that they are dominated by the intermediate zone.

\paragraph{\small Punctures: intermediate zones.}\hspace{-.3cm}
Only the $w=0$ sector has stationary phase. Focusing on the Gaussian zone of the intermediate zone, the $w=0$ sector gives according to \eqref{eq:gaussian_rank1_gen_int} (in the limit where the cut-offs are removed $\epsilon_\sigma\to0,\,\Lambda_\sigma\to\infty$)
\begin{equation}
\begin{split}
    Z^{\text{el}_{\,}}_{p_1^\text{g.z}}(q)&\approx \beta^{-3/4}\frac{\Gamma(1/8)}{\Gamma(7/8)}\iint \frac{2\pi i\,\mathrm{d}u_\text{gz}\wedge\mathrm{d}\bar u_\text{gz}}{2} \ e^{4\pi^2(u_\text{gz}^2-\bar u_\text{gz}^2)}(16\pi^2\beta u_\text{gz}\bar u_\text{gz})\ (16\pi^2{\beta}u_\text{gz}\bar u_\text{gz})^{-3/4}\\
    &=\beta^{-1/2}\,\frac{\Gamma(1/8)}{\Gamma(7/8)}\,{2\pi^{3/2}}\!\int_{r=0}^\infty\int_{\theta=0}^{2\pi} e^{4\pi^2\, 2i\,r^2\,\sin2\theta}r^{1/2}\,2r\mathrm{d}r\mathrm{d}\theta\,\\
    &=\beta^{-1/2}\,\frac{\Gamma(1/8)\Gamma(5/8)}{2\sqrt{2}\Gamma(3/8)\Gamma(7/8)}\,.
    \end{split}
\end{equation}
This is a factor of 2 smaller than \eqref{eq:mag_appetizer_asy}, as can be seen via the Legendre duplication formula~\eqref{eq:LagrangeDuplication}. This factor is recovered by noting that an equal contribution comes from the Gaussian zone of the puncture around $y=-1$: 
\begin{equation}
    Z^{\text{el}_{\,}}_{p_1^\text{g.z}}(q)+Z^{\text{el}_{\,}}_{p_2^\text{g.z}}(q)\approx Z^\text{mag}(q)\,.
\end{equation}

\vspace{.3cm}


\addtocontents{toc}{\protect\setcounter{tocdepth}{3}}

\section{Non-perturbative saddle screenings from factorization}\label{sec:factorization&screening}

Suppression of the gauge-enhancing or infinite-flux saddles in non-abelian examples by powers of $\beta$ can be demonstrated using the methods discussed above. One has to keep track of higher and higher order terms in the small-$\beta$ expansions and establish their order by order cancellations. These are essentially perturbative calculations in the 2d Kaluza-Klein (or ``high-temperature'') EFT.

To demonstrate the exponential (or beyond-all-orders) suppression of these saddles, more powerful tools are needed, which we shall develop and deploy in this section.

Our focus will be on the second-sheet index. On the first sheet there are branch-cut obstructions to the contour crossing step of our Lorentzian factorization, as mentioned in Footnote~\ref{fn:crossing} and elaborated on in Section~\ref{subsubsec:cut_obstruction}. However, it is worth noting that the obstruction on the first sheet can be obviated by turning on suitable flavor fugacities. In other words, instead of $R$-twisted boundary condition around the circle, one can use flavor-twisted boundary conditions, and again benefit from the contour crossing and the subsequent Lorentzian factorization.

\subsection{Aspects of Lorentzian factorization and analyticity in magnetic flux}\label{subsec:lorentzian_factorization}

Our technical maneuvers in this section are possible due to \emph{analytic properties of the integrand away from the semiclassical holonomy contour}. In a first step, we shall extend the finite holonomy contours to infinite contours starting and ending in infinite imaginary-holonomy regions. In a second step, the completed contours are rotated (or ``crossed'') to vertical contours safely away from the chiral multiplet singularities (or the branch cuts in the Cardy limit); we can hence use Poisson resummation. In each winding sector, the resulting (Euclidean) surface integral can then, in a third step, be written in a Lorentzian factorized form.

We emphasize that we go through this three-step ordeal to be able to write the full index in factorized form (though we need to factorize in each winding sector and then sum over windings). Writing merely the integrand in factorized form is quite trivial in comparison and has already been done in Eqs.~\eqref{eq:CS_fac}--\eqref{eq:Ch_fac}.

\subsubsection{Large-flux asymptotics and holonomy contour extensions to infinity}


\subsubsection*{Large imaginary holonomy asymptotics}

We wish to evaluate the asymptotics of the integrand of Eq.~\eqref{eq:indexsumint}, for the special case of non-chiral theories, as $\mathrm{Im}\left(x_a\right)\xrightarrow[]{}\pm \infty$. Since we are mainly interested in non-abelian cases, we also assume $k_{gR}=0.$ Note from Eqs.~\eqref{eq:uDef}--\eqref{eq:Ch_fac} that, by absorbing $-4\pi\mathrm{Im}x/\beta$ into $m$, we can think of the limit $\mathrm{Im}\left(x_a\right)\xrightarrow[]{}\pm \infty$ alternatively as the large-flux limit of the integrand.

Let us recall the integrand below for the reader's convenience, omitting multiplicative factors that are irrelevant to our discussion:
\begin{equation*}
\begin{split}
\mathcal{J}_{\mathrm{Non-Chiral}}\propto  z^{k^+_a(m)}_{a}
\prod_{\alpha_+} (1-q^{-\alpha_+(m)/2}z^{\pm\alpha_+})\prod_{\Phi}\prod_{\rho\in \mathcal{R}_\Phi}\frac{(e^{-i\pi\nu r_{\Phi}}z^{-\rho}q^{-\rho(m)/2+1-r_\Phi/2}\,;q)}{(e^{i\pi\nu r_{\Phi}}z^{\rho}q^{-\rho(m)/2+r_\Phi/2};q)}\,.
    \end{split}
\end{equation*}
Crucially, we need the asymptotics of the $q$-Pochhammer symbols. From the Jacobi triple product formula, and an application of its Poisson resummation, one can obtain:
\begin{equation}
\label{eq:modulartheta}
    \theta_0(e^{2\pi iu};e^{2\pi i\tau})=e^{-\frac{i\pi}{\tau}(u^2+u+\frac{1}{6})+i\pi(u+\frac{1}{2})-\frac{i\pi\tau}{6}}\theta_0(e^{2\pi i\frac{u}{\tau}};e^{-2\pi i/\tau})\,,
\end{equation}
where $\theta_0(z;q)=(e^{2\pi i u};q)(qe^{-2\pi i u};q)$. For $\mathrm{Im}\left(u\right)\xrightarrow[]{}-\infty$, it is clear from the definition of the $q$-Pochhammer symbols that $\left(qe^{-2\pi i u}; q\right)\xrightarrow[]{}1$, and therefore the formula above reduces to
\begin{equation}
    (e^{2\pi i u};e^{2\pi i \tau})\xrightarrow{\mathrm{Im}u\to-\infty} e^{-\frac{i\pi}{\tau}(u^2+u+\frac{1}{6})+i\pi (u+\frac{1}{2})-\frac{i\pi\tau}{6}} \,
    \theta_0(e^{2\pi i\frac{u}{\tau}};e^{-2\pi i/\tau})
    \,.\label{eq:theta0modSimpForPoch}
\end{equation}
Note that the dependence on $\mathrm{Im}\left(u\right)$ is periodic for the theta function appearing on the RHS of the equation above, using the fact that $\mathrm{Re}\left(\tau\right)=\mathrm{Re}\left(\frac{i\beta}{2\pi}\right)=0$. Therefore the theta function on the RHS gives a bounded contribution. Thus, the asymptotics of the $q$-Pochhammer symbol is determined by the exponential factor multiplying the theta function on the RHS of Eq.~\eqref{eq:theta0modSimpForPoch}. An important subtlety is that the above exponential fall-off is valid away from the poles of the Pochhammers, which we discuss more in depth below.

For a general non-Abelian non-chiral theory, we evaluate the asymptotics in the large imaginary part of any holonomy variable $|\mathrm{Im}(x_{A})| \xrightarrow[]{}\infty$, for any sheet $\nu$, to be given by
\begin{equation}
    \label{eq:nonchiralasy}
    \mathcal{J}_{\mathrm{Non-Chiral}}\xrightarrow[]{|\mathrm{Im}\left(x_A\right)|\xrightarrow[]{}+\infty\,}\, e^{-2\pi R_m(\mathrm{Im}(x_A))}\,, 
\end{equation}
where
\begin{equation}
    R(m)=-\sum_{\Phi}(r_\Phi-1)\sum_{\rho^+}|\rho^\Phi(m)|-\sum_{\alpha_+}|\alpha_+(m)|,
\end{equation}
is the BPS monopole $R$-charge.
Importantly, note that there is no dependence on other $x_{a\neq A}$ in \eqref{eq:nonchiralasy}, a feature of non-chiral theories, allowing us to independently study the asymptotics for large $\mathrm{Im}(x_a)$ independently of one another.

Since the theory is non-chiral, the BPS monopoles are gauge-invariant. The large-$|\mathrm{Im}x|$ asymptotics of the integrand is hence exponentially suppressed if the $R$-charges of BPS monopoles are positive, which is required by unitarity in case the $R$-charge corresponds to the superconformal $R$-current. More generally, we assume strictly positive $R$-charge for gauge-invariant BPS monopoles throughout this work. This guarantees the exponential suppression of the integrand of non-chiral theories at large imaginary holonomies.

As an example, for the SU(2) SQCD with $N_f=3$ and $R_\Phi=1/3$, the asymptotics can be worked out for $|\mathrm{Im}\left(x\right)|\xrightarrow[]{}+\infty$ to be
\begin{equation}
    \exp\left(-4\pi |\mathrm{Im}(x)| + O(|\mathrm{Im}(x)|^0)\right)\,,
\end{equation}
since $R(m=1)=2$ in this case.
\subsubsection*{Contour extension to infinity}

Let us illustrate the approach in the case of SQED. Generalization to other non-chiral theories is straightforward. 

This example is easy as all the poles of the integrand on the LHS of \eqref{eq:SQED=XYZ} are simple. The case with non-simple poles will follow from a generalization of this example.
The integrand for the electric index in the duality is
\begin{equation}
    \label{eq:SQEDIntegrand}
    \mathcal{J}_{\mathrm{SQED}}=\left(-z\right)^m\frac{\left(q^{\frac{5}{6}\mp\frac{m}{2}}z^{\mp};q\right)}{\left(q^{\frac{1}{6}\mp\frac{m}{2}}z^{\pm};q\right)}\,.
\end{equation}
There is an obvious symmetry that the integrand possesses which makes non-chiral theories particularly nice. Indeed, we have seen that non-chiral theories may have their contours closed in either half-plane (lower or upper) in the complex plane of the holonomy variable $x$. The symmetry in question is $\left(z,m\right)\xrightarrow{}\left(z^{-1},-m\right)$. The positions of the poles of the above integrand are\footnote{Although we will not need them, the corresponding residues are
\begin{align*}
    \begin{split}
\mathrm{R}^+_{k,m}
=\frac{\left(q^{1+k-m};q\right)\left(q^{\frac{2}{3}-k+m};q\right)}{\left(q^k;q\right)_k\left(q^{\frac{1}{3}+k};q\right)\left(q;q\right)}\left(-q^{-k-\frac{1}{6}+\frac{m}{2}}\right)^m\,,\quad
\mathrm{R}^-_{k,m}
=-\frac{\left(q^{1+k+m};q\right)\left(q^{\frac{2}{3}-k-m};q\right)}{\left(q^k;q\right)_k\left(q^{\frac{1}{3}+k};q\right)\left(q;q\right)}\left(-q^{+k+\frac{1}{6}+\frac{m}{2}}\right)^m\,,
    \end{split}
\end{align*}
on the 1st sheet, and may be found on higher sheets via $q\to q e^{2\pi i\nu}$.}
\begin{align}
\begin{split}
    Z_+\left(k,m\right)&=q^{-k+m/2-\frac{r}{2}}e^{-\pi i\nu r}\Longrightarrow X_+(k,m)=i\frac{\beta}{2\pi}(-k+\frac{m}{2}-\frac{r}{2})-\frac{\nu\,r}{2}, \\
    Z_-\left(k,m\right)&=q^{k+m/2+\frac{r}{2}}e^{+\pi i\nu r}\Longrightarrow X_-(k,m)=i\frac{\beta}{2\pi}(k+\frac{m}{2}+\frac{r}{2})+\frac{\nu\,r}{2},
\end{split}
    \label{eq:PolesSQED}
\end{align}
for $k=0,1,\dots$. Note that $Z_+\left(k,m\right)=\left(Z_-\left(k,-m\right)\right)^{-1}$. Alternatively, defining $X_\pm\left(k,m\right)=\frac{1}{2\pi i}\log\left(Z_\pm\left(k,m\right)\right)$, we have $X_+\left(k,m\right)=-X_-\left(k,-m\right)$. 

The zeros of the numerator of \eqref{eq:SQEDIntegrand} are at
\begin{align}
\begin{split}
    \tilde Z_+\left(\tilde k,m\right)&=q^{-\tilde k-m/2-1+\frac{r}{2}}e^{\pi i\nu r}\Longrightarrow \tilde X_+(\tilde k,m)=i\frac{\beta}{2\pi}(-\tilde k-\frac{m}{2}-1+\frac{r}{2})+\frac{\nu\,r}{2}, \\
    \tilde Z_-\left(\tilde k,m\right)&=q^{\tilde k-m/2+1-\frac{r}{2}}e^{-\pi i\nu r}\Longrightarrow \tilde X_-(\tilde k,m)=i\frac{\beta}{2\pi}(\tilde k-\frac{m}{2}+1-\frac{r}{2})-\frac{\nu\,r}{2},
\end{split}
    \label{eq:ZerosSQED}
\end{align}
with $\tilde k=0,1,\dots$. We thus see that for any fixed $m,$ any $X_-$ pole labeled by $k$ is canceled by an $\tilde X_+$ zero labeled by $\tilde k$ if
\begin{equation}
    \tilde k=-k-m-1\ge0\Longrightarrow 0\le k\le-m-1\,.\label{eq:k_canceled_poles}
\end{equation}
We now want to argue that this shows that the $X_-$ poles below the real axis, and the $X_+$ poles above the real axis, are canceled against zeros.

Let us focus on the $X_-\left(k,m\right)$ poles that lie below the real $x$ axis. These have
\begin{equation}
    \mathrm{Im} X_-\left(k,m\right)\propto{m}{}+{r}{}+2k< 0\xrightarrow{m+2k\,<\,-r\ \,\Longrightarrow\ \leq\,-1}0\leq k\leq -m/2 -1/2 \,.
\end{equation}
Crucially, for $m<0,$ this is a strict subset of \eqref{eq:k_canceled_poles}. Therefore all $X_-$ poles below the real axis are canceled against zeros of the numerator of \eqref{eq:SQEDIntegrand}. A similar argument applies to the $X_+$ poles above the real axis.

Considering the second sheet, $\nu=1$, for concreteness, we conclude that for any fixed $m$, \emph{there are no poles in the bottom-right and upper-left quadrants of the $x$-plane.} This statement has a simple outer-patch 2d EFT counterpart in the Cardy limit: the twisted superpotentials arising from integrating out heavy chiral multiplets have branch cuts emanating from the unit circle in the $y$ plane, as expected from \eqref{eq:heavy_W}.\\

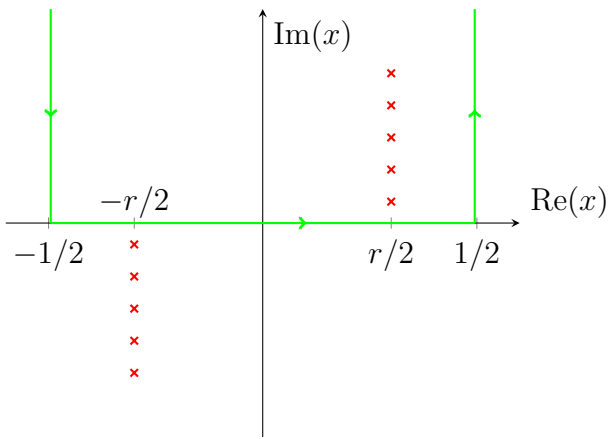
\begin{figure}[h!]
\centering
\begin{tikzpicture}
  \begin{axis}[
    axis lines=middle,
    xmin=-2.4, xmax=2.4,
    ymin=-2, ymax=2,
    xlabel={$\mathrm{Re}(x)$},
    xlabel style={at={(axis description cs:1,.55)}, anchor=west},
    ylabel={$\mathrm{Im}(x)$},
    xtick       = {-2, 0, 1.2, 2},              
    xticklabels = {$-1/2$, $0$, $r/2$, $1/2$},  
    extra x ticks={-1.2},                       
    extra x tick labels={$-r/2$},               
    extra x tick style={tick label style={anchor=south}}, 
    ytick       = \empty,
    axis equal image,
    clip=false
  ]

  \draw[green, thick, <-] (axis cs:-1.98, 1) -- (axis cs:-1.98, 2); \draw[green, thick, -<] (axis cs:-1.98, 0) -- (axis cs:-1.98, 1.05); \draw[green, thick, ->] (axis cs:-1.98, 0) -- (axis cs:0.4, 0); \draw[green, thick, >-] (axis cs:0.35, 0) -- (axis cs:1.98, 0);
  \draw[green, thick, >-] (axis cs:1.98, 1) -- (axis cs:1.98, 2); \draw[green, thick, ->] (axis cs:1.98, 0) -- (axis cs:1.98, 1.05);

  \addplot[only marks, mark=x, mark options={red, thick}, forget plot]
    coordinates {
      (-1.2, -0.2)
      (-1.2, -0.5)
      (-1.2, -0.8)
      (-1.2, -1.1)
      (-1.2, -1.4)
    };

  \addplot[only marks, mark=x, mark options={red, thick}, forget plot]
    coordinates {
      (1.2, 0.2)
      (1.2, 0.5)
      (1.2, 0.8)
      (1.2, 1.1)
      (1.2, 1.4)
    };

  \end{axis}
\end{tikzpicture}
\caption{Contour used to evaluate the index on the second sheet ($\nu=1$) of a non-chiral theory in complex $x$ plane. The red crosses represent poles.}
\label{fig:Secondsheetcontour}
\end{figure}

Having studied the pole structure, we now move on to the evaluation of the index. We consider again $\nu=1$ for concreteness. Taking advantage of 
\begin{itemize}
    \item[$i$)]  periodicity under $x\to x+1$, and
    \item[$ii$)] large-$\mathrm{Im}\,x$ suppression,
\end{itemize}
for any fixed $m$ we complete the original $x\in (-\frac{1}{2},\frac{1}{2}]$ contour, to one as in Fig.~\ref{fig:Secondsheetcontour}.

Note that the procedure outlined above for SQED, can be applied to more general non-chiral gauge theories as well.

\subsubsection{Lorentzian factorization
via crossing to vertical holonomy contours}

\begin{figure}[h!]
    \centering
\begin{tikzpicture}
  \begin{axis}[
    axis lines=middle,
    xmin=-2.4, xmax=2.4,
    ymin=-2, ymax=2,
    xlabel={$\mathrm{Re}(x)$},
    xlabel style={at={(axis description cs:1,.55)}, anchor=west},
    ylabel={$\mathrm{Im}(x)$},
    xtick       = {-2, 0, 1.2, 2},              
    xticklabels = {$-1/2$, $0$, $r/2$, $1/2$},  
    extra x ticks={-1.2},                       
    extra x tick labels={$-r/2$},               
    extra x tick style={tick label style={anchor=south}}, 
    ytick       = \empty,
    axis equal image,
    clip=false
  ]

  \draw[green, thick, <-] (axis cs:0.03, -.65) -- (axis cs:0.03, 2);
  \draw[green, thick, >-] (axis cs:0.03, -.6) -- (axis cs:0.03, -2);
  \draw[green, thick, ->] (axis cs:1.98, -2) -- (axis cs:1.98, .65);
  \draw[green, thick, >-] (axis cs:1.98, .6) -- (axis cs:1.98, 2);

  \addplot[only marks, mark=x, mark options={red, thick}, forget plot]
    coordinates {
      (-1.2, -0.2)
      (-1.2, -0.5)
      (-1.2, -0.8)
      (-1.2, -1.1)
      (-1.2, -1.4)
    };

  \addplot[only marks, mark=x, mark options={red, thick}, forget plot]
    coordinates {
      (1.2, 0.2)
      (1.2, 0.5)
      (1.2, 0.8)
      (1.2, 1.1)
      (1.2, 1.4)
    };

  \end{axis}
\end{tikzpicture}
\caption{Deformed contour of a non-chiral theory on the second sheet in complex $x$-plan. We have used the residue theorem to move the contours about.}
    \label{fig:2ndsheetdeformed}
\end{figure}
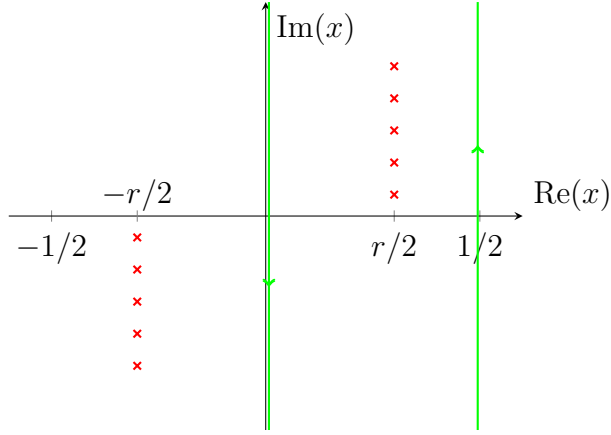

Since the contributions from the lines at $\mathrm{Re}\left(x\right)=\pm \frac{1}{2}$ cancel each other out by the periodic property of the the variable $x\sim x+1$, the index on the second sheet is just the sum over one set of poles with the appropriate sign. 
Due to the well-behaved asymptotics of the non-chiral theory at positive and negative large imaginary $x$, we may instead consider the deformed contour as in Fig.~\ref{fig:2ndsheetdeformed}. 

This leads to the following equation for the the index on a sheet $\nu \frac{r}{2} \neq 0 \, \mathrm{mod}\, 1$,
\begin{equation}
    Z_{\nu+1}^{\mathrm{SQED}}= \sum_{m \in \mathbb{Z}} \int^{\frac{1}{2}}_{\frac{1}{2}} \mathrm dx\, \mathcal{J}_{\mathrm{SQED}}\left(x,m\right)= -i \sum_{m}\int_{-\infty}^{\infty} \mathrm dx \big[\mathcal{J}_{\mathrm{SQED}} \left(ix+x_\ell, m\right)-\mathcal{J}_{\mathrm{SQED}} \left(ix+x_r, m\right)\big]\,.
\end{equation}

Since the integrand factorizes as
\begin{equation}
    \mathcal{J}_{\text{SQED}}(x,m)=f\big(x-i\frac{\beta m}{4\pi}\big)\,\tilde f\big(x+i\frac{\beta m}{4\pi}\big)=f(u)\,\tilde f(\bar u),
\end{equation}
we can write the sum-integrals with rotated holonomy contour as
\begin{equation}
    -i\sum_w\int\mathrm{d}m\int\mathrm{d}x \,\Big(\big(f(iu^-+x_\ell)e^{-\frac{4\pi^2}{\beta}w(iu^-+x_\ell)}\,\tilde f(iu^++x_\ell)e^{\frac{4\pi^2}{\beta}w(iu^++x_\ell)}\big)-\big(\ell\leftrightarrow r\big)\Big)\,,
\end{equation}
where $u^\pm=x\pm\frac{\beta m}{4\pi}$. The choices $x_\ell=0,\, x_r=\frac{1}{2}$ are particularly convenient, and we stick to them throughout.
Upon replacing $\mathrm{d}m\mathrm{d}x=\frac{2\pi}{\beta}\mathrm{d}u^+\mathrm{d}u^-$, each term factorizes and we get
\begin{equation}
    -\frac{2\pi i}{\beta}\sum_{w\in\mathbb{Z}}\Big(B^{(w)}_\ell(q)\,\widetilde{B^{(w)}_\ell}(q)-(\ell\leftrightarrow r)\Big),\label{eq:fully_factorized}
\end{equation}
with
\begin{equation}
    B^{(w)}_\ell(q):=\int_\ell\mathrm{d}u^- f(u^-)\,e^{-\frac{4\pi^2}{\beta}wu^-},\qquad \widetilde{B^{(w)}_\ell}(q):=\int_\ell\mathrm{d}u^+ g(u^+)\,e^{\frac{4\pi^2}{\beta}wu^+},
\end{equation}
the holomorphic blocks, and similarly for $B_r^{(w)},\,\widetilde{B_r^{(w)}}$.

\subsubsection{Branch-cut obstructions to crossing and Dotsenko-Fateev contours}
\label{subsubsec:cut_obstruction}

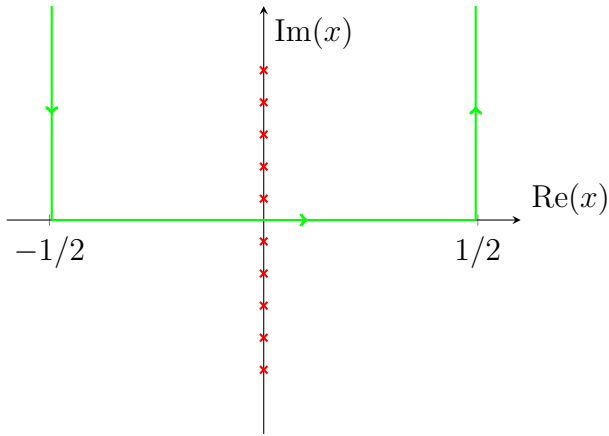
\begin{figure}[h!]
\centering
\begin{tikzpicture}
  \begin{axis}[
    axis lines=middle,
    xmin=-2.4, xmax=2.4,
    ymin=-2, ymax=2,
    xlabel={$\mathrm{Re}(x)$},
    xlabel style={at={(axis description cs:1,.55)}, anchor=west},
    ylabel={$\mathrm{Im}(x)$},
    xtick       = {-2, 0, 2},              
    xticklabels = {$-1/2$, $0$, $1/2$},  
    ytick       = \empty,
    axis equal image,
    clip=false
  ]

  \draw[green, thick, <-] (axis cs:-1.98, 1) -- (axis cs:-1.98, 2); \draw[green, thick, -<] (axis cs:-1.98, 0) -- (axis cs:-1.98, 1.05); \draw[green, thick, ->] (axis cs:-1.98, 0) -- (axis cs:0.4, 0); \draw[green, thick, >-] (axis cs:0.35, 0) -- (axis cs:1.98, 0);
  \draw[green, thick, >-] (axis cs:1.98, 1) -- (axis cs:1.98, 2); \draw[green, thick, ->] (axis cs:1.98, 0) -- (axis cs:1.98, 1.05);

  \addplot[only marks, mark=x, mark options={red, thick}, forget plot]
    coordinates {
      (0, -0.2)
      (0, -0.5)
      (0, -0.8)
      (0, -1.1)
      (0, -1.4)
    };

  \addplot[only marks, mark=x, mark options={red, thick}, forget plot]
    coordinates {
      (0, 0.2)
      (0, 0.5)
      (0, 0.8)
      (0, 1.1)
      (0, 1.4)
    };

  \end{axis}
\end{tikzpicture}
\caption{Contour computing the index for a non-chiral theory on the first sheet, with the pole structure denoted with red crosses in the complex $x$-plane}
    \label{fig:firstsheetcontour}
\end{figure}

On the first sheet of nonchiral theories, the situation is more subtle. For nonchiral theories with $\nu r= 0\, \mathrm{mod}\, 2$, we end up with a pole structure that looks like Fig. $\ref{fig:firstsheetcontour}$.  

There is no deformation of this contour that allows us to perform the Lorentzian factorization that we desire, with directed contours running parallel to the imaginary $x$-axis. Note that in the Cardy limit where the poles condense to form branch cuts, the fact that the holonomy contour is pinched by branch cuts is the reason we have obstruction to contour deformation. It is possible of course to promote $\nu$ to a continuous variable, do the factorization for some $\nu>0,$ and then use analyticity in $\nu$ to take the $\nu\to0$ limit of the factorized expressions. We do not pursue this approach here.

Instead, we excise $\epsilon$ neighborhoods of singularities (the origin in Figure~\ref{fig:firstsheetcontour}), and separately consider   the Cardy limit of the contribution from the punctured surface and the puncture, as in Section~\ref{subsec:power_law}. The contribution from the punctured surface, after Poisson resummation, is dominated by the zero winding sector. This then takes the form of Dotsenko-Fateev integrals, as seen in various examples in the previous section. In cases these integrals are convergent in the limit of shrinking the punctures, their evaluation actually goes through a Lorentzian factorization argument, albeit the resulting contours are not as straightforward as those in e.g. Figure~\ref{fig:2ndsheetdeformed}. See \cite{dotsenko1988lectures}.

\subsection{Saddle screenings through the lens of holomorphic blocks}\label{subsec:block_screening}

We first treat gauge-enhancing saddles, then the more subtle infinite-flux saddles, in preparation for which we also discuss the contribution from saddle-free (or non-stationary) contours.

\subsubsection{Gauge-enhanced saddles}\label{subsec:gauge-enhancing_saddles}

Among the various saddle points one may encounter on the second sheet, there exists a special class that are gauge-enhancing, meaning they are mapped to themselves under the action of a subgroup of the Weyl group. Specializing to the case of $G = \mathrm{SU}(2)$, the full Weyl group is $\mathbb{Z}_2$ and acts as $u \to -u$. Thus, there are two fixed points, one at $u = 0$ and the other at $u=\frac{1}{2}$. On the second sheet, with generic $R$-charges, the points $u = 0$ and $u=\frac{1}{2}$ are away from any light matter.
Below we first consider the contribution from a neighborhood of $u=0,$ and then comment on that of $u=\frac{1}{2}.$

After factorization, we have the holomorphic factor (or block):
\begin{equation}
    \begin{split}
    B_\ell^{(w)}(q)=\int_\ell \mathrm{d}u\ e^{W_{\text{CS}^+}\,-\,\frac{4\pi^2}{\beta}\,w\,u}\, \prod_{\Phi}\prod_{\rho\in \mathcal{R}_\Phi}\frac{1}{(y^{\rho}e^{i\pi r_{\Phi}}q^{\frac{r_\Phi}{2}};q)}\,\prod_{\alpha_+}\, (y^{-\alpha_+
/2}-y^{\alpha_+
/2})\,,   
    \end{split}\label{eq:B}
\end{equation}
and similarly for the anti-holomorphic factor and the $r$ contour blocks.

In reference to the saddles, we have to consider
\begin{equation}
    W^{(w)} = -\frac{4\pi^2}{\beta}\left({\frac{1}{2}k^+_{gg}\big(u^2}+2u\big)+\frac{1}{2}k^+_{gR}u+w(u)+\sum_{\Phi}\sum_{\rho\in \mathcal{R}_\Phi}\frac{1}{-4\pi^2}\mathrm{Li}_2\big(e^{2\pi i \rho(u)+i\pi r_\Phi}\big)\,\right).\label{eq:Ww_nonchiral}
\end{equation}
We notice that for there to be a Bethe vacuum at $u = 0$, we necessarily need $k^+_{gg} + \frac{1}{2}k^+_{gR} \in \mathbb{Z}$,  since then by taking 
\begin{equation}
    w=w_0:=-(k^+_{gg} + \frac{1}{2}k^+_{gR}),
\end{equation}
we can kill the only term in the above for which the derivative at $u = 0$ is non-zero. (Recall that we are assuming vectorlike matter content.) Proceeding with this condition enforced, we see that $W_{\text{CS}^+}\,-\,\frac{4\pi^2}{\beta}\,w\,u$ is even under $u \to -u$. So is the chiral multiplet product in \eqref{eq:B}, since we are assuming vectorlike matter content.

There remains the product over the positive roots in \eqref{eq:B}, which is odd under $u\to -u$ for any non-abelian gauge group. The integrand of $B_\ell^{(w_0)}(q)$ is thus odd under $u\to -u$ and therefore 
\begin{equation}
    B_\ell^{(w_0)}(q)=0\,,
\end{equation}
for a non-chiral theory with 
non-abelian gauge group, assuming that there are no light charged chirals at $u=0$.
This implies, in particular, that the saddle at $u=0$ is suppressed beyond all orders.

Similar considerations apply near $u=\frac{1}{2},$ as can be seen via a change of variables $u=\frac{1}{2}-u'$. The linear term in $u'$ in the twisted superpotential is then seen to cancel if $\frac{3}{2}k^+_{gg} + \frac{1}{2}k^+_{gR} \in \mathbb{Z}$,  so we can take 
\begin{equation}
    w=w_{1/2}:=w_0-k^+_{gg}/2\,.
\end{equation}
Then, similarly to how we argued above, we have
\begin{equation}
    B_r^{(w_{1/2})}(q)=0\,,
\end{equation}
for a non-chiral theory with 
non-abelian gauge group, assuming that there are no light charged chirals at $u=1/2$. This implies, in particular, that the saddle at $u = 1/2$ is suppressed beyond all orders.

\subsubsection{Non-stationary contours and infinite-flux saddles}\label{subsec:non-contours}

We shall heavily rely on \eqref{eq:no_stationary_asy} for both non-stationary contours and infinite-flux saddles. The discussion of the infinite-flux saddles is particularly technical in this approach. We suspect that more elegant arguments through contour deformations is possible, with a rewriting of the form:
\begin{equation}
    B^{(w)}_\ell\,\widetilde{B^{(w)}_\ell}-(\ell\leftrightarrow r)=(B^{(w)}_\ell+B^{(w)}_r)(\widetilde{B^{(w)}_\ell}+\widetilde{B^{(w)}_r})-\big(B^{(w)}_\ell(\widetilde{B^{(w)}_\ell}+\widetilde{B^{(w)}_r})+\widetilde{B^{(w)}_r}(B^{(w)}_\ell+B^{(w)}_r)\big),
\end{equation}
so that $\ell$ and $r$ contours can be combined, but we leave that for future work.

\paragraph{\small Non-stationary contours:}\hspace{-.3cm}

We refer to $r,\ell$ contours not passing through saddles as \emph{non-stationary contours}. Using all-order expansions, which are valid since our contours are away from singularities, we can write $B_{\ell,r}^{(w)}$ as
\begin{equation}
    \int_{\ell,r}\mathrm{d}u\ e^{W^{(w)}}h(u,\beta),
\end{equation}
where $W^{(w)}\propto\frac{1}{\beta},$ and $h(u,\beta)$ encodes the rest of the terms arising from all-order expansions.\footnote{That the integrand of the block admits such an all-order expansion in $\beta$, with the coefficients holomorphic functions of $u$ on the contour, seems to require tools from several complex variables to demonstrate. We leave a careful demonstration of this to future work.} These contours are at $\mathrm{Re}\,u=0$ or $\mathrm{Re}\,u=1/2,$ where they have constant $\text{Re}\,{W}^{(w)}$. (This can be shown by considering $W^{(w)}+\overline{W^{(w)}}$ and using~\eqref{eq:WH_pair}.) We can thus use Eq.~\eqref{eq:no_stationary_asy}, combined with the large-$\mathrm{Im}\,u$ exponential suppression of our integrands as in Eq.~\eqref{eq:nonchiralasy}, to deduce suppression to all orders in $\beta$.

\paragraph{\small Saddles at infinity:}\hspace{-.3cm}

The preceding argument can not apply if $W^{(w)}{}'$ decays as $|\mathrm{Im}u|\to\infty,$ in which case we have saddles at infinity on the $u$-strip.

\paragraph{\small Determining the winding sectors with saddles at infinity.}\hspace{-.3cm}Let us consider $\mathrm{Im}u\to+\infty$ for concreteness. In this case, of the dilogarithm terms in \eqref{eq:Ww_nonchiral} only those with $\rho(u)<0$ contribute. Using
\begin{equation}
    \begin{split}
    \mathrm{Li}_2(z)&=-\frac{\pi^2}{6}-\frac{\big(\log(-z)\big)^2}{2}+\mathrm{Li}_2\big(\frac{1}{z}\big)\\
    &\xrightarrow{z\to\infty}-\frac{\pi^2}{6}-\frac{\big(\log(-z)\big)^2}{2}+\mathcal{O}\big(\frac{1}{z}\big),\label{eq:Li2_large_z}
    \end{split}
\end{equation}
we find that $W^{(w)}{}'$ decays as $\mathrm{Im}u\to+\infty$ if $k_{gg}=0$ and
\begin{equation}
    w=w^{\ell,r}_{i\infty}=\frac{1}{2}\sum_{\Phi}\sum_{\rho_-\in\mathcal{R}_\Phi}(r_\Phi\mp1)\rho_--(k^+_{gg} + \frac{1}{2}k^+_{gR})\,.
\end{equation}
Similarly there is a saddle at $\mathrm{Im}u\to-\infty$ if $k_{gg}=0$ and
\begin{equation}
    w=w^{r,\ell}_{-i\infty}=\frac{1}{2}\sum_{\Phi}\sum_{\rho_+\in\mathcal{R}_\Phi}(r_\Phi\mp1)\rho_+-(k^+_{gg} + \frac{1}{2}k^+_{gR})=-w^{\ell,r}_{i\infty}-2(k^+_{gg} + \frac{1}{2}k^+_{gR})\,.
\end{equation}
Note that for our SQCD example where $k^+_{gg}=6$ and $k^+_{gR}=0$ we get from the above relations $w^{\ell,r}_{i\infty}=-4,-10$ and $w^{r,\ell}_{-i\infty}=-8,-2$, compatibly with Figure~\ref{fig:infinity_saddles}.

\paragraph{\small First look at the asymptotics from saddles at infinity.}\hspace{-.3cm}In the SQCD example, as mentioned in the previous section, the saddles at $\mathrm{Im}u=\pm\infty$ yield power-law growth, because e.g.
\begin{equation}
    W^{(-8)}-\overline{W^{(-8)}}\xrightarrow{u\to-i\infty}0.
\end{equation}
Since on the magnetic side of the duality there is no saddle with power-law growth, we expect that on the electric side the saddles at $\mathrm{Im}u=\pm\infty$ are suppressed not just by powers of $\beta$, but fully, analogously to the gauge-enhancing saddles discussed earlier.

\paragraph{\small Demonstrating all-order screening for saddles at infinity.}\hspace{-.3cm}We shall demonstrate the screening to all orders using the holomorphic factorization equation, Eq.$\left(\ref{eq:fully_factorized_intro}\right)$. As we will argue shortly, the screening is a result of all-order in $\beta$ cancellations between the contributions from contours $l$ and $r$, rather than a symmetry argument as for the gauge-enhancing saddle points. Let us first consider the saddle at $\mathrm{Im}u\xrightarrow[]{}-\infty$ for $w=-8$. Our argument proceeds in four steps: $i$) decomposing the integration domain of the block $B^{-8}_\ell$ receiving a contribution from the saddle at infinity into a large-$u$ piece and the rest, $ii$) arguing that the large-$u$ piece is negligible due to the exponential decay of the integrand, $iii$) evaluating the contribution from the remaining piece of the contour, $iv$) putting $B^{(-8)}_\ell, \widetilde{B^{(-8)}_\ell}$ together and arguing for all-order cancelation against $B^{(-8)}_r, \widetilde{B^{(-8)}_r}$. A similar argument establishes cancelation of the $B^{(-4)}_\ell$, $\widetilde{B^{(-4)}_\ell}$ contribution relevant to the $\mathrm{Im}u\to+\infty$ saddles against $B^{(-10)}_r$, $\widetilde{B^{(-10)}_r}$. The reader uninterested in the technical details can skip the rest of this section.

\paragraph{\small Step 1: contour decomposition.}\hspace{-.3cm}We now split the $B^{(-8)}_\ell$ integration into
\begin{equation}
    B^{(-8)}_\ell(q)=\left(\int_{-\infty}^{-\frac{1}{\sqrt{\beta}}}+\int_{-\frac{1}{\sqrt{\beta}}}^\infty\right)\ \mathrm{d}u^- \,f(iu^-;\beta)\,e^{-\frac{4\pi^2}{\beta}i\cdot (-8)\cdot u^-}.\label{eq:Bl_dec_sec422}
\end{equation}
Instead of $-\frac{1}{\sqrt{\beta}},$ we could have chosen $-\frac{1}{\beta^\gamma}$ for any $\gamma\in(0,1).$

\paragraph{\small Step 2: arguing that the large-$u$ piece is negligible.}\hspace{-.3cm}The exponential decay of the integrand as $u^-\to-\infty$ implies that the first integral is $\mathcal{O}(e^{-1/\sqrt{\beta}})$ and thus negligible.

\paragraph{\small Step 3: evaluating the contribution from the rest of the contour via repeated integration-by-parts.}\hspace{-.3cm}The second integral is amenable to the treatment of \cite{miller2006applied}, since it contains no stationary-phase points in that integration region. We write the second integral as
\begin{equation}
\begin{split}
e^{\mathrm{Re}W^{(-8)}}\int_{-\frac{1}{\sqrt{\beta}}}^{\infty}\mathrm{d}u^- \,\left(f(iu^-;\beta)\,e^{-\frac{4\pi^2}{\beta}i\cdot (-8)\cdot u^-}e^{-W^{(-8)}}\right)&\,e^{i\,\mathrm{Im}W^{(-8)}}\\
&\hspace{-4cm}=e^{\mathrm{Re}W^{(-8)}}\int_{-\frac{1}{\sqrt{\beta}}}^{\infty}\mathrm{d}u^- \,\left(f(iu^-;\beta)\,e^{-W}\right)\,e^{i\,\mathrm{Im}W^{(-8)}},
\end{split}
\end{equation}
where we have multiplied the integrand by $e^{W^{(-8)}-W^{(-8)}}$, and taken advantage of the fact that $\mathrm{Re}W^{(w)}$ is constant on our $\ell,r$ contours to factor out the $e^{\mathrm{Re}W^{(-8)}}$ piece. Now, noting that $W^{(-8)}\propto 1/\beta,$ and writing $f(iu^-;\beta)\,e^{-\frac{4\pi^2}{\beta}i\cdot (-8)\cdot u^-}e^{-W^{(-8)}}$ as an all-order expansion in $\beta$ with coefficients holomorphic functions of $u^-$, we can use \eqref{eq:no_stationary_asy}:
\begin{equation}
e^{\mathrm{Re}W^{(-8)}}\int_{-\frac{1}{\sqrt{\beta}}}^{\infty}\mathrm{d}u^- \,\left(f(iu^-;\beta)\,e^{-W}\right)\,e^{i\,\mathrm{Im}W^{(-8)}}=e^{W^{(-8)}}\sum_{n=1}^\infty(-i\beta)^{n} \frac{g_{n-1}}{\beta\,W^{(-8)}{}'} \Bigg|_{-\frac{1}{\sqrt{\beta}}}^{\infty},\label{eq:miller_used_422}
\end{equation}
\begin{equation}
    \text{with}\quad g^{}_{n}=-\frac{d}{du}\Big(\frac{g^{}_{n-1}}{\beta\,{W^{(-8)}{}'}}\Big)\,\quad \text{and} \quad g^{}_{0}=f(iu^-;\beta)\,e^{-W}\,.
\end{equation}
The upper limit $u^-\to\infty$ gives zero, since $W^{(-8)}{}'$ does not decay there (as there is no saddle for $B^{(-8)}_\ell$ at $-\infty$), and $g_n$ decay (as calculations very similar to those reported below for near $-\infty$ indicate). To evaluate the contribution from the lower limit $u^-\to-\frac{1}{\sqrt{\beta}}\to-\infty,$ we start from\footnote{On the second line, we use \eqref{eq:theta0modSimpForPoch} with $u\to i\rho_- u^-+\frac{r}{2}+\frac{r}{2}\frac{i\beta}{2\pi}.$ Since this $u$ has positive real part, we can replace $\theta_0(e^{2\pi i\frac{u}{\tau}};e^{-2\pi i/\tau})$ in \eqref{eq:theta0modSimpForPoch} with $-e^{2\pi i\frac{u}{\tau}}$ as $\beta\to0$ up to $\mathcal{O}(e^{-\#/\beta})$ relative error. The $e^{-\frac{\mathrm{Li}_2}{\beta}}$ terms are Taylor expanded, either directly (for $\rho<0$) or after applying \eqref{eq:Li2_large_z} (for $\rho>0$).}
\begin{equation}
\begin{split}
   f(iu^-;\beta)\,e^{-W}&=\prod_{\alpha_+}\, (y_-^{-\alpha_+
/2}-y_-^{\alpha_+
/2})\ \frac{e^{-\frac{\sum_{\Phi,\,\rho_\Phi}\mathrm{Li}_2(y_-^{\,\rho}\,e^{i\pi r_\Phi})}{\beta}}}{\prod_{\Phi}\prod_{\rho\in \mathcal{R}_\Phi}(y_-^{\rho}e^{i\pi r_{\Phi}}q^{\frac{r_\Phi}{2}};q)}\Bigg|_{y_-=e^{-2\pi u^-}}\\
&\hspace{-2cm}\xrightarrow[\eqref{eq:theta0modSimpForPoch},\,\eqref{eq:Li2_large_z}]{u^-\to-\infty,\,\beta\to0}e^{-\pi R_m(u^-)+i\frac{\pi}{4}\mathrm{Tr}_cR^2+\beta(\frac{\mathrm{Tr}_c1}{48}-\frac{\mathrm{Tr}_cR^2}{16})}\ \big(y^{-\alpha_+}-1\big)\\
&\left(1+\sum_{n=1}^\infty\frac{\big(\frac{-\sum_{\rho<0}\sum_{k=1}^\infty (y_-^{\rho}e^{i\pi r_\Phi})^k}{\beta k^2}\big)^n}{n!}\right)\left(1+\sum_{n=1}^\infty\frac{\big(\frac{\sum_{\rho>0}\sum_{k=1}^\infty (y_-^{-\rho}e^{-i\pi r_\Phi})^k}{\beta k^2}\big)^n}{n!}\right)\,,\label{eq:g0_limit}
\end{split}
\end{equation}
up to $\mathcal{O}(e^{-\#/\beta})$ relative error. The $c$ subscripts indicate that the contributions come from charged chiral multiplets.
In our example $R_m(1)=2,$ so $g_0$ is proportional to $e^{2\pi u^-}=y_-^{-1}$ in the limit. Explicit evaluation shows that $\beta W^{(-8)}{}'$ is also proportional to $y_-^{-1}$, up to $\mathcal{O}(e^{-1/\sqrt{\beta}})$ relative error, as $u^-\to-\frac{1}{\sqrt{\beta}}\to-\infty$.
We then obtain (through repeated differentiation of the result for $n=1$) that the coefficient of $(-i\beta)^n$ in \eqref{eq:miller_used_422} decays to a constant as $u^-\to-\frac{1}{\sqrt{\beta}}\to-\infty$:
\begin{equation}
  \frac{g_{n-1}}{\beta\,{W}^{(-8)}{}'}\xrightarrow{u^-\to-\frac{1}{\sqrt{\beta}}\to-\infty\,}  C_{n}+\mathcal{O}\left(e^{-1/\sqrt{\beta}}\right)\,.
\end{equation}
Note that $C_n$ is a constant in the sense of being independent of $u^-$, but it has dependence on $\beta$ of the form $e^{\beta(\frac{\mathrm{Tr}_c1}{48}-\frac{\mathrm{Tr}_cR^2}{16})}P_n(\frac{1}{\beta})$, with $P_n(\frac{1}{\beta})$ a polynomial in $1/\beta.$ The presence of $P_n(1/\beta)$ implies that whether the series in \eqref{eq:miller_used_422} has well-defined coefficients as a formal power series in $\beta$ is a subtle question. We assume the coefficients are well-defined and proceed.

\paragraph{\small Step 4: combining $B,\widetilde{B}$, comparison between $\ell,r$, and establishing cancelations.}\hspace{-.3cm}A similar argument applies to $\widetilde{B_\ell^{(-8)}}$, with analogous ``constants'' $\tilde C_n$. We next note that when evaluating the combination
\begin{equation}
    B^{(-8)}_\ell\,\widetilde{B^{(-8)}_\ell}-B^{(-8)}_r\,\widetilde{B^{(-8)}_r},
\end{equation}
the fact that $C_n,\tilde C_n$ are independent of $u^{\mp},$ implies that the shifts in $u^\mp$ taking us from the $1/\sqrt{\beta}$ boundary points of $B^{(-8)}_\ell,\,\widetilde{B^{(-8)}_\ell}$ to the $1/\sqrt{\beta}$ boundary points of $B^{(-8)}_r,\,\widetilde{B^{(-8)}_r}$ have negligible effect up to $\mathcal{O}(e^{-1/\sqrt{\beta}})$ error. We thus get
\begin{equation}
    B^{(-8)}_\ell\,\widetilde{B^{(-8)}_\ell}-B^{(-8)}_r\,\widetilde{B^{(-8)}_r}=\mathcal{O}(e^{-1/\sqrt{\beta}}),
\end{equation}
demonstrating that the contours passing through the infinite-flux saddles cancel each other's contributions beyond all orders in $\beta$.

\section{Discussion and open problems}\label{sec:open_problems}

\paragraph{Conceptual overview.} We believe we have finally tamed the Cardy limit of the superconformal index of 3d $\mathcal{N}=2$ gauge theories at finite rank $N$. Although our analysis in this paper focuses on rank-one theories, the central tools we develop appear sufficiently robust to extend to higher—though still finite—ranks.\footnote{Large-$N$ indices relevant for holography require complementary techniques; cf.~\cite{Choi:2019dfu,GonzalezLezcano:2022hcf,Bobev:2022wem,BenettiGenolini:2023rkq,Bobev:2024mqw,Hosseini:2025jxb}.}

The main idea of our work is to \textbf{\small divide and conquer the BPS moduli space} appearing in the Coulomb branch localization formula, analogously to what was done for the 4d index in \cite{Ardehali:2021irq}. We \textbf{\small divide the moduli space according to the scale} of the distance to the ``matter singularities'', where there is light charged matter in the 2d Kaluza--Klein EFT. The \textbf{\small important regions are the matter near zones, the matter Gaussian zones, and the gauge patch}, corresponding respectively to $\mathcal O(\beta)$,
$\mathcal O(\sqrt\beta)$, and $\mathcal O(\beta^0)$ distances to the matter singularities on the $u$-strip (or the $y$-plane).

The \emph{matter Gaussian zones} and the \emph{gauge patch} are conquered via \textbf{\small Poisson resummation and the method of multidimensional stationary phase}. In non-abelian cases, certain special stationary-phase loci (gauge enhanced or infinite flux) are screened to all orders in $\beta$. This screening can be demonstrated for non-chiral theories via the \textbf{\small Lorentzian factorization} method of Section~\ref{sec:factorization&screening}. This novel Coulomb branch factorization technique also offers a promising route to further understanding monopole-vortex correspondences in 3d $\mathcal{N}=2$ theories \cite{Aharony:1997bx,Intriligator:2013lca,Intriligator:2014fda}, when combined with earlier Higgs branch factorization results in \cite{Hwang:2012jh,Hwang:2015wna,Hwang:2017kmk,Crew:2020psc,Bullimore:2020jdq}.

On the \emph{matter near zones} the integrand has (gamma function) poles, so Poisson resummation is not allowed. These zones are conquered on a case-by-case basis; their analysis often simplifies since the $\beta$ scaling of the integrand there can be factored out. 

The road now seems paved for revisiting the 3d$\to$2d reduction of supersymmetric gauge theories by leveraging the Cardy limit of the 3d superconformal index, in close analogy with the 4d$\to$3d reductions studied in \cite{Aharony:2013dha,Aharony:2013kma,ArabiArdehali:2015ybk,ArabiArdehali:2019zac,ArabiArdehali:2024ysy,ArabiArdehali:2026kvt}. Although the pioneering work of \cite{Aharony:2017adm} has initiated the study of four-supercharge 3d$\to$2d reductions with the aid of partition functions, subsequent progress has so far remained comparatively preliminary relative to the now well-developed 4d$\to$3d literature.\footnote{We thank S.~Razamat for discussions on this point.}

\paragraph{Technical overview and limitations.}
We summarize here our main technical contributions and the scope of their applicability. The results naturally group into three themes.
(1) \emph{Non-perturbative saddle screening}, relevant to the exponential asymptotics on the second sheet.
(2) \emph{Multiscale analysis of the BPS moduli space}, relevant to the power-law asymptotics typical of the first sheet. (On the second sheet, by contrast, a single scale usually suffices; see the last column of Table~\ref{tab:patches_intro}.)
(3) \emph{Poisson resummation}, which is how we tame the sum over monopole sectors.
\begin{itemize}
    \item[(1)] We developed Lorentzian factorization tools (of potentially broader applicability) that allow demonstrating non-perturbative, or exponential, suppression of the gauge-enhanced saddles on the 2nd sheet of non-chiral gauge theories. We have exemplified our approach in the case of $\mathrm{SU}(2)$ SQCD with $N_f=3$ in Section~\ref{subsec:nonabelian}, where we have also demonstrated beyond-all-orders suppression of the infinite-flux saddles via factorization.\\
    \textbf{\small Limitation:} nonzero $k_{gg}$ or non-vectorlike matter changes the asymptotics of the integrand, so that the validity of the contour deformation \eqref{eq:cont_crossing} is no longer guaranteed. See Problem~\textbf{1}.
    
    \item[(2)] We introduced a decomposition scheme for dividing the BPS moduli-space arising in Coulomb branch localization. We get an outer patch, and various inner patches---themselves decomposed into near zones and intermediate zones. On the first sheet of non-chiral theories we conjecture that only the near zones and the outer patch may dominate the asymptotics, and have found their $\beta$-scaling at rank one; see \eqref{eq:cases_pp_nz}. On the first sheet of more general theories we conjecture that only the Gaussian subzone of the intermediate zones enter the competition (with the near zones and the outer patch), and have found their $\beta$-scaling at rank one assuming vectorlike matter and $k_{gg}\in2\mathbb{Z}$; see \eqref{eq:gaussian_rank1_gen_rep}.\\
    \textbf{\small Limitation:} near zones on the first sheet of chiral theories have not been treated systematically (the only such example that we treated above was the electric side of elementary mirror symmetry, for which we borrowed case-specific results from \cite{Benini:2012ui}); non-Gaussian parts of the intermediate zones also remain to be demonstrated negligible in general. See Problems~\textbf{4}, \textbf{5}, and \textbf{7}.

    \item[(3)] Regarding Poisson resummation, we have demonstrated how its application on the gauge patch turns the sum-integral expression for the 3d index into a sum of \emph{integrals over punctured surfaces}, amenable to the multidimensional stationary phase method. On the outer patch, the latter sum is interpreted as being over the winding sectors $w$ of the 2d dual photon multiplet. Since only finitely many of these sectors host stationary-phase loci, the sum over $w$ (unlike the original infinite sum over $m$) is not a hindrance to the asymptotic analysis.\\
    \textbf{\small Limitation:} on gauge inner patches (where there is gauge enhancement but no light charged chiral multiplets in the 2d EFT), the interpretation of the sum $\sum_w$ arising from Poisson resummation is not fully clear. See Problem~\textbf{6}.
\end{itemize}

\paragraph{Open problems.} We similarly divide some of the remaining problems into the same three categories.

\vspace{.1cm}
The following two problems concern \textbf{\small non-perturbative saddle screenings}.
\vspace{-.2cm}
\begin{itemize}
    \item \textbf{\small Problem~\textbf{1}.~Generalizing the block decomposition.} Generalize the factorization formula \eqref{eq:fully_factorized_intro} to $\mathrm{SU}(2)$ theories with a nonzero tree-level CS coupling. In particular, factorize the 2nd-sheet index of the electric side of the duality appetizer, in order to demonstrate the non-perturbative screening of its gauge-enhancing Bethe roots.
    \vspace{-.2cm}
    \item \textbf{\small Problem~\textbf{2}.~Screened saddles and EFT.} The full screening of the gauge-enhanced saddles on the 2nd sheet follows from the oddity of the integrand after factorization. This suggests that a $\mathbb{Z}_2$ anomaly (possibly mixed with a symmetry tied to factorization) underlies the cancelation. Identify that anomaly in the 2d $\mathcal{N}=(2,2)$ EFT. The anomaly might be an obstruction to 2d gauge-enhancement (or to 3d deconfinement).
\end{itemize}


\vspace{.1cm}
The next three problems concern \textbf{\small multiscale analysis of the BPS moduli-space}.
\vspace{-.2cm}
\begin{itemize}
    \item \textbf{\small Problem~\textbf{3}.~Patch competition on the first sheet and vortex-instantons.} On the first sheet of non-chiral theories, a puncture (in fact, the puncture's near zone) dominates over the punctured plane if the $S^2$ partition function of the near zone EFT is finite as its cut-offs $\Lambda,\,\Lambda_m$ are sent to infinity. At rank one, the convergence criterion was formulated in \eqref{eq:inner_div_criter_rank1} as $\mathrm{Tr}_{L^c_p}R<-2.$ Reformulate this as a criterion on the $R$-charge of the vortex-instanton on the 2d $\mathcal{N}=(2,2)$ Coulomb branch. Use that reformulation to generalize the result to higher ranks, and also to link the patch competition to the 2d gauge dynamics induced by vortex-instanton superpotentials \cite{Hori:2000kt}.\footnote{The connection between convergence of $S^2$ partition functions and the $R$-charge of 2d vortex-instantons would be analogous to the one between convergence of $S^3$ partition functions and the $R$-charge of 3d monopole-instantons (see e.g.~\cite{ArabiArdehali:2015ybk}). The link between patch competition on the first sheet of the 3d index and the 2d gauge dynamics induced by vortex-instanton superpotentials would be analogous to the one between patch competition on the first sheet of the 4d index and the 3d gauge dynamics induced by monopole-instanton superpotentials \cite{ArabiArdehali:2019zac,ArabiArdehali:2024ysy,Affleck:1982as}.}
    \vspace{-.2cm}
    \item \textbf{\small Problem~\textbf{4}.~Non-Gaussian zones.} Find an example where the index is dominated by a non-Gaussian subzone of an intermediate zone, or prove that is impossible.
    \vspace{-.2cm}
    \item\textbf{\small Problem~\textbf{5}.~Multiscale resolution of the $q$-Pochhammer symbol.} Obtain a full multiscale resolution of the estimate \eqref{eq:qPoch_master_estimate_body} for the $q$-Pochhammer symbol, up to $o(\beta^0)$ error, analogous to what \eqref{eq:ftoy_near}--\eqref{eq:ftoy_far} achieve for \eqref{eq:ftoy}.
\end{itemize}


\vspace{.1cm}
The last two problems concern \textbf{\small Poisson resummation and 2d duality}.
\vspace{-.2cm}
\begin{itemize}
    \item \textbf{\small Problem~\textbf{6}.~Poisson resummation on gauge inner patches and EFT.} As mentioned in Section~\ref{subsec:surface_integral}, Poisson resummation on the (complexified) outer patch corresponds to dualizing the 2d photon and then summing over the winding sectors of the dual-photon superfield. Gauge inner patches are also amenable to Poisson resummation. Since their effective gauge group is non-abelian, however, the corresponding 2d duality is less clear. It would be interesting to clarify this and make contact with the presumably related proposals in \cite{Aharony:2016jki,Gu:2018fpm}.
    \item \textbf{\small Problem~\textbf{7}.~Treating the punctures with GLSM $\to$ Landau-Ginzburg.} We have systematically treated the punctured $y$-plane via Poisson resummation. But our treatment of the punctures has not been fully systematic. We have focused only on their near zones and Gaussian zones, and have treated those on a case-by-case basis. GLSM/Landau-Ginzburg dualities might be an accurate and generalizable approach. It would be nice to systematically complement the Poisson resummation on the punctured $y$-plane with a GLSM/LG treatment of the punctures, at least in some examples. That may allow, among other things, a systematic lifting of GLSM/LG dualities to three dimensions.
\end{itemize}

\begin{acknowledgments}
    This project began as conversations between Z.~Komargodski, S.~Razamat, and the first author. AA would like to thank them, as well as A.~Cabo-Bizet and S.~Murthy for related discussions during the early stages of this work. We also thank M.~Ansari, P.~Benetti~Genolini,  C.~Closset, G.~Cuomo, A.~Deb, S.~Fallahdoost, D.~Gang, J.~Hong, Z.~Komargodski, H.~Rosengren, M.~Sacchi, S.~Shao, and A.~Sharon for related conversations and/or comments on the draft. SF would like to thank M. Litvinov for discussions on the inter-play of branch cuts and saddle point analysis. The work of MB is supported by the FRQNT doctoral training scholarship.
\end{acknowledgments}

\appendix

\section{Some asymptotic analysis}\label{app:AA}

\subsection{Estimating $\mathrm{Li}_2(z\!\to\!1)$, $\Gamma(z\!\to\!\infty)$, and $(z;q\!\to\!1)$}\label{app:estimates}

\subsection*{Asymptotics of $\mathrm{Li}_2(z)$ as $z\to1$}\label{app:Li2Asy}

In the main text we need the small-$x$ asymptotic of $\mathrm{Li}_2(e^{-x})$.
Since $\mathrm{Li}_2'(z)=-\log(1-z)/z$ is singular at $z=1$, naive Taylor expansion is not effective here. We use instead a technique described in \cite{zagier2006mellin}.

Noting $\mathrm{Li}_2(z)=\sum_{n=1}^\infty z^2/n^2$, we write
\begin{equation}
    \frac{\mathrm{Li}_2(e^{-x})}{x^2}=\sum_{n=1}^\infty\left(\frac{1}{(nx)^2}+\frac{e^{-nx}}{(nx)^2}-\frac{1}{(nx)^2}\right)=\frac{\pi^2}{6x^2}+\sum_{n=1}^\infty f(nx),\label{eq:Li2subSum}
\end{equation}
where
\begin{equation}
    f(x)=\frac{e^{-x}-1}{x^2}.
\end{equation}
We have added and subtracted $1/(nx)^2$ in the summand in the middle of \eqref{eq:Li2subSum} in order for the resulting $f$ to have the small-$x$ expansion:
\begin{equation}
    f(x)=\frac{b_{-1}}{x}+b_0+\mathcal{O}(x),
\end{equation}
so we can use the result in \cite{zagier2006mellin} that
\begin{equation}
    \sum_{n=1}^\infty f(nx)=\frac{1}{x}\left(b_{-1}\log(1/x)+I^\ast_f\right)-\frac{b_0}{2}+\mathcal{O}(x),\label{eq:ZagierLemma}
\end{equation}
with $I^\ast_f:=\int_0^\infty \big(f(y)-b_{-1}e^{-y}/y\big)\,\mathrm{d}y.$

Combining \eqref{eq:ZagierLemma} and \eqref{eq:Li2subSum}, and using $b_{-1}=-1$, $b_0=1/2$, $I^\ast_f=-1,$ we get
\begin{equation}
    \mathrm{Li}_2(e^{-x})=\frac{\pi^2}{6}+x\left(\log(x)-1\right)-\frac{x^2}{4}+\mathcal{O}(x^3).\label{eq:Li2_est_app}
\end{equation}

\subsection*{Asymptotics of $\Gamma(z)$ as $z\to\infty$}\label{app:gammaAsy}

We use
\begin{equation}
    \Gamma(z)=\sqrt{2\pi}\,z^{\,z-\frac{1}{2}}\,e^{-z}\cdot e^{\,\mu(z)},
\end{equation}
with
\begin{equation}
    \mu(z)=\sum_{k=0}^\infty\left(\big(z+k+\frac{1}{2}\big)\ln(1+\frac{1}{z+k})-1\right)=\mathcal{O}\Big(\frac{1}{|z|}\Big),
\end{equation}
valid for $|\mathrm{arg}\,z|<\pi.$ We immediately obtain
\begin{equation}
    \Gamma(z)=\sqrt{2\pi}\,z^{\,z-\frac{1}{2}}\,e^{-z}\,\big(1+\mathcal{O}(1/|z|)\big).\label{eq:gamma_asy}
\end{equation}

\subsection*{Asymptotics of $(z;q)$ as $q\to1$}\label{app:PochAsy}

Our starting point is the identity \cite{ArabiArdehali:2026ddr}:
\begin{equation}
\begin{split}
    \frac{1}{\prod_{r=0}^\infty (1-e^{-\beta X}e^{-\beta r})}=\text{p.v.}\prod_{n\in\mathbb
    {Z}}\bigg(\frac{\Gamma(X+\frac{2\pi i n}{\beta})}{\sqrt{2\pi}}\,\cdot\,& (X+\frac{2\pi i n}{\beta})^{-(X+\frac{2\pi i n}{\beta}-\frac{1}{2})}\cdot e^{X+\frac{2\pi i n}{\beta}}\bigg)\,\\
    & \ e^{\frac{\mathrm{Li}_2(e^{-\beta X})}{\beta}}\, (1-e^{-\beta X})^{-1/2}\, \,e^{-\beta/24},\label{eq:invPochProdHjalmar}
\end{split}
\end{equation}
where all the multi-valued functions on the RHS are evaluated on their principal sheet, and p.v. stands for the principal value: $\text{p.v.}\prod_{n\in\mathbb{Z}}:=\lim_{\Lambda\to\infty}\prod_{n=-\Lambda}^\Lambda.$

Applying \eqref{eq:gamma_asy} to all but one factor on the RHS of \eqref{eq:invPochProdHjalmar} we obtain\footnote{This approach to the $q\to1$ asymptotics of $(z;q)$ is better suited to our purposes than the one in \cite{Garoufalidis:2018qds}, since we do not need to impose constraints analogous to their $|w|<1$ or $\nu\beta=o(1).$}
\begin{equation}
    (q^X;q)= \left(\frac{\sqrt{2\pi}}{\Gamma(\tilde{X})}\cdot e^{\,\tilde X\,(\log \tilde X-1)}\cdot e^{-\frac{1}{2}\log\tilde X}\right)\,
    \times \left( e^{-\frac{\mathrm{Li}_2(e^{-\beta X})}{\beta}}\, (1-e^{-\beta X})^{1/2} \right)\times \big(1+ \mathcal{O}(\beta)\big).\label{eq:master_estimate_app}
\end{equation}
Here $\tilde{X}$ stands for the representative of $X$ inside the strip $-\frac{\pi}{\beta}\le\mathrm{Im}X<\frac{\pi}{\beta},$ via a vertical shift by an integer multiple of $\frac{2\pi i}{\beta}.$

The right-hand side of \eqref{eq:master_estimate_app} is still too complicated: it has different small-$\beta$ expansions for e.g. $\beta X$ of order one, versus $\beta X$ of order $\beta$ (or more generally $\beta \tilde X$ of order $\beta$). Simplifications of it arise in two regimes of particular interest.

The first is the \emph{far} regime $|\tilde X|\gg1$. Then the first parenthesis on the RHS of the estimate \eqref{eq:master_estimate_app} tends to unity according to the Stirling approximation $\Gamma(x)\xrightarrow{x\to\infty}\sqrt{2\pi}\,x^{\,x-\frac{1}{2}}\,e^{-x}$, and we get the simpler estimate\footnote{See \eqref{eq:outer_est_Poch_main} for a further simplification of this in cases where $X$ contains an $O(\frac{1}{\beta})$ as well as an $O(\beta^0)$ piece.}
\begin{equation}
    (q^X;q) \approx e^{-\frac{\mathrm{Li}_2(e^{-\beta X})}{\beta}}\, (1-e^{-\beta X})^{1/2},\qquad\text{\textbf{far} (Gaussian zone and outer patch)}.\label{eq:outer_estimate_app}
\end{equation}
The right-hand side now is of a form amenable to standard stationary phase or saddle point techniques, if we change variables from $X$ to $e^{-\beta X}$.

The second is the \emph{near} regime where $|\tilde{X}|$ is of order one (less than some $\Lambda$), and hence $|\beta\tilde{X}|$ is of order $\beta$. Then on the one hand \eqref{eq:Li2_est_app} gives $    e^{-\frac{1}{\beta}\mathrm{Li}_2(e^{-\beta X})}\approx e^{-\frac{\pi^2}{6\beta}-X\,(\log X-1)-X\log\beta }$,
and on the other hand $(1-e^{-\beta X})\approx \beta X$, both valid for $X$ of order one. We then deduce from \eqref{eq:master_estimate_app} that
\begin{equation}
    ( q^X;q)\xrightarrow{\beta\to0}e^{-\frac{\pi^2}{6\beta}}\,\beta^{\frac{1}{2}-\tilde{X}}\,\frac{\sqrt{2\pi}}{\Gamma(\tilde{X})}\, \big(1+o(\beta^0)\big),\qquad\qquad\text{\textbf{near} (near zone)}.\label{eq:qPochNearEst_app}
\end{equation}
This is particularly useful if the $\beta^{-\tilde X}$ factor cancels against similar contributions from other chiral multiplets, so that the dependence on $\beta$ and $\tilde{X}$ completely disentangle.

\subsection{The method of stationary phase for multidimensional integrals}\label{app:stationary_phase}

This appendix closely follows Section~5.7 in \cite{miller2006applied}.

Consider an integral
\begin{equation}
    F(\beta):=\int e^{\frac{i I(\boldsymbol{x})}{\beta}}g(\boldsymbol{x}),
\end{equation}
over a nice enough measurable domain in $\mathbb{R}^d$, with $I(\boldsymbol{x})$ and $g(\boldsymbol{x})$ sufficiently smooth functions and $I(\boldsymbol{x})$ real-valued.

First assume that the integration domain (in our applications either a punctured $u$-plane, or a cover thereof) does not contain any stationary point of $I(\boldsymbol{x})$. Then repeated use of the Divergence Theorem (higher-dimensional integration by parts) yields \cite{miller2006applied}:
\begin{equation}
    F(\beta)\sim\sum_{n=1}^{\infty} ({-i\beta})^n \int_{\partial}\  e^{\frac{i I(\boldsymbol{x})}{\beta}}\ \frac{g_{n-1}(\boldsymbol{x})\,\mathbf{n}(\boldsymbol{x})^T\,\nabla I(\boldsymbol{x})}{|\nabla I(\boldsymbol{x})|^2},\label{eq:no_stationary_asy}
\end{equation}
where $g_{n-1}(\boldsymbol{x})$ are defined iteratively as
\begin{equation}
    g_{0}(\boldsymbol{x}):=g(\boldsymbol{x})\,,\qquad g_{n}(\boldsymbol{x}):=-\nabla\cdot\Big[\frac{g_{n-1}(\boldsymbol{x})\nabla I(\boldsymbol{x})}{|\nabla I(\boldsymbol{x})|^2}\Big]\,,
\end{equation}
and $\mathbf{n}(\boldsymbol{x})$ is an exterior normal unit vector.

Even though the leading asymptotic of $F(\beta)$ appears to be $\mathcal O(\beta)$ from the $n=1$ term in \eqref{eq:no_stationary_asy}, the boundary integral itself ought to be treated via a stationary point analysis. Often the latter yields an additional suppression by $\beta^{\frac{d-1}{2}}$ due to the presence of isolated stationary points on the boundary (see e.g. page~185 in \cite{miller2006applied}, or Eq.~\eqref{eq:stationary_asy} below with $d\to d-1$ to go to the boundary). In such cases, the leading asymptotic becomes
\begin{equation}
    F(\beta)=\beta^{\frac{d-1}{2}}\mathcal{O}(\beta)=\mathcal{O}(\beta^{\frac{d+1}{2}})\,.\label{eq:no_stationary_asy_d}
\end{equation}
In our rank-1 examples where $d=2$, this yields $F(\beta)=\mathcal{O}(\beta^{3/2}).$

Now assume there exists a unique isolated stationary-phase point in the interior of the domain of integration. Without loss of generality we can take it to be at the origin $\boldsymbol{x}_0$. We also assume, for simplicity (and all our relevant examples in the main text are simple in this sense), that at the isolated stationary-phase point the Hessian matrix $\mathbf{H}$ of $I(\boldsymbol{x})$ has $q$ strictly positive eigenvalues and $q':=d-q$ strictly negative eigenvalues. The leading asymptotic contribution from near the origin is then \cite{miller2006applied}:
\begin{equation}
    F(\beta)= e^{i\pi(q-q')/4}\,\frac{e^{\frac{iI(\boldsymbol{0})}{\beta}}}{\sqrt{|\mathrm{det}\,\mathbf{H}|}}\,(2\pi\beta)^{d/2}\, g(\boldsymbol{0})\,\big(1+\mathcal{O}(\beta)\big),\label{eq:stationary_asy}
\end{equation}
up to a sign arising from the determinant of the diagonalization transformation so that the positive eigenvalues come first.

The case of multiple isolated stationary-phase points in the interior of the domain is a rather trivial generalization.

\bibliographystyle{JHEP}
\bibliography{refs}

\end{document}